\documentclass[aps,pre, superscriptaddress, altaffilletter, twocolumn]{revtex4-1}
\usepackage{hyperref}
\usepackage{graphicx,epsfig,subfig}
\usepackage{epstopdf}
\usepackage{amssymb,amsthm,amssymb}
\usepackage{color}
\usepackage{caption}
\usepackage{amsmath}
\usepackage{hyperref}
\usepackage{soul}
\usepackage{ulem}
\usepackage{wasysym}
\usepackage{graphicx,epsfig,tikz}
\theoremstyle{definition}

\begin{document}
\title{Nonmodal stability analysis of miscible viscous fingering with non-monotonic viscosity profiles}
\author{Tapan Kumar Hota}
\author{Manoranjan Mishra}
\altaffiliation[Also affiliated to:~]{Department of Chemical Engineering, \\ Indian Institute of Technology Ropar,  Rupnagar - 140001, Punjab, India}
\affiliation{Department of Mathematics, \\ Indian Institute of Technology Ropar, Rupnagar-140001, Punjab, India}


\date{\today}

\begin{abstract}
A non-modal linear stability analysis (NMA) of the miscible viscous fingering in a porous medium is studied for a toy model of non-monotonic viscosity variation. The onset of instability and its physical mechanism are captured in terms of the singular values of the propagator matrix corresponding to the non-autonomous linear equations. We discuss two types of non-monotonic viscosity profiles, namely, with unfavorable (when a less viscous fluid displaces a high viscous fluid) and with favorable (when a more viscous fluid displaces a less viscous fluid) end-point viscosities. A linear stability analysis yields instabilities for such viscosity variations. Using the optimal perturbation structure, we are able to show that an initially unconditional stable state becomes unstable corresponding to the most unstable initial disturbance. In addition, we also show that to understand the spatiotemporal evolution of the perturbations it is necessary to analyse the viscosity gradient with respect to the concentration and the location of the maximum concentration $c_m$. For the favorable end-point viscosities, a weak transient instability is observed when the viscosity maximum moves close to the pure invading or defending fluid. This instability is attributed to an interplay between the sharp viscosity gradient and the favorable end-point viscosity contrast. Further, the usefulness of the non-modal analysis demonstrating the physical mechanism of the quadruple structure of the perturbations from the optimal concentration disturbances is discussed. We demonstrate the dissimilarity between the quasi-steady-state approach and NMA in finding the correct perturbation structure and the onset, for both the favorable and unfavorable viscosity profiles. The correctness of the linear perturbation structure obtained from the non-modal stability analysis is validated through nonlinear simulations. We have found that the nonlinear simulations and NMA results are in good agreement. In summary, a non-monotonic variation of the viscosity of a miscible fluid pair is seen to have a larger influence on the onset of fingering instabilities, than the corresponding Arrhenius type relationship. 
\end{abstract}
\maketitle
\section{Introduction}\label{sec:introduction_chap5}
Flow stability in displacement processes in porous media has been the subject of numerous past investigations in the petroleum industry, solute transportation in aquifers and packed bed regeneration, to name a few. In particular, when a fluid of lower viscosity displaces a fluid of higher viscosity in a porous medium, the interface between the fluids becomes unstable and the resulting displacement pattern is known as viscous fingering (VF) \citep{Saffman1958, Homsy1987}. In the area of miscible displacement, there have been many theoretical and experimental studies addressing the onset of instability and the subsequent growth of unstable disturbances (viscous fingers). Several studies have been concerned with the determination of conditions leading to the onset of instability, essentially employing modifications of \cite{Tan1986} theory for miscible viscous instability in porous media. Further, attempts are also made to develop simplified predictive schemes for the description of finger growth \citep{Kim2012, Pramanik2013}.

The vast majority of previous studies \citep{Homsy1987, Ben2002, Pritchard2009, Mishra2010, Hejazi2010} on the miscible viscous fingering have focused on a monotonic viscosity-concentration relationship of Arrhenius type, that is, $\mu(c) = \exp(Rc)$. Here $\displaystyle R = \ln(\mu_2/\mu_1)$ is the log-mobility ratio where $\mu_1$ and $\mu_2$ correspond to the viscosity of the displacing and displaced fluids, respectively. For such monotonic profiles, it is well known that the instability criteria  for an unfavourable viscosity contrast (when less viscous fluid displaces a high viscous fluid) is determined from the end-point viscosities, equivalently when $R>0$. However, in practice the monotonic relationship may not always represents close approximation of the miscible fluid combinations. For example, exploratory investigations of enhanced oil recovery process have employed slugs of alcohol or alcohol mixtures that separate the oil from the water, which is used as the driving fluid \citep{Latil1980}. Since different kinds of alcohol are generally miscible with each other, as well with water and oil, the dependence of the viscosity of these mixtures on the respective concentrations will affect the overall dynamics of the displacement process. In general, the viscosity, $\mu(c)$, will depend on the fluid pair employed and can neither be linear nor exponential. Some fluid pairs like isopropyl alcohol and water employed in laboratory experiments have a non-monotonic viscosity-concentration relationship \citep{Weast1990}.

Recently, it has been observed experimentally \cite{Riolfo2012, Nagatsu2014} and theoretically \cite{Hejazi2010} that due to a miscible chemical reaction there could be a buildup of non-monotonic viscosity profiles, i.e. the relationship between viscosity $\mu$ and concentration $c$ need not be monotonic, rather display a maximum viscosity at the intermediate concentration value. Similarly, in the application of chromatographic column \cite{Haudin2016} shown that a non-monotonic viscosity profile can be observed due to non-ideal mixing properties of certain alcohols in a porous medium. Further, some enhanced oil recovery schemes such as  water-alternating-gas (WAG) have the potential to introduce mobility non-monotonicities by exploiting the dependence of the oil's mobility on the amount of dissolved gas. Blunt and Christie\cite{Blunt1991} have simulated a tertiary  WAG process in which the oil saturation profile and consequently the mobility profile are non-monotonic. Hickernell and Yortsos\cite{Hickernell1986} shown that in the absence of physical dispersion, any rectilinear miscible displacement with a locally unfavourable viscosity profile is unstable. Later, this observation is confirmed by Chikhliwala \textit{et al.}\cite{Chikhliwala1988}, who analyze the immiscible displacements considering the non-capillary displacements that are equivalent to miscible displacements without physical dispersion. Bacri \textit{et al.}\cite{Bacri1992a} extend these investigations by including the dispersion effect. For the step profile associated with time $t = 0$, they identify a single stability parameter with arbitrary viscosity-concentration profile. Manickan and Homsy \cite{Manickam1993} pointed out that in the case of non-monotonic viscosity-concentration profiles, the stability of the system is depends on the end-point derivatives of the viscosity-concentration relationship. They performed the linear stability analysis based on a quasi-steady-state approximation (QSSA) and noted that at early times QSSA is questionable where the base state changes rapidly. Later, Pankiewitz and Meiburg \cite{Pankiewitz1999} analyze the influence of a non-monotonic viscosity-concentration relationship on miscible displacements in porous media for radial source flows and the quarter five-spot configuration. Schafroth \textit{et al.}\cite{Schafroth2007} also extended the work of Manickan and Homsy \cite{Manickam1993} for miscible displacement in a Hele-Shaw cell with Stokes equation governing the flow. Further, such non-monotonic viscosity profiles can also typically be obtained in the study of reaction diffusion problem \cite{Hejazi2010, Nagatsu2011}, double-diffusion problems \cite{Pritchard2009, Mishra2010} and effect of nano-particles in miscible VF \cite{Dastvareh2017}. Moreover, a detailed discussions of the non-monotonic viscosity profile and stabilization in a radial Hele-Shaw cell has been presented by Li-Cheih Wang \cite{Wang2014}. He presented a non-linear simulation to study the radial injection-driven miscible flow and lifting radial Hele-Shaw cell, both with the monotonic and non-monotonic viscosity profile.

Although considerable analysis have been made for monotonic viscosity-concentration profiles, it is evident from the existing literature that, only few attempts \citep{Manickam1993, Manickam1994, Kim2011} have been made to address the linear stability analysis for non-monotonic viscosity-profile in miscible rectilinear displacements. Utilizing the self-similarity in the concentration base-state, Kim and Choi\cite{Kim2011} employed an eigenanalysis in a self-similar coordinate by which transient nature of the base-state can be removed, albeit, the linearized operator remain time-dependent. But, the eigenanalysis presented by Kim and Choi\cite{Kim2011} fails to demonstrate the quadruple structure and hence the physical mechanism of perturbation growth which was shown by Manickan and Homy\cite{Manickam1993} by the means of the vorticity perturbations. The methods demonstrated in the works of Manickam and Homsy\cite{Manickam1993} and Kim and Choi\cite{Kim2011} neither consider the energy amplification nor the effect of viscosity-concentration parameters on the perturbation structures. Furthermore, for different parameters in viscosity-concentration relationships proposed by Manickam and Homsy\cite{Manickam1993} (\textit{viz.}, end-point viscosity, maximum concentration and maximum viscosity) one may result in different vorticity configurations. It has been observed that the global dynamics of the fingers and the entire displacement, will be strongly affected by this base-flow vorticity \cite{Manickam1993,Manickam1994}. Thus, it is important to realize that the base-flow vorticity is a function of the viscosity profile itself. Hence, when analysing displacements the nature of the viscosity profile is expected not just to affect the fingering process directly, but also indirectly through the base-flow vorticity. Thus, it is necessary to carry a  linear stability analysis which can capture the disturbance structure and  determine the onset of instability accurately. Moreover, in miscible VF, the governing linearized equations are time-dependent. Owing to the non-autonomous nature of the linear stability matrix, we have adopted the non-modal analysis based on propagator matrix approach. The stability of the dynamical system is then described in terms of the singular values of the propagator matrix. This approach can address the time evolving modes and their spatial structure more appropriately than QSSA or eigen-analysis. Hence, our goal is to illustrate the advantages of NMA and the physical mechanism of stability based on the optimal structure of the concentration perturbations. The novelty of the present analysis is that we can determine the onset \& describe the effects of the non-monotonic viscosity-concentration parameters  on stability without invoking the stream function-vorticity formulation.

The organization of the paper is as follows. In section \ref{sec:formulation_chap5}, the governing equation and linearized perturbation equations are derived for a general viscosity-concentration profile. Then, we describe a parametric study that demonstrates the non-monotonic dependence of the viscosity-concentration relations. To conclude this section, we have summarized the non-modal analysis. In section \ref{sec:results_chap5}, the nonmodal stability results are discussed and a comparison is made with nonlinear simulations and QSSA in self-similar coordinate followed by conclusion in section \ref{sec:conclusion_chap5}.

\section{Mathematical formulation}\label{sec:formulation_chap5}
\begin{figure}
	\centering
	\includegraphics[width=3.5in, keepaspectratio=true, angle=0]{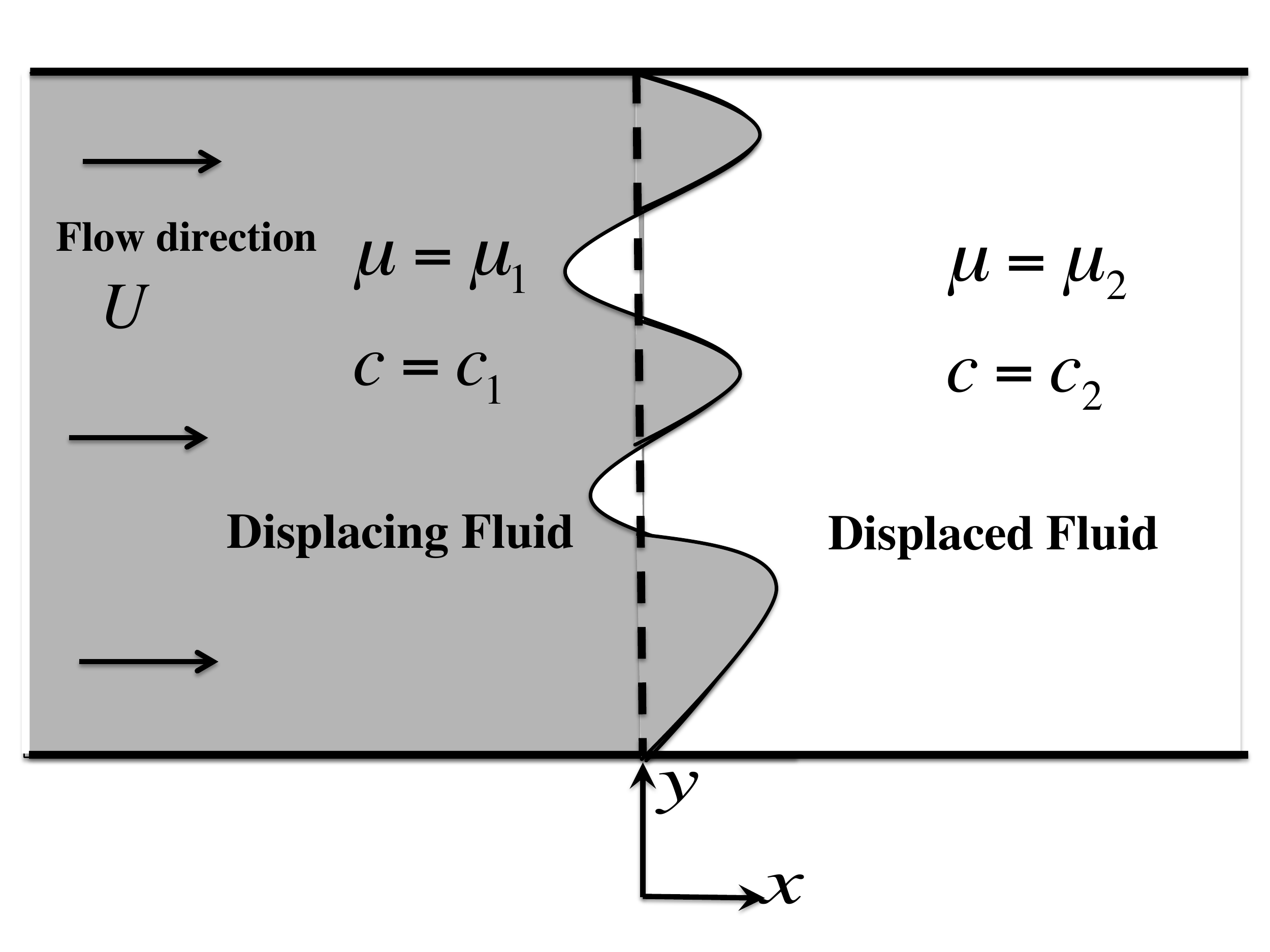}
	\caption{Schematic of the flow configuration with coordinate system. Initially the interface is located at $x=0$ shown as dashed line.}
	\label{fig:Schematic_chap5}
\end{figure}
Consider the miscible displacement flow in a porous media as shown in Fig. \ref{fig:Schematic_chap5}. The fluid of viscosity $\mu_1$ displaces a fluid of viscosity, $\mu_2$ with a uniform velocity $U$. The fluids are assumed to be Newtonian, non-reactive and neutrally buoyant, and the porous media is homogeneous with a constant permeability and isotropic dispersion. Fluid flow and mass transport in the porous medium are governed by the equations for conservation of
mass, conservation of momentum in the form of Darcy's law, and volume averaged mass balance equation in the form of a convection-diffusion equation and are given by
\begin{eqnarray} 
\label{eq:cont_eqn_ND}
&& \nabla \cdot \vec{u} = 0,\\ 
\label{eq:Darcy_1_ND}
&& \nabla p = -\mu\vec{u}, \\ 
\label{eq:convec_diffuse_ND}
&& \frac{\partial c} {\partial t} + \vec{u} \cdot \nabla c = D \nabla^2 c,\\
\label{eq:visco_ND}
&& \mu = \mu(c), 
\end{eqnarray} 
where $\vec{u} = (u,v)$ is the two-dimensional Darcy velocity, $c$ is the solute concentration, $p$ is the fluid pressure, $\mu$ is the dynamic viscosity and $D$ is the isotropic dispersion coefficient taken throughout to be constant. At the initial time $t=0$, the concentration and viscosity of the displacing fluid are $c_1$ and $\mu_1$, respectively whereas the displaced  fluid have concentration and viscosity as $c_2$ and $\mu_2$, respectively.

In the present work, we have used the Darcy’s law to describe the flow in a porous media
or equivalently Hele-Shaw flow with-in thin gap approximation. In particular, equation \eqref{eq:Darcy_1_ND} is obtained by averaging the parabolic velocity profile in between the parallel plates. The mathematical analysis of Darcy’s law in thin regions have been studied extensively, e.g.for averaging of creeping flow \cite{Bayada1986} and for averaging of Navier-Stokes system \cite{Nazarov1990}. Validity of Darcy’s law, in continuum modelling of the flow in porous media, implicitly assumed that the viscosity varies over the macro-scopic scale and can be treated as constant on the micro-scale on which the permeability is computed. Similarly Zick and Homsy \cite{Zick1982} observed that the viscosity variation is slow relative to the grain size, so that the force is determined to a good approximation by a constant viscous stress. In addition Nagatsu and De Wit \cite{Nagatsu2011} and Riolfo \textit{et al.} \cite{Riolfo2012} successfully shown that the experimental findings are meticulously agreeing with the numerical simulation of reaction-
driven viscous fingering based on the Darcy’s law model.

In order to make the governing equations dimensionless, characteristic scales have to be introduced. We note that the set of equations \eqref{eq:cont_eqn_ND}-\eqref{eq:visco_ND} involves neither a characteristic time-scale nor a length-scale; so the equations are made dimensionless using diffusive scaling, i.e., we considered the characteristic length $D/U$ and time $D/U^2$, where $D$ is the isotropic dispersion coefficient. Thus the nondimensional form of equations that govern the two-dimensional flow in a reference moving with the constant injection velocity $U$ are described by  \citep{Manickam1993}
\begin{eqnarray} 
\label{eq:cont_eqn}
& & \nabla \cdot \vec{u} = 0,\\ 
\label{eq:Darcy_1}
& & \nabla p = -\mu(c) (\vec{u} + \hat{i}), \\ 
\label{eq:convec_diffuse}
& & \frac{\partial c} {\partial t} + \vec{u} \cdot \nabla c = \nabla^2 c, 
\end{eqnarray}
where velocity, concentration, viscosity  and  pressure are nondimensionalized with $U$, $\displaystyle \frac{c-c_1}{c_2 - c_1}$, $\mu_1$,  and  $p/\mu_1 D$, respectively. Here $\hat{i}$ is the unit vector along the main flow direction, i.e., $x$ direction.\\

The initial and boundary conditions associated with the coupled equations \eqref{eq:cont_eqn}-\eqref{eq:convec_diffuse} are given by  \citep{Nield1992}\\
Initial conditions: 
\begin{eqnarray}
\label{eq:IC_1}
& & \vec{u}(x, y, t = 0) = (u, v)(x, y, t=0)= (0,0), \\
\label{eq:IC_2}
& & \mbox{and} ~\forall y, ~~ c(x, y, t=0) = \begin{cases}
1, &  x < 0\\
0, &  x \geq 0.
\end{cases}
\end{eqnarray}  
Boundary conditions:
\begin{eqnarray}
\label{eq:BC_1}
& & \vec{u} = (0, 0), ~~~ \frac{\partial c}{\partial x} \rightarrow 0, ~~~ |x| \rightarrow \infty, \\
\label{BC_2}
& &   u = 0, ~~ \frac{\partial v}{\partial y} \rightarrow 0,~~~\frac{\partial c}{\partial y} \rightarrow 0, ~~~ |y| \rightarrow \infty,
\end{eqnarray} 
where $u$ and $v$ are the axial and transversal-component of two-dimensional velocity $\vec{u}$. Further, the coupled equations \eqref{eq:cont_eqn}-\eqref{eq:convec_diffuse} admit the following transient base state
\begin{eqnarray}\label{eq:basecon_chap5}
\vec{u}_b = \vec{0}, c_b = \frac{1}{2}\text{erfc}\left(\frac{x}{2\sqrt{t}}\right), p_b=-\int_{-\infty}^{x} \mu_b(s,t)ds, 
\end{eqnarray}
by solving the following equations
\begin{eqnarray}\nonumber
\frac{\partial p_b}{\partial x} = - \mu ~~ \mbox {and} ~~ \frac{\partial c_b}{\partial t} = \frac{\partial^2 c_b}{\partial x^2}.
\end{eqnarray}
Here erfc$(\cdot)$ is the complimentary error function and the subscript $b$ stands for the base-state. 

\subsection{Non-monotonic viscosity relationship}\label{subsec:nonmon_visco_chap5}
The exact nature of the of relationship between the dynamic viscosity $\mu$ and concentration $c$ will depend on the particular combination of fluids under consideration \citep{Pankiewitz1999,Haudin2016}. It has been observed that there exists alcohol-water pair, for which viscosity $\mu(c)$ can achieve a maximum at some intermediate local concentration composition $c$ \citep{Manickam1993, Manickam1994, Schafroth2007, Haudin2016}. In other words, the flow develops a potentially unstable region followed downstream by a potentially stable region or vice-versa. In this scenario, the stable region acts as a barrier to the growth of  fingers, thereby providing a mechanism to control the VF. Hence, a fundamental understanding of the finger propagation in flows with non-monotonic viscosity profiles  is essential to develop methodologies aimed at controlling the growth of viscous fingers. In order to understand the influence of non-monotonic viscosity profiles, we focused on the non-monotonic viscosity-concentration model proposed by \cite{Manickam1993}. 

In order to allow comparisons with earlier studies of rectilinear displacements, we employ the non-monotonic class of viscosity-concentration profiles used by Manickam and Homsy\cite{Manickam1993, Manickam1994} and Pankiewitz and Meiburg\cite{Pankiewitz1999}. The viscosity-concentration profiles are sine functions  modified through a sequence of transformations which is well suited for the non-monotonic profiles of alcohol mixtures \citep{Weast1990} and given by the expressions
\begin{eqnarray}
\label{eq:visco-concen_chap5}
& & \mu(c) = \mu_m \sin(\zeta),\\
& & \zeta  = \zeta_0 (1 - \eta) + \zeta_1 \eta, \\
&& \eta = \frac{(1+a)c}{1+ac},
\end{eqnarray}
where
\begin{equation}\label{eq:constants_chap5}
\left.\begin{aligned}
\zeta_0 &= \arcsin (\alpha / \mu_m),\\
\zeta_1 &= \pi - \arcsin(1/\mu_m),\\
a &= \frac{c_m - \eta_m}{c_m(\eta_m -1)}, \\
\eta_m &= \frac{\pi/2 - \zeta_0}{\zeta_1 - \zeta_0}, ~~ \mbox{and} ~~ \alpha = \mu_2/\mu_1.
\end{aligned}
\right\}
\qquad 
\end{equation}
The transformations are defined in such a way that the end point values are $\mu(0) = \alpha$, $\mu(1) = 1$ and attains a maximum value $\mu_m$ at $c=c_m$. 
\begin{figure}
	\centering
	(a)\hspace{1.5in}(b)\\
	\includegraphics[width=1.6in, keepaspectratio=true, angle=0]{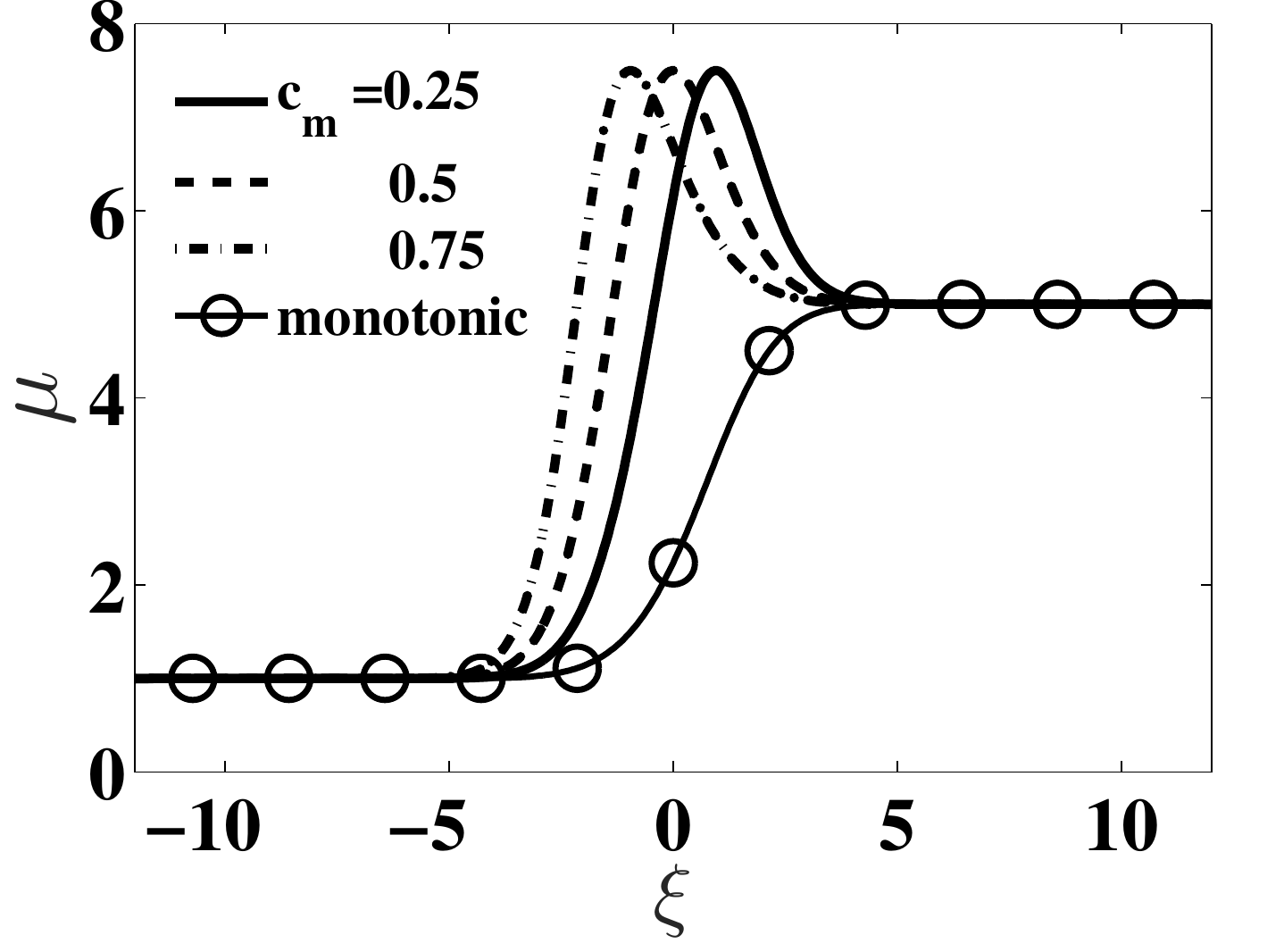}
	\includegraphics[width=1.6in, keepaspectratio=true, angle=0]{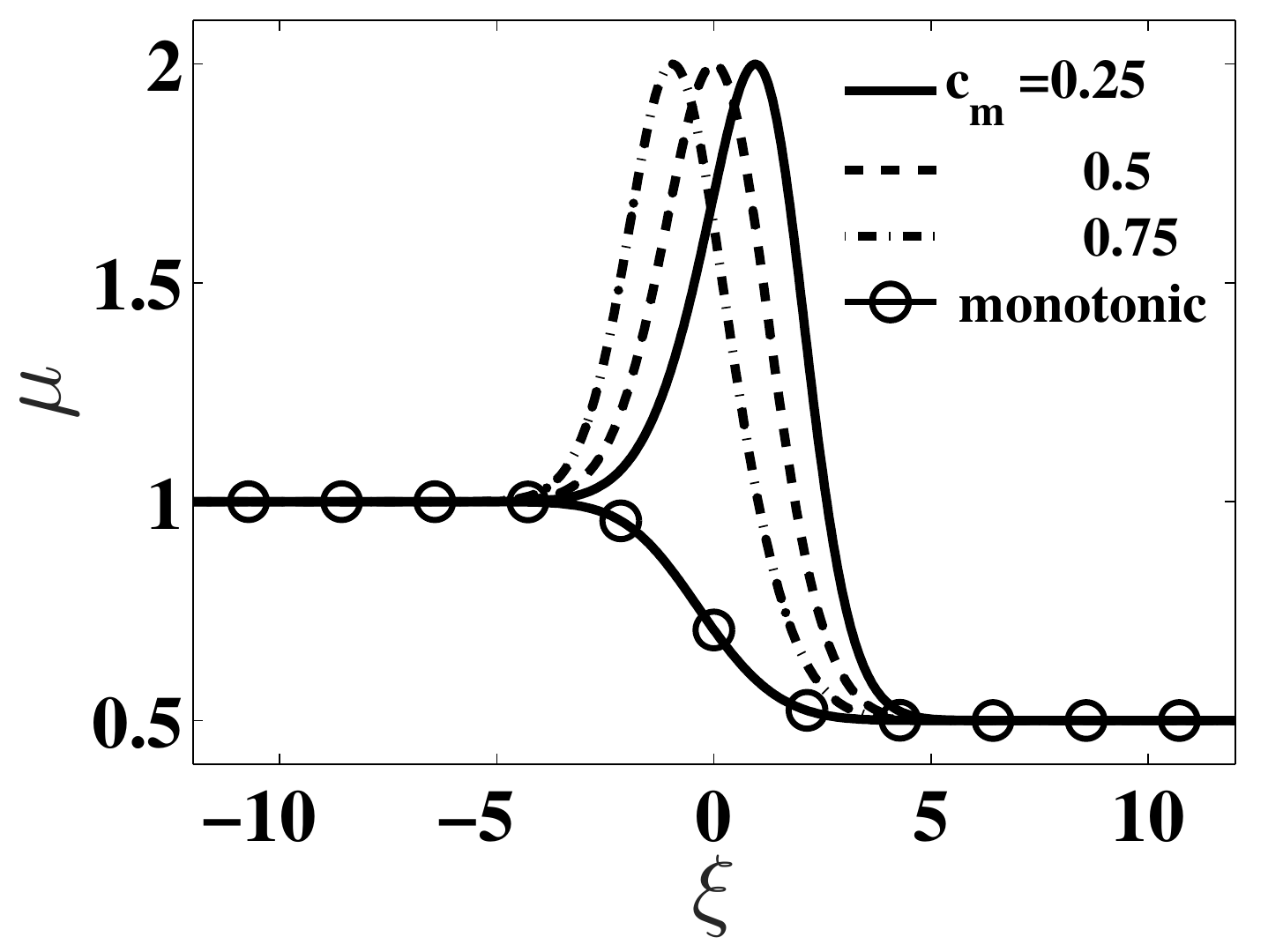}\\
	\caption{Spatial variation of the viscosity relation equation \eqref{eq:visco-concen_chap5} in a self-similar coordinate $\xi = x/\sqrt{t}$ for (a) $ \alpha =5,  \mu_m = 7.5$ and (b) $ \alpha =0.5,  \mu_m = 2$. The monotonic viscosity is $\mu(c) = \exp(\ln(\alpha)(1- c))$.}
	\label{fig:visco-concen_chap5}
\end{figure}

The non-montontic profile of dynamic viscosity is characterized by a family of curves described by three parameters, namely, $\alpha, c_m$ and $\mu_m$. The parameter $\alpha$ is the ratio of the end-point viscosities, i.e., $\alpha = \mu_2/\mu_1$, the traditional measure of stability in viscous fingering. In particular, for $\alpha < 1$ the flow is said to have a favorable viscosity contrast, and when $\alpha > 1$, the flow is said to have an unfavorable viscosity contrast. Figs. \ref{fig:visco-concen_chap5}(a) and (b) demonstrate the spatial variation of $\mu(c)$ for $\alpha=5, \mu_m=7.5$ and $\alpha =0.5, \mu_m = 2$, for different values of $c_m$. It can be noted that for $c_m > 0.5$, $\mu_m$ is located closer to the displacing fluid, while it is near to the displaced fluid, if $c_m < 0.5$. Manickam and Homsy\cite{Manickam1993} showed that the stability criterion for non-monotonic profile [equation \eqref{eq:visco-concen_chap5}] is determined from a parameter $\chi $ that relates the end point slopes of the viscosity profile which defined as
\begin{eqnarray}\label{eq:chi_chap5}
&& \chi =  -\left(\displaystyle \frac{ \displaystyle \frac{\mbox{d} \displaystyle \mu}{\mbox{d}c}\bigg|_{ \displaystyle c=0} + \displaystyle \frac{\mbox{d} \displaystyle\mu}{\mbox{d}c}\bigg|_{\displaystyle  c=1}}{\displaystyle  \alpha + 1}\right).
\end{eqnarray}
In particular, the system is stable if $\chi <0$, otherwise it is unstable. For the monotonic case, i.e., $\mu(c) = \exp(R(1-c)), R = \ln (\alpha)$, we have $\chi = R $. In contrast, in non-monotonic case the sign of $\chi$ depends on the magnitude of the gradient at the end points. Thus, in non-monotonic case, the stability depends not on the end-point viscosities but the  derivative of viscosity with respect to concentration at end-points.

\subsection{Linear stability analysis}\label{subsec:LSA_chap5}
In order to carry the linear stability analysis, we introduce an infinitesimal perturbation to the base state, equation \eqref{eq:basecon_chap5}. Then, we linearize the equations \eqref{eq:cont_eqn}-\eqref{eq:convec_diffuse} about the base-state \eqref{eq:basecon_chap5} and eliminate the pressure and transverse velocity disturbances by taking the curl of Darcy's law and utilizing the continuity equation. The final form of the linearized perturbation equations are given by \citep{Manickam1993}
\begin{eqnarray}
\label{eq:perturbeqn_xt1_chap5}
M_1 u'  = M_2 c',~~ \frac{\partial c'}{\partial t}  = M_3 c' + M_4 u',
\end{eqnarray}
where $c'$ and $u'$ denote the perturbation quantities representing the concentration and the axial velocity component, respectively and
\begin{equation} \label{eq:LSAxteqns1A_chap5}
\left.\begin{aligned}
M_1& = \mathcal{D}_x^2 + \frac{1}{\mu_b} \left(\mathcal{D}_x c_b\right) \mathcal{D}_x  +\mu_b  \mathcal{D}_y^2, M_2 = - R(\mu_b)\mathcal{D}_y^2,\\
M_3& = \mathcal{D}_x^2 + \mathcal{D}_y^2,~~~M_4 = -\mathcal{D}_x c_b,
\end{aligned}
~~\right\}
\qquad 
\end{equation}
$\mathcal{D}_x^n = \displaystyle \frac{\partial^n}{\partial x^n}, \mathcal{D}_y^n = \displaystyle \frac{\partial^n}{\partial y^n}, n =1, 2$. Since the coefficients of the above equations are independent of $y$, the disturbances are decomposed in terms of Fourier component in the $y$ direction 
\begin{eqnarray}\label{eq:normalmode_chap5}
\left[c', u'\right](x,y,t) = \left[\phi_c, \phi_u \right](x,t)e^{iky},~~ i = \sqrt{-1},
\end{eqnarray}
where $k$ is the non-dimensional wave number. Using equation \eqref{eq:normalmode_chap5}, the operators in equation \eqref{eq:LSAxteqns1A_chap5} can be recast as
\begin{equation} \label{eq:LSAxteqns2A_chap5}
\left.\begin{aligned}
M_1 & = \mathcal{D}_x^2 + \frac{1}{\mu_b} \left(\mathcal{D}_x c_b\right) \mathcal{D}_x  - k^2\mu_b\mathcal{I},~~~M_2 = k^2 R(\mu_b)\mathcal{I},\\
M_3 &= \mathcal{D}_x^2 - k^2\mathcal{I},~~~M_4 = -\mathcal{D}_x c_b,
\end{aligned}
\right\}
\qquad 
\end{equation}
where $\mathcal{I}$ is the identity operator and $R(\mu_b)$ is the viscosity-related parameter given by
\begin{eqnarray}\label{eq:viscosity-parameter_chap5}
R(\mu_b) &=& \frac{1}{\mu_b} \frac{\mbox{d} \mu_b}{\mbox{d} c_b}= \frac{(1+a)(\zeta_1 - \zeta_0)}{(1+ac_b)^2} \cot(\zeta).
\end{eqnarray}
Here, $\mu_b$ and $c_b$ are the viscosity and concentration base states, respectively. It can easily verifiable that for monotonic viscosity-concentration profiles, $R(\mu_b) = R = \ln(\alpha)$, the log-mobility ratio. Now, the linearized equation \eqref{eq:perturbeqn_xt1_chap5} can be recast as an initial-boundary value problem
\begin{eqnarray}\label{eq:IBVPxt_chap5}
\frac{\partial \phi_c }{\partial t}  & = &\widetilde{\mathcal{L}} \phi_c, 
\end{eqnarray}
where $\widetilde{\mathcal{L}} = M_3 + M_4 M_1^{-1}M_2$, and $M_i$'s are as in equation \eqref{eq:LSAxteqns2A_chap5}. The associated boundary conditions are $(\phi_c, \phi_u) \rightarrow 0,$ as $x \rightarrow \pm \infty$ and a random initial condition. Manickam and Homsy\cite{Manickam1993, Manickam1994} analyse the stability of equation\eqref{eq:IBVPxt_chap5} by using a quasi-steady-state approximation (QSSA) and compare linear stability results with nonlinear simulations. They have shown that the validity of QSSA is questionable at short times where the base state changes rapidly. A fundamental problem with such approach is that the concentration eigenfunctions are spanned all over the spatial domain, i.e. the eigenfunctions of the operator $\mathcal{D}_x ^2= \displaystyle \frac{\partial^2}{\partial x^2},  x \in (-\infty, \infty)$ are global modes. Hence, they do not provide an appropriate basis for streamwise perturbations\cite{Ben2002}. As the present problem can have a small onset time ($t_{\rm} \approx \mathcal{O}(1)$) for instability, the QSSA as such is ill-suited to the task of resolving the early-time behavior. To overcome this difficulty, Kim and Choi\cite{Kim2011} uses the self-similar property of the base-state, i.e. they transform the $(x,t)$ co-ordinate to a self-similar co-ordinate $(\xi,t)$, $\xi = x/\sqrt{t}$ such that the base-state, equation\eqref{eq:basecon_chap5} becomes $\displaystyle c_b(\xi) = \frac{1}{2}\bigg[\text{erfc}\left(\frac{\xi}{2}\right)\bigg]$. In the self-similar co-ordinate $(\xi,t)$, the streamwise operator
\begin{equation} \label{eq:streamwise_operator_I_chap5}
\mathbf{T}= \frac{\partial^2}{\partial \xi^2}+ \frac{\xi}{2}\frac{\partial}{\partial \xi},
\end{equation}
satisfies the following eigenvalue problem
\begin{align}\label{eq:streamwise_operator_II_chap5} \nonumber
\mathbf{T}e_n(\xi)&= \lambda_n e_n(\xi)\\
&~= \lambda_n a_n\mathcal{H}_n(\xi/2)\exp(-\xi^2/4), n=0,1, 2, \ldots,
\end{align}
where $\mathcal{H}_n(\xi)$ are the $n^{\rm th}$ Hermite polynomial, $a_n$ are positive constants and $\displaystyle \lambda_n = -\frac{n+1}{2}$. The Hermite function based eigenfucntions, being localized around the bast state, provide an optimal basis for streamwise perturbations. This suggests that the numerical simulation of the miscible viscous fingering dynamics in an unbounded domain is best done in the $(\xi, t)$ coordinates. Due to this reason we have investigated the stability analysis of miscible viscous fingering in $(\xi, t)$ coordinates. On rewriting equation \eqref{eq:IBVPxt_chap5} in transformed co-ordinates $(\xi,t)$, we have
\begin{eqnarray} \label{eq:IBVP_chap5}
\frac{\partial \phi_c}{\partial t}  & = &\mathcal{L}(t) \phi_c, 
\end{eqnarray}
where $\mathcal{L}(t) = T_3 + T_4 T_1^{-1}T_2$, and $T_i$'s are as follows

\begin{equation} \label{eq:LSAxiteqnA_chap5}
\left.\begin{aligned}
T_1 &= \mathcal{D'}^2  + \frac{1}{\mu_b} \frac{\mbox{d} \mu_b}{\mbox{d} \xi} \mathcal{D}' - k^2 t \mathcal{I}, ~~T_2 = k^2 t R(\mu_b)\mathcal{I},\\
T_3 &= \frac{1}{t}\mathcal{D'}^2 + \frac{\xi}{2t}\mathcal{D}' - k^2\mathcal{I} ,~~ T_4 = -\frac{1}{\sqrt{t}}\frac{\mbox{d}c_b}{\mbox{d}\xi} \mathcal{I},
\end{aligned}
\right\}
\qquad 
\end{equation}
$\displaystyle\mathcal{D'}^n = \frac{\partial^n}{\partial \xi^n}, n =1, 2 $ and $R(\mu_b)$ as in equation \eqref{eq:viscosity-parameter_chap5}. Even though two sets of equations \eqref{eq:IBVPxt_chap5} and  \eqref{eq:IBVP_chap5} are mathematically equivalent, there is one restriction. It is observed that the transformation to the self-similar coordinates $(x,t)\rightarrow (\xi ,t)$ is singular at $t=0$. Hence, we must restrict our evolution away from this singular limit $t=0$ and omission of this singular limit is not practically important. Further, we have presented the relationship between the onset time and energy of the perturbations that obtained from both the coordinate in Appendix \ref{App-B}.

\subsection{Non-modal analysis}\label{subsec:NMA_chap5}
As the linear stability operator, $\mathcal{L}(t)$, in equation \eqref{eq:IBVP_chap5} is non-autonomous, we have employed the non-modal analysis (NMA) described by Schmid\cite{Schmid2007}. We first discretise the linearized disturbance system, equation \eqref{eq:IBVP_chap5} to get an initial value problem (IVP)
\begin{eqnarray} \label{eq:IVP_chap5}
\frac{\mbox{d} \phi_c}{\mbox{d} t}  & = &\mathcal{L}(t) \phi_c.
\end{eqnarray}
For a chosen time interval $[t_p, t_f]$, let $\Phi(t_p;t_f)$ be a formal solution of equation \eqref{eq:IVP_chap5}, where $\phi_c(t_f) = \Phi(t_p;t_f)\phi_c(t_p)$, $\phi_c(t_p)$ being an arbitrary initial condition. Substitute this value of $\phi_c(t_f)$ in equation \eqref{eq:IVP_chap5} to get a matrix-valued IVP
\begin{equation}
\frac{\text{d}}{\text{d}t}\Phi(t_p;t_f) = \mathcal{L}(t_f)\Phi(t_p;t_f),
\end{equation}
with initial condition $\Phi(t_p;t_p) = \mathcal{I}$, where $\mathcal{I}$ is the identity matrix. Since, the operator $\Phi(t_p;t_f)$ propagating the information from initial perturbation time, $t_p$ (time at which the perturbation is introduced to the base-state) to final time $t_f$, it is known as propagator operator. It can be noted that by the existence and uniqueness of solution to equation \eqref{eq:IVP_chap5} can be established under the hypothesis that the map $t \rightarrow \mathcal{L}(t)$ is continuous from $\mathbb{R}^+$ to the set of $n \times n$ matrix over real numbers $\mathbb{R}$ \citep{Vidyasagar2002}. Further, assuming that $\phi_c$ is square integrable over $\mathbb{R}$ we wish to find the maximum perturbation energy gain that is,
\begin{align}\label{eq:optimal_amp}\nonumber
G(t_f)=G(t_f, k, \text{Pe}, R, \delta)&:=\max_{\phi_c(t_p)} \frac{E_{\phi_c}(t_f)}{E_{\phi_c}(t_p)}  \\ \nonumber
&~= \max_{\phi_c(t_p)} \| \Phi(t_p;t_f)\phi_c(t_p)\|_2\\
&~= \|\Phi(t_p; t_f)\| = \displaystyle \sup_{j} s_j(t_f),
\end{align}
where $s_j$'s are the singular values of $\Phi(t_p; t_f)$, in other words, the eigenvalues of the self-adjoint matrix $\Phi^*(t_p; t_f) \Phi(t_p; t_f)$ can be found by singular value decomposition (SVD) of $\Phi(t_p; t_f)$. Here $E_{g}(t_f) := \|g(t_f)\|_2^2=\displaystyle \int_{-\infty}^{\infty} |g(w,t_f)|^2 \text{d} w$. 

\subsubsection{Numerical solution of the stability problem}\label{subsubsec:Nnumerical_chap5}
In the stability equation \eqref{eq:IBVP_chap5}, we have used central finite difference scheme with uniform grids for all spatial derivatives and the fourth order Runge-Kutta method for time integration. The advantage of finite difference technique is that we can study the complete spectrum of the eigenvalues. As the appropriate boundary conditions for all disturbances is that they must goes to zero far from the front. Mathematically, this implies the appropriate eigenfunctions for the instability are localized and must be zero away from the interface. It has been observed from the standard spectral theory for an unbounded domain that due to this far-field boundary conditions for disturbances, we only need to deal with the discrete eigenspectrum, instead of the essential modes whose eigenfunctions do not decay at the infinities, of the governing operator $\mathcal{L}$ \citep{Pego1994, Chang1998}. This fact is also verified by Manickam and Homsy\cite{Manickam1993}. In our analysis, the infinite streamwise domain is truncated into a finite computational domain such that it can fully capture all the decaying discrete-eigenfunctions. In order to check the validity of the results, the code was tested for several domain size and spatial step size. The results were reported if the obtained eigenvalues and eigenvectors are independent of domain and spatial step size. It is observed that the discrete eigenvalues associated to the velocity perturbations are sensitive to the width of the domain. We have performed numerical simulations by taking step size $h=0.1, 0.15$ and $0.2$. The relative error between the singular-vectors have been calculated corresponding to all three simulations  and it is found that the maximum relative error is of order $\mathcal{O}(10^{-2})$. The error has been calculated in terms of the standard Euclidean norm in $\mathbb{R}^n$, defined as, $\parallel \cdot \parallel^2 = \displaystyle \sum_{j=1}^n \left( \cdot \right)^2$. Hence, for all the simulation the spatial step size is taken to be $0.2$ with the computational domain $[-110, 90]$. Fig. \ref{fig:perturb_struct_chap5} shows that this domain length is good enough for our analysis. Further, the solution procedure has been validated by comparing with linear stability results of Hota \textit{et. al}\cite{Hota2015a} for Arrhenius type viscosity-concentration profiles.

\section{Results and discussion}\label{sec:results_chap5}
In order to understand the fundamental features and onset of instability in miscible displacements with non-monotonic viscosity profiles, we have studied two different cases, namely $\alpha$ less than $1$ and greater than $1$. Using NMA, we have shown that the singular vectors carry the information about coherent optimal perturbation structures and their temporal evolution. We validate our numerical findings by comparing with nonlinear simulations (NLS) performed using Fourier pseudospectral method. Further, our results are in contrast to the existing linear stability analyses \citep{Manickam1993, Kim2011}. 

\subsection{Stability analysis for unfavorable end-point viscosity contrast}
\begin{figure}
	\centering
	(a)\hspace{1.5in}(b)\\
	\includegraphics[width=1.6in, keepaspectratio=true, angle=0]{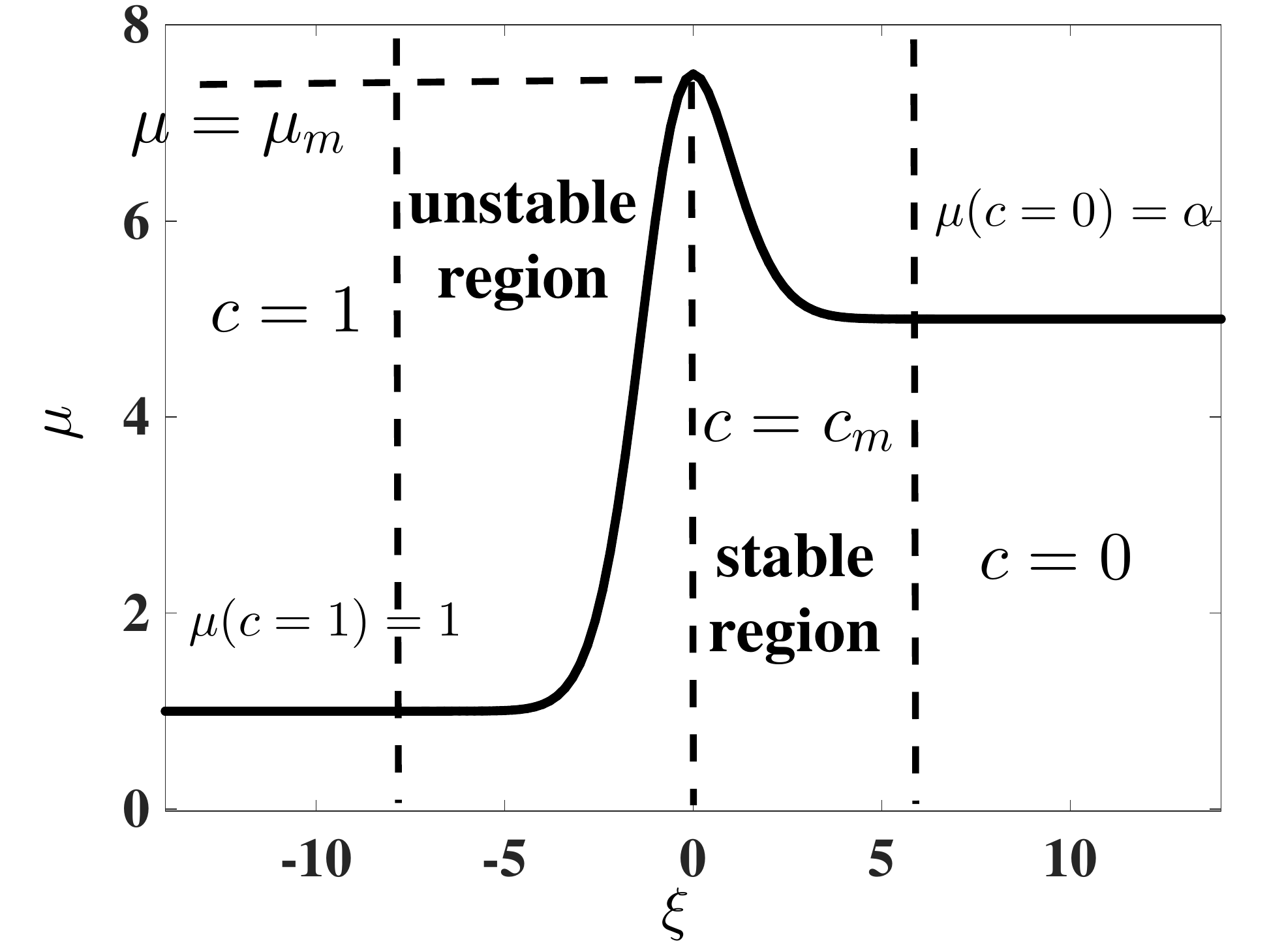}
	\includegraphics[width=1.6in, keepaspectratio=true, angle=0]{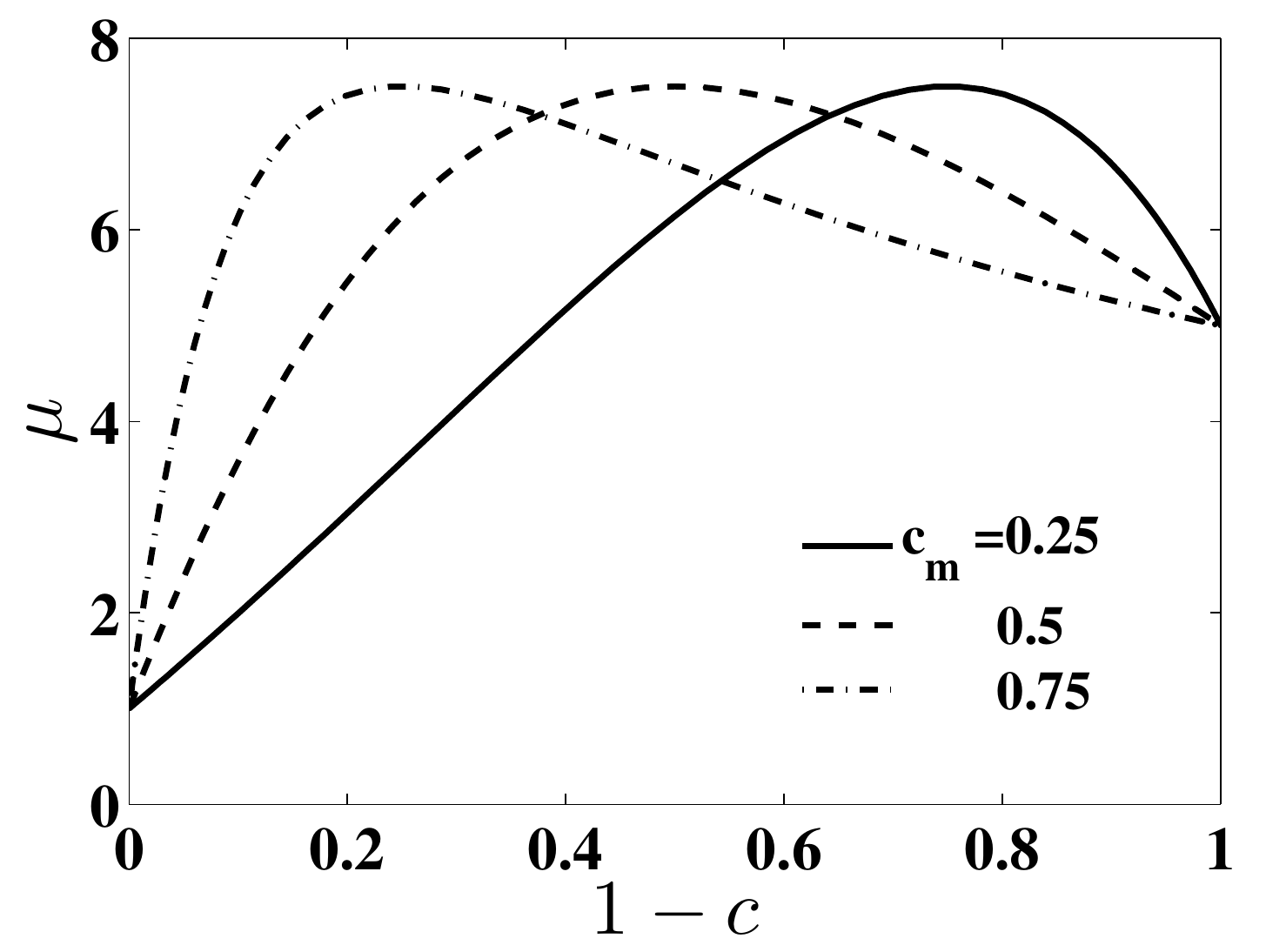}\\
	\caption{(a) Schematic of the spatial variation of viscosity for a diffused concentration profile and $\alpha >1$. (b) Viscosity profiles for $\alpha =5, \mu_m =7.5$  and various values of $c_m$. As $c_m$ increases the viscosity gradient in the unstable region steepens. }
	\label{fig:viscosity_concen_chap5}
\end{figure}
In this case, we have $\alpha >1$, equivalently, $\mu_1  < \mu_2$. The spatial variation of viscosity profile and the corresponding variation with concentration is given in Fig. \ref{fig:viscosity_concen_chap5}(a) and (b), respectively. From Fig. \ref{fig:viscosity_concen_chap5} there would develop a potentially unstable region, where the viscosity increases in the flow direction, followed in the downstream direction by a potentially stable region, where the viscosity decreases in the flow direction. In order to compare our results with the existing literature, we choose the following parameters: $\alpha =5, \mu_m =7.5$ and $c_m = 0.25, 0.5$ and $0.75$. With this configuration, the viscosity ratio within the unstable zone grows $7.5$ times, while it decreases moderately by a factor $7.5/5 = 1.5$ within the stable zone.

\subsubsection{Optimal amplification} \label{subsec:Optimal_amp_chap5}
\begin{figure}
	\centering
	(a)\hspace{1.5in}(b)\\
	\includegraphics[width=1.6in, keepaspectratio=true, angle=0]{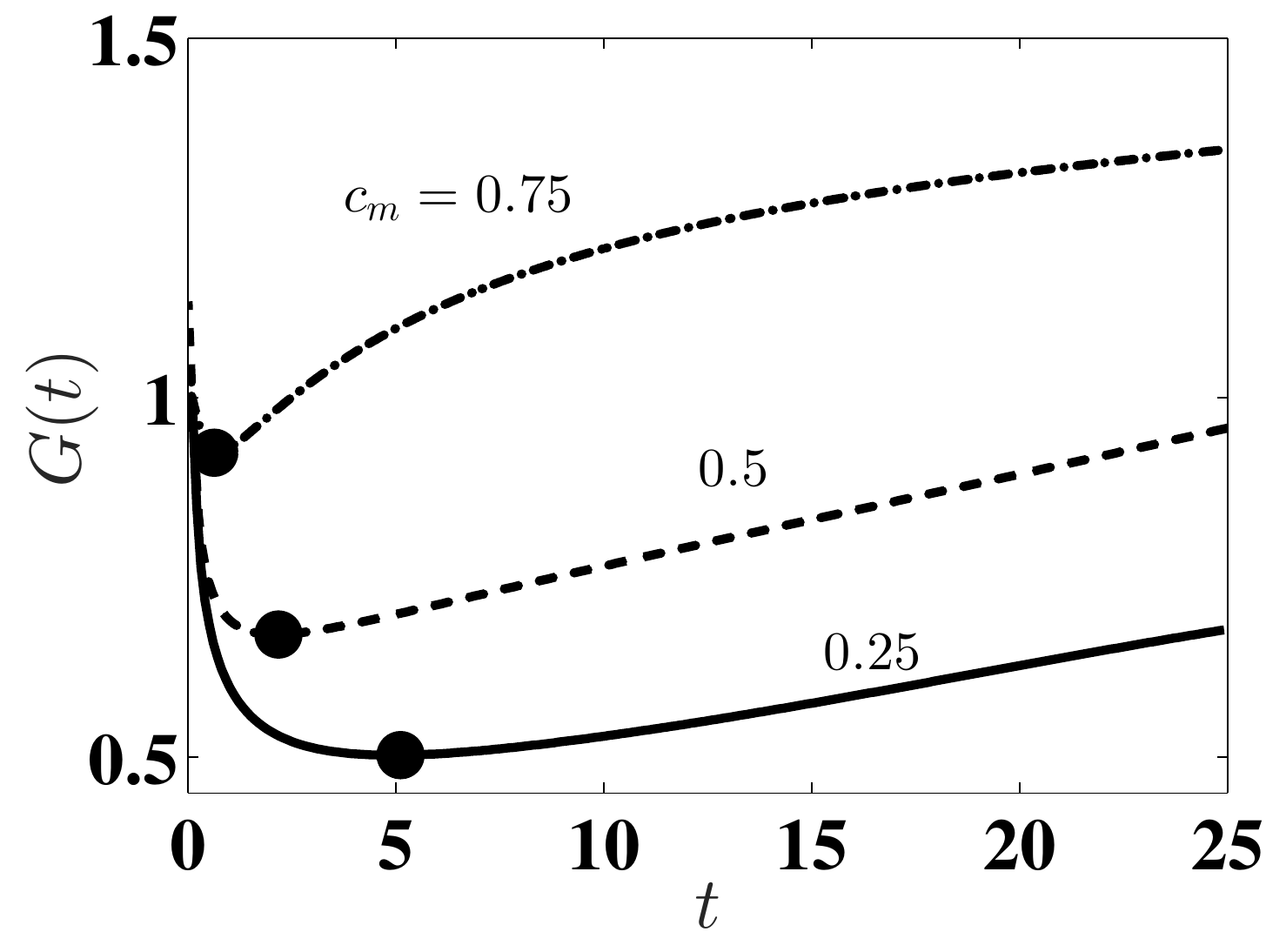}
	\includegraphics[width=1.6in, keepaspectratio=true, angle=0]{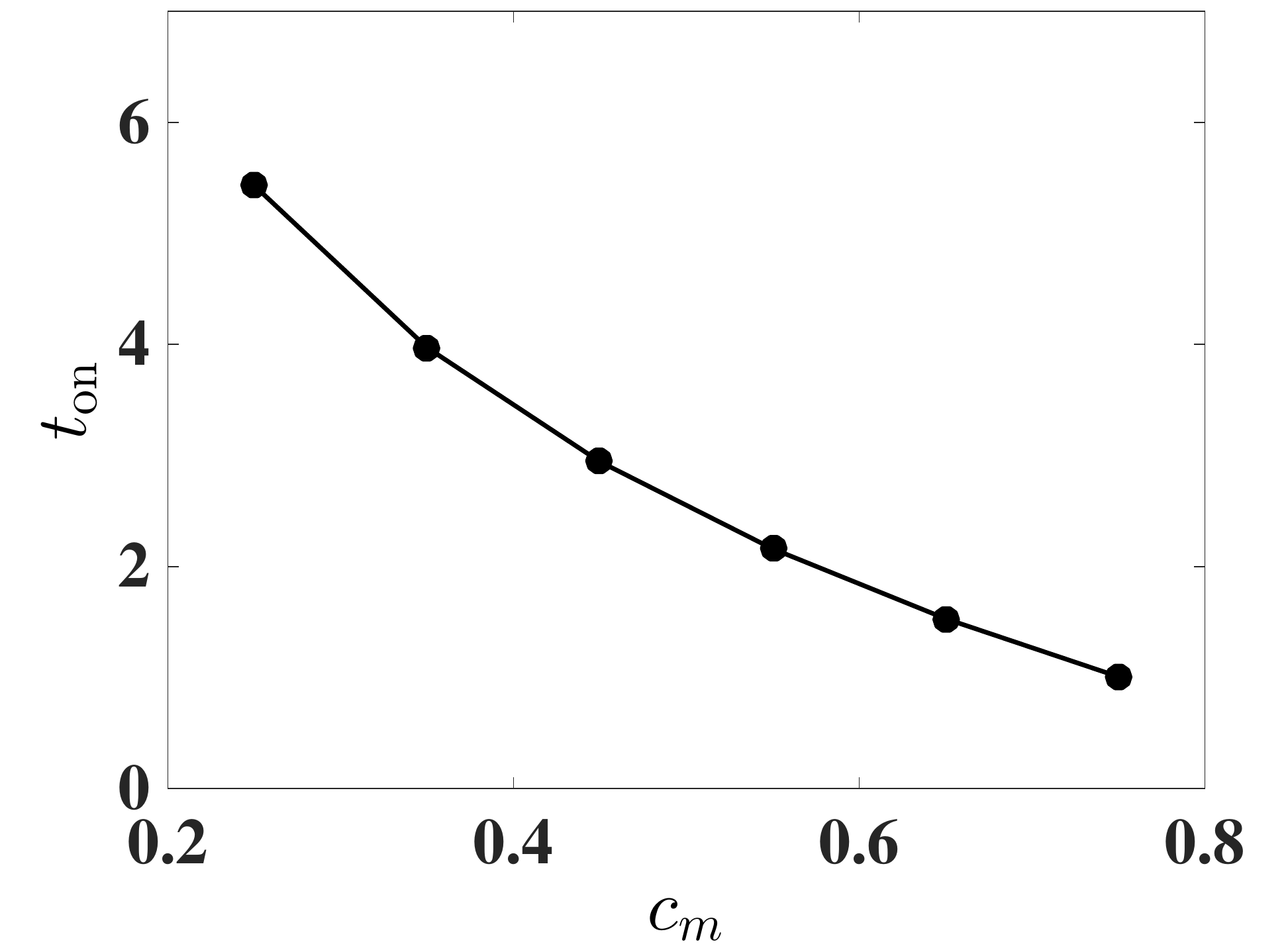} \\
	\caption{(a) Optimal amplification, $G(t)$, for $\alpha =5$ and $\mu_m =7.5$ for three different values of $c_m$. The black dots ($\CIRCLE$) denote the onset of instability, $t_{\rm on}$. (b) Onset time, $t_{\rm on}$ is monotonically decreasing with increase in $c_m$.}
	\label{fig:optimal_ampl_chap5}
\end{figure}

The optimal amplification, $G(t)$ is the maximum possible energy that a perturbation can have, incorporating all possible initial conditions. From $G(t)$, the onset of instability can be obtained as follows
\begin{eqnarray}
t_{\rm on} = \displaystyle \min \left\{ t>0: \frac{\mbox{d}G(t)}{\mbox{d}t} =0 \right\}.
\end{eqnarray}
Fig. \ref{fig:optimal_ampl_chap5}(a) shows the optimal amplification, $G(t)$ [see equation \eqref{eq:optimal_amp}], for $\alpha =5$ and $\mu_m =7.5$ and initial perturbation time, $t_p = 0.01$. The initial perturbation time, $t_p$ is chosen to be atleast one order of magnitude smaller than the onset of instability, $t_{\rm on}$. We have shown in Fig. \ref{fig:optimal_ampl_chap5}(b)that the onset time, $t_{\rm on}$, for $c_m = 0.25, 0.5$ and $0.75$ are found approximately to be $5.11, 2.17$ and $0.63$, respectively. In other words, we have found that for the parameters considered, $t_{\rm on}$ is a decreasing function of $c_m$. Further, the energy is most amplified for $c_m =0.75$ in comparison to $c_m =0.25$ and $0.5$, which is in contrast to the findings of Manickam and Homsy \cite{Manickam1993} and Kim and Choi\cite{Kim2011}. These authors have shown that the instabilities set in the unstable region and propagate downstream into the stable barrier. Fig. \ref{fig:viscosity_concen_chap5}(a) illustrates that the concentration $c_m$, at which the viscosity reaches a maximum determines the length of the stable zone. Further, larger the value of $c_m$, longer the stable zone and consequently its effectiveness in stunting the downstream propagation of the viscous fingers. 

\begin{table*}
	\centering
	\begin{tabular}{ccc|cc}
		\hline
		\hline
		& ~~~~~~~$\alpha = 5, \mu_m =7.5$ &  ~~& ~~$\alpha= 0.5, \mu_m=2$  &  \\
		\cline{2-5}
		$c_m$ ~~ &\multicolumn{1}{c}{unstable interval}   ~~ &\multicolumn{1}{c}{stable interval} ~~ &\multicolumn{1}{c}{unstable interval}   ~~ &\multicolumn{1}{c}{stable interval}\\ 
		$0.25$ ~~ & $[-6, 0.95]$ ~~& $[0.95, 5.8]$ ~~ & $[-4.7,1]$ ~~ & $[1, 6.5]$  \\
		$0.5$ ~~ & $[-6,0]$ ~~& $[0, 5.5]$  ~~ & $[-5,0]$ ~~ & $[0,6]$ \\
		$0.75$ ~~ & $[-6.5, -0.95]$ ~~& $[-0.95, 5.1]$ ~~ &  $[-5.7, -1]$ ~~ & $[-1, 5.8]$ \\
		\hline
	\end{tabular}
	\caption{\label{table:stable-unstablezone} The unstable and stable interval for  $\left(\alpha, \mu_m\right) = (5, 7.5)$ and $(0.5, 2)$ (See Fig. \ref{fig:viscosity_concen_chap5}(a) and \ref{fig:viscosity_concen_alphaless1}(a)). The initial unperturbed interface is located at $\xi =0$. In each case the left hand end-points are obtained when the value of $\mu$ is equals to $\mu(c=1) =1$ and similarly, the right hand end-points are obtained when the value equals to $\mu(c=0) = \alpha$, with an absolute error of order $\mathcal{O}(10^{-12})$. It is evident that the stable region is increasing with increase in the value of $c_m$.}
\end{table*}
From Table \ref{table:stable-unstablezone} and Fig. \ref{fig:viscosity_concen_chap5}(b), it can be observed that although with the increase in $c_m$, one has larger stable zone, but the corresponding viscosity gradient is larger. In this light, the present contrasting results from NMA can be explained from the structure of viscosity profiles. For $\alpha =5, \mu_m =7.5$, Fig. \ref{fig:viscosity_concen_chap5}(b) depicts that as $c_m$ increases the resident fluid experiences an increase in viscosity. This enhances the energy of the perturbations and thus results in a early onset of instability. Note that, the values of $\chi$ (as defined in equation \eqref{eq:chi_chap5}) for $c_m = 0.25, 0.5$ and $0.75$ are $-2.12, 3.58$ and $14.05$, respectively. This implied that there is a strong influence of the concentration gradient with respect to concentration in non-monotonic viscosity-profiles. Hence, it is shown that the end-point viscosity gradient, along with unfavorable end-point viscosity contrast are having a substantial effect on the growth rate. The NMA findings are also supported by the findings of Wang\cite{Wang2014} on the radial miscible flow in a Hele-Shaw cell. Wang\cite{Wang2014} found that the time evolution of the interfacial length (which is a global measurement of the onset of fingering) suggests that for unfavorable end-point viscosity a higher value of $c_m$ leads to more unstable interface at early times.

\begin{figure}
	\centering
	(a1)\hspace{1.5in}(b1)\\
	\includegraphics[width=1.6in, keepaspectratio=true, angle=0]{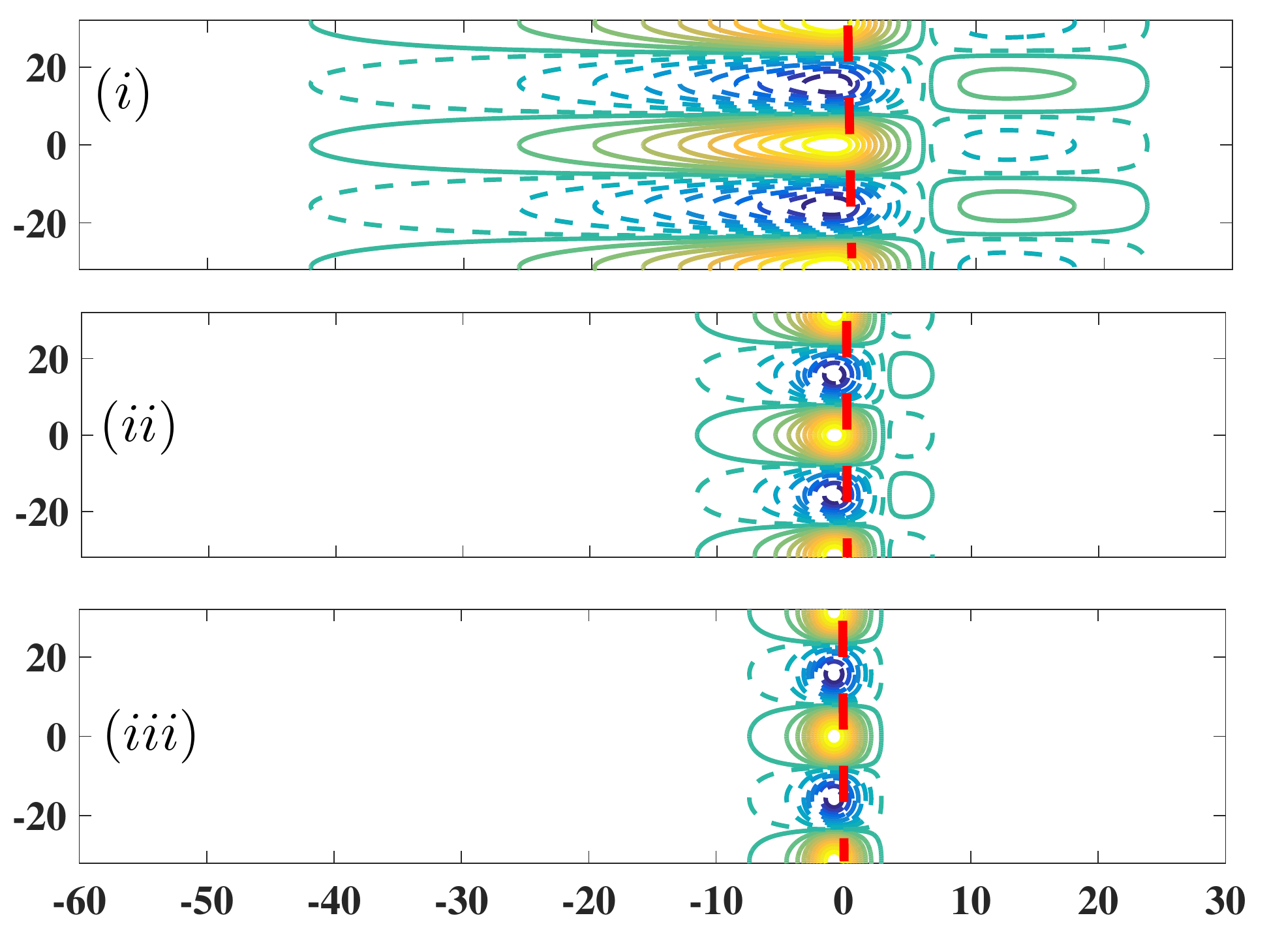}
	\includegraphics[width=1.6in, keepaspectratio=true, angle=0]{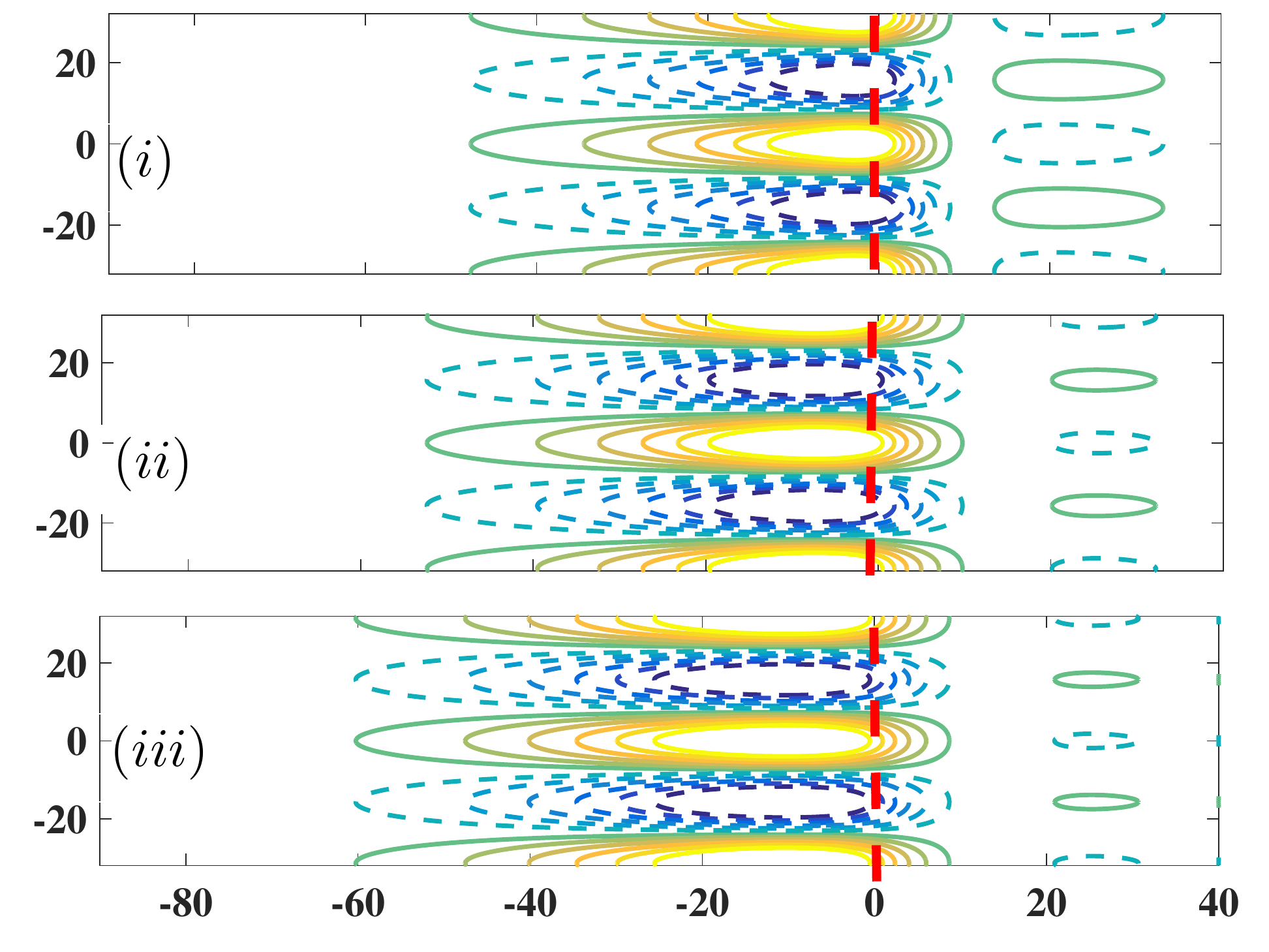}\\
	(a2)\hspace{1.5in}(b2)\\
	\includegraphics[width=1.6in, keepaspectratio=true, angle=0]{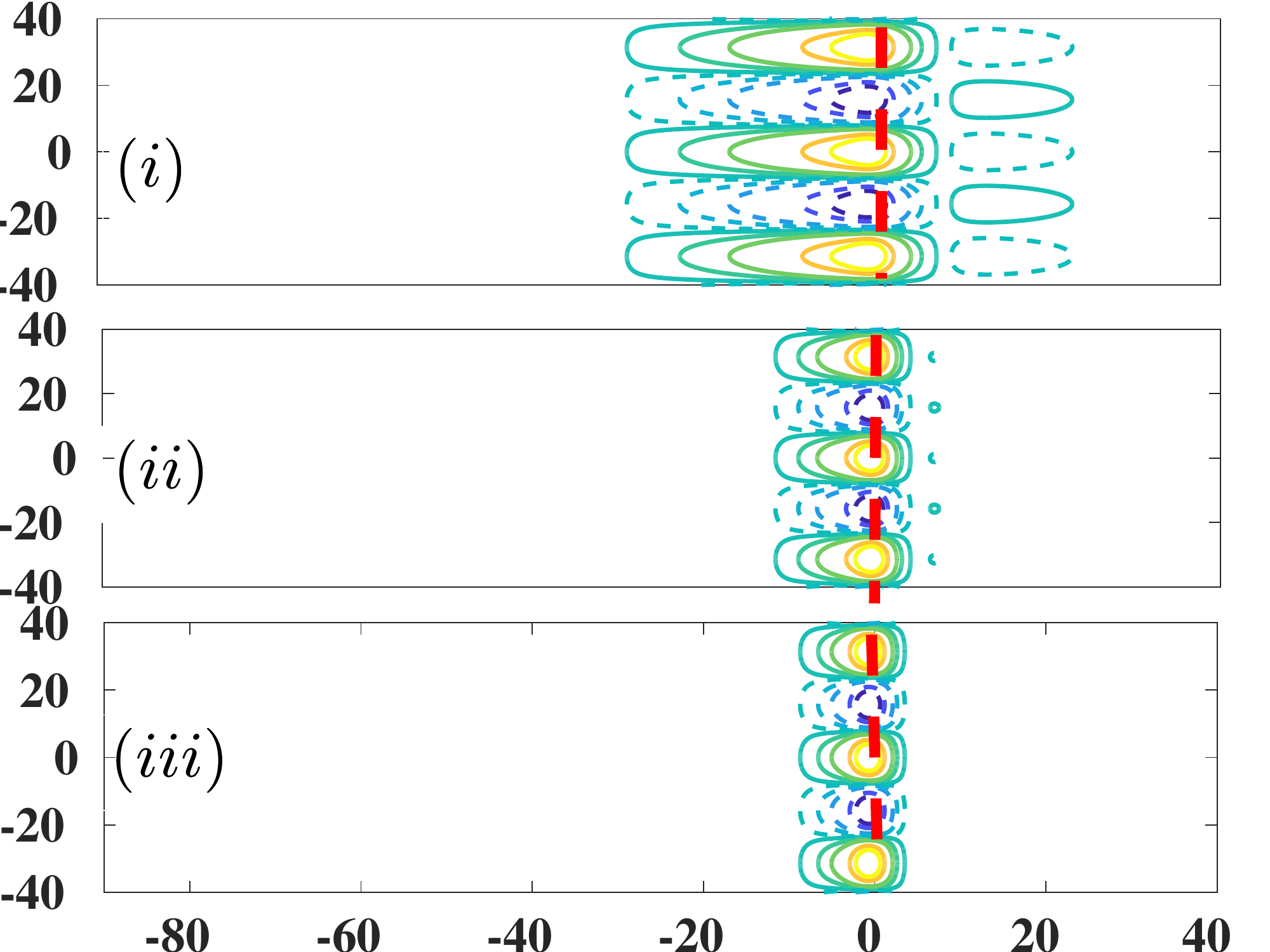}
	\includegraphics[width=1.6in, keepaspectratio=true, angle=0]{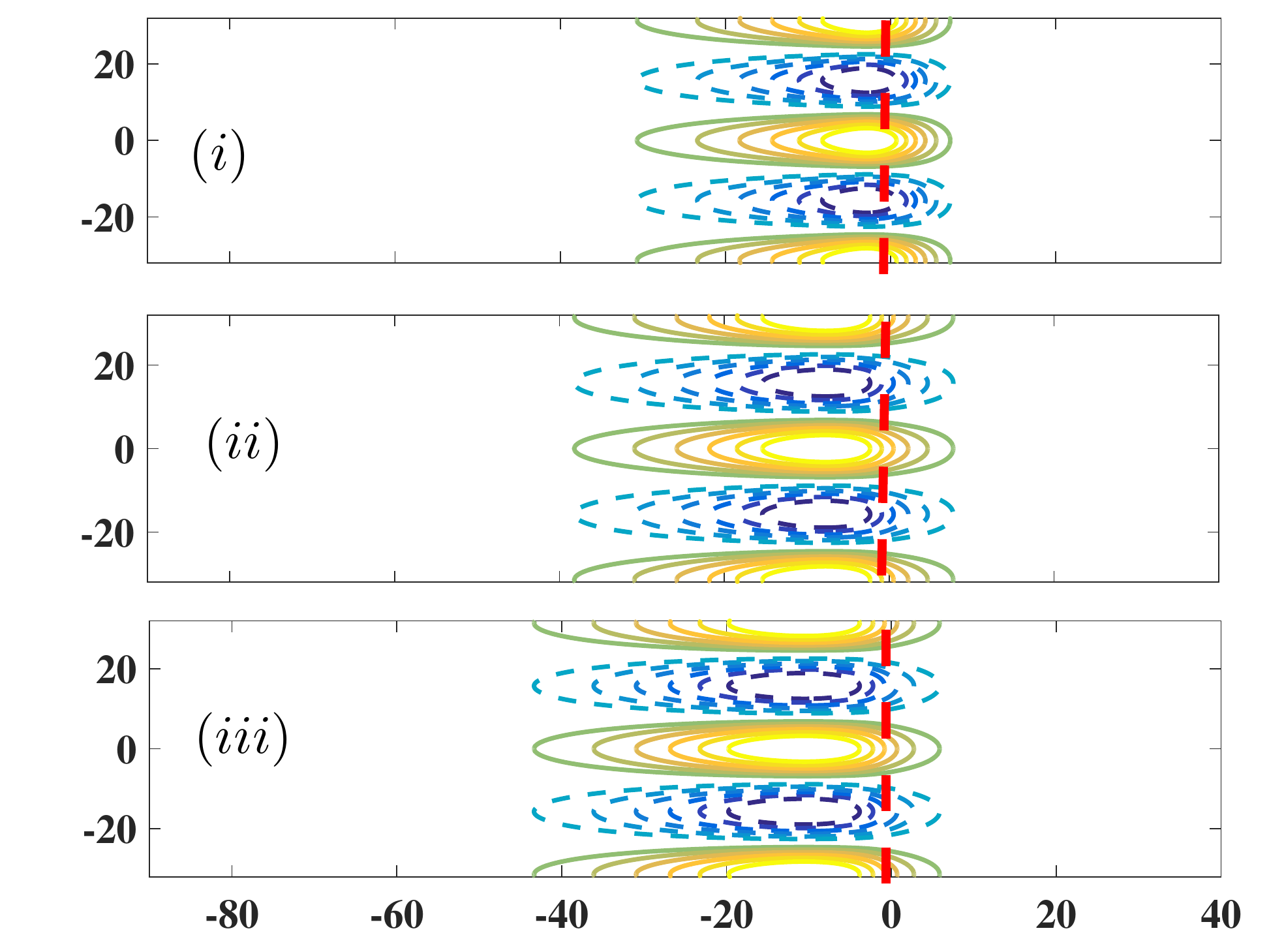}\\
	(a3)\hspace{1.5in}(b3)\\
	\includegraphics[width=1.65in, keepaspectratio=true, angle=0]{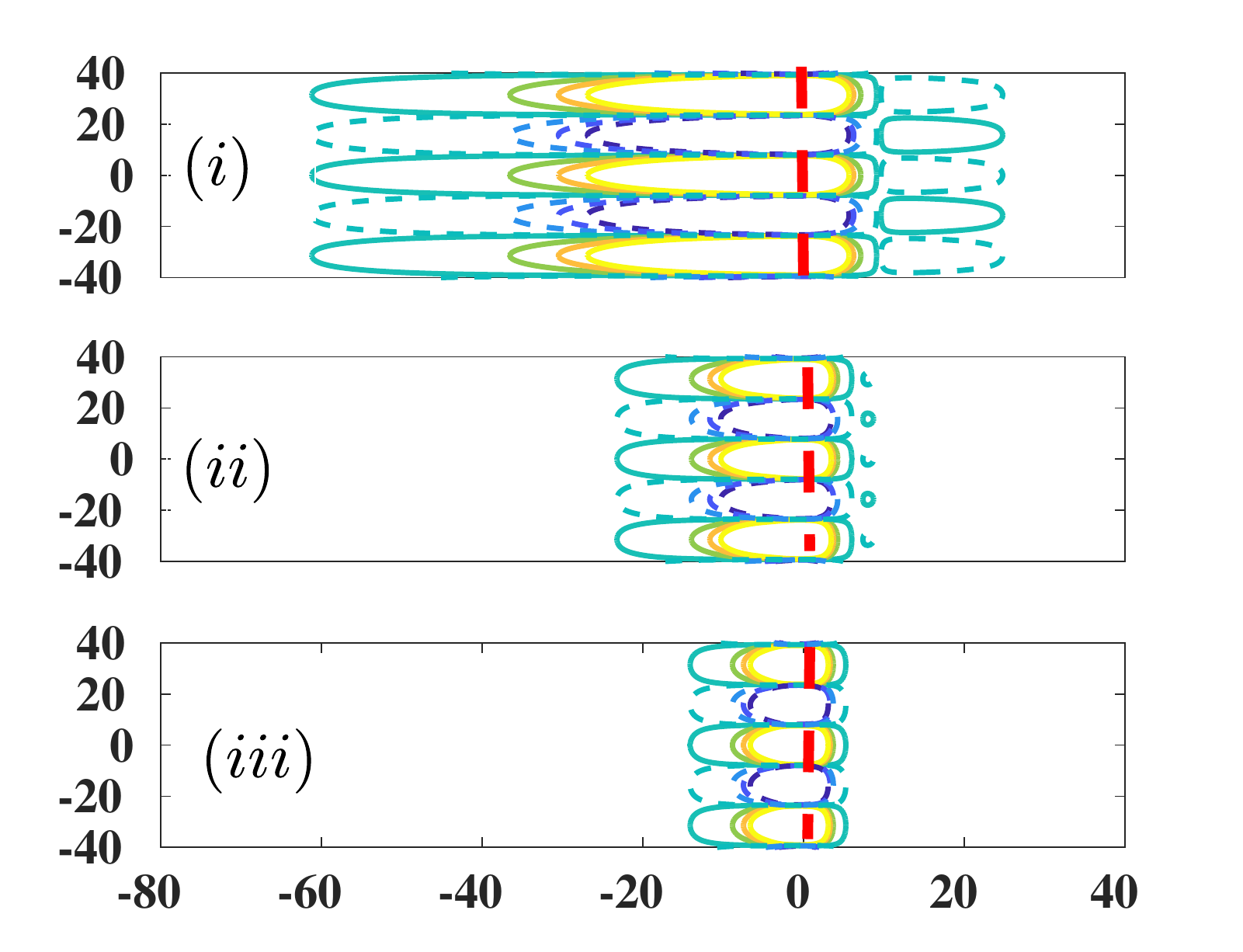}
	\includegraphics[width=1.65in, keepaspectratio=true, angle=0]{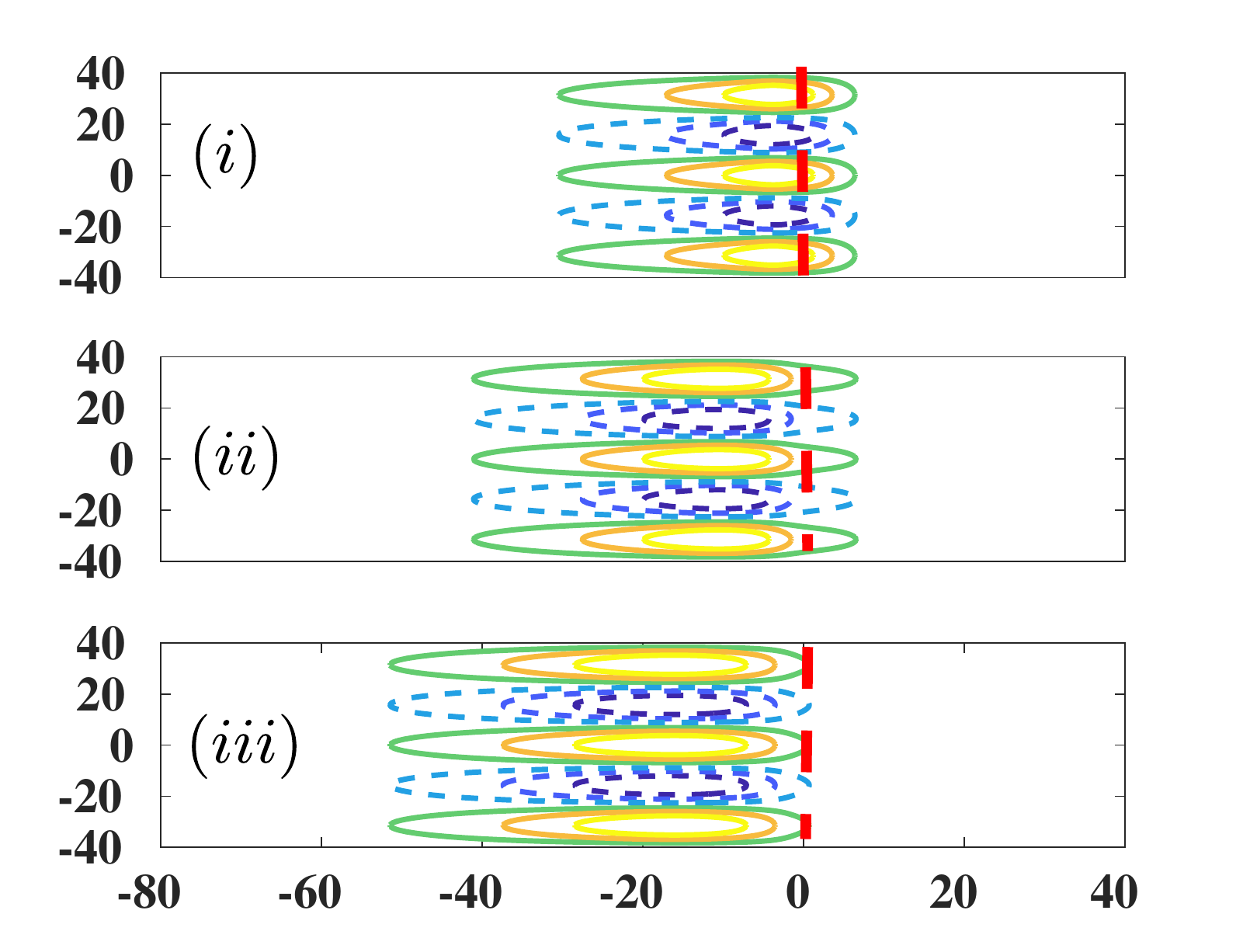}\\
	\caption{ For $\alpha =5, \mu_m= 7.5, k=0.2$: left column (a) the optimal initial perturbations, $c'_p= V_{\rm opt} \cos(ky)$  and right column (b) the corresponding evolved state, $c' = U_{\rm opt} \cos(ky)$. From top to bottom $c_m =0.25, 0.5$ and $0.75$. Here the time integration intervals are $(i)=[0.01, 0.5], (ii)= [0.01, 4]$ and $(iii)= [0.01, 10]$. Both $c_p'$ and $c'$ are normalized with respect to sup-norm. The dashed lines correspond to the negative contours and continuous lines correspond to the positive contours. The vertical dashed lines show the initial fluid-fluid interface. The concentration contours shown : (a1) \& (b1) $0.02$ to $0.7$ with eight equal increments, (a2) \& (b2) $0.01$ to $0.6$ with five equal increments (a3) \& (b3) span from $0.01$ to $0.8$ with four equal increments.}
	\label{fig:perturb_struct_chap5}
\end{figure}
\subsubsection{Structure of optimal perturbations} \label{subsec:Optimal_perturb_chap5}
For miscible displacements with non-monotonic viscosity-concentration profile, it is found that even for a favorable (unfavorable) endpoint viscosity contrast the displacement can be unstable (stable) \cite{Bacri1992a, Bacri1992b, Manickam1993, Manickam1994}. Manickam and Homsy \cite{Manickam1993, Manickam1994} analyze the physical mechanism of the stability of the flow, in terms of the structure of the vorticity and stream function fields. It is observed that the vorticity field has a quadruple structure which can give rise to dynamics fundamentally different from the dipole structure [see Fig. $5$ of Manickam and Homsy\cite{Manickam1993} and Fig. $7$ of Hota \textit{et al.}\cite{Hota2015a}] that dominates the evolution of monotonic displacements.

The structure of the optimal perturbations can be analyzed by the singular value decomposition (SVD) of the propagator matrix $\Phi(t_p;t_f)$. The SVD of $\Phi(t_p;t_f)$ at a given final time, $t_f$, is given by 
\begin{equation}\label{eq:SVD_1_chap5}
\Phi(t_p; t_f) = \mathbf{U}_{[t_p;t_f]}\mathbf{\Sigma}_{[t_p;t_f]}\mathbf{V^*}_{[t_p;t_f]},
\end{equation} 
where $\mathbf{U}$ and $\mathbf{V}$ are the right and left-singular values of $\Phi(t_p;t_f)$ and the super-script star ($*$) denotes the Hermitian.

For $\alpha =5, \mu_m =7.5, k=0.2$ and $c_m = 0.25, 0.5$ and $0.75$, the left column(a)'s and right column(b)'s in Fig. \ref{fig:perturb_struct_chap5} show the contours of initial optimal perturbations, $c_p' = V_{\rm opt} \cos(ky)$ and corresponding evolved state, $c' = U_{\rm opt} \cos(ky)$, respectively. The vertical dashed line shows the initial interface position, which is $\xi =0$ in all our simulations. It is observed that displacements characterized by non-monotonic viscosity profiles typically lead to quadruple structures of the flow field, as opposed to the dipoles observed for monotonic profiles \citep{Manickam1993, Hota2015a}. It is evident from Fig. \ref{fig:perturb_struct_chap5}(a1)-(a3) that the optimal perturbations have two columns of isocontours and the contours right to the dashed line being situated in the stable region have larger impact than the column of contours on the left.

In Figs. \ref{fig:perturb_struct_chap5}(b1)-(b3) illustrates the optimal output associated with the initial optimal perturbations shown in Figs.  \ref{fig:perturb_struct_chap5}(a1)-(a3). In  Figs. \ref{fig:perturb_struct_chap5}(b1)-(b3), the left column contours are destabilizing as they move low viscosity fluid to the high viscosity regions in the direction of the flow and move high viscosity fluid to the region of lower viscosity, against the direction of the flow. On the other hand, the right column contours do exactly opposite. Formation of quadruple contours for the perturbations is an unique feature of non-monotonic viscosity profile. It is observed from Figs. \ref{fig:perturb_struct_chap5}(b1)-(b3) that the optimal perturbations are spread farther in the backward than in the forward direction and this spreading is more for higher value of $c_m$, i.e., 0.75. The flow is unstable if the destabilizing left columns are stronger than the right column contours, otherwise stable. Further, as the value of maximum concentration, $c_m$ increases, the diffusive region connecting the perturbations and the resident fluid experiences an increase in viscosity. As a result, the left column contours diminishes which cause to set the instability early with increase in $c_m$. Thus, the onset time, $t_{\rm on}$, of a flow with non-monotonic viscosity-concentration profile with unfavorable end-point viscosity contrast is monotonically decreasing function of $c_m$ due to the relative strengths of left and right column of isocontours. These results are consistent with the findings of Manickam and Homsy\cite{Manickam1993} based on the the vorticity perturbation equations. Thus, the optimal perturbation obtained from NMA captured the effect of non-monotonic viscosity-concentration effect without invoking vorticity perturbation equations.

\subsubsection{Comparison with nonlinear simulations}\label{subsec:comapre_NLS_chap5}
Here, we  compare our NMA results with nonlinear simulations. As the flow is two-dimensional, we use stream-function formulation, $\psi$. Further, writing the unknown variables as, $\psi'(x,y,t) = \psi(x,y,t) - \psi_b(x,t), ~ c'(x,y,t) = c(x,y,t) - c_b(x,t)$, and solve the nonlinear equations \eqref{eq:cont_eqn}-\eqref{eq:convec_diffuse} using Fourier pseduospectral method proposed by Tand and Homsy\cite{Tan1988} and the detailed algorithm one can found in Manickam and Homsy\cite{Manickam1994} and Pramanik \textit{et al.} \cite{Pramanik2015JFM}. The non-dimensional width of the computational domain is $\mbox{Pe} = UH/D$, the P\'eclet number and the corresponding length of the domain is $A\cdot \mbox{Pe}$. Here, $A = L/H$ is the aspect ratio and $L, H$ being the dimensional length and width of the computational domain, respectively. The length of the computational domain is taken large enough so that we can accommodate the viscous fingers. To obtained the nonlinear growth of perturbations, the computational domain is chosen to be Pe $= 512$ and $A = 4$ with $1024\times 256$ grid points for discretizing the domain. The time integration is performed by taking time stepping $\triangle t = 10^{-3}$. Convergence study has been carried out by taking spatial discretization steps ($\triangle x$, $\triangle y$) = $(4, 4)$, ($\triangle x$, $\triangle y$) = $(2, 4)$ and ($\triangle x$, $\triangle y$) = $(2, 2)$ in a computational domain $[0, 2048] \times [0, 512]$. Relative error between the transversely averaged concentration profiles $\bar{c}(x,t) = \displaystyle \frac{1}{\mbox{Pe}}\int_{0}^{\mbox{Pe}} c(x,y,t, \mbox{d}y)$ has been calculated corresponding to both the simulations and it is found that with respect to  Euclidean norm, $\displaystyle \parallel \cdot \parallel^2=\sum_{j=1}^n \left( \cdot \right)^2$, the maximum relative error is of order $\mathcal{O}(10^{-2})$. To get optimal result, thus ($\triangle x$, $\triangle y$) =(4, 4), with $\triangle t = 0.1 $ has been chosen.
\begin{figure}
	\centering
	\includegraphics[width=3.5in, keepaspectratio=true, angle=0]{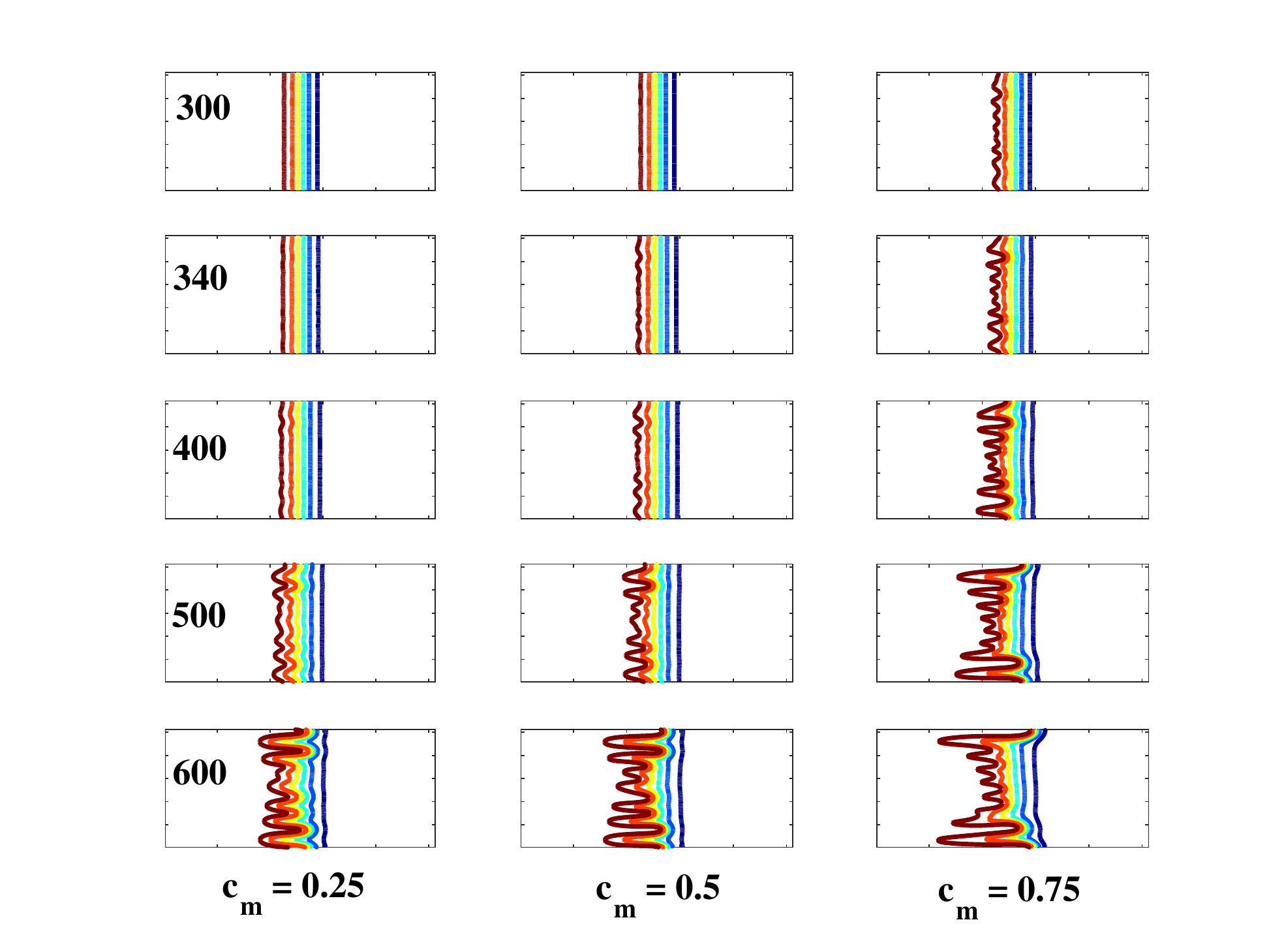}
	\caption{Finger propagation for the set of non-monotonic profiles with $\alpha=5, \mu_m=7.5$, and (a) $c_m=0.25$, (b) $c_m=0.5$, and (c) $c_m=0.75$. All the simulations are with aspect ratio $A=4$. The concentration contours shown span from $c'=0.1$ to $c'=0.9$ with six equal increments.}
	\label{fig:NLS_chap5}
\end{figure}

Fig. \ref{fig:NLS_chap5} shows the NLS results for $\alpha =5, \mu_m =7.5$ and various values of maximum concentration, $c_m$. It is observed that at time, $t=300$, the fingers are visible for $c_m = 0.75$ , whereas the time at which the fingers are visible for $c_m =0.5$ and $c_m=0.25$ are $t = 340$ and $400$, respectively. From this important visual observation it can clearly be noted that the onset of fingers is early for the higher value of $c_m$ in comparison to smaller values of $c_m$, or in other words, onset of fingers is a monotonicaly decreasing function of $c_m$. These results are consistent with the onset of instability in the linear regime determined from NMA [see Fig. \ref{fig:optimal_ampl_chap5}]. Further, in contrast to the monotonic viscosity-concentration profiles, in the case of non-monotonic viscosity-concentration, the fingers propagate faster in the backward direction than in the forward direction.  This is illustrated in Fig. \ref{fig:NLS_chap5} where the fingers have spread farther to the left than to the right, especially for the case, $c_m =0.75$. This result is consistent with nonlinear simulations of Manickam and Homsy\cite{Manickam1994}  where they refer to this phenomenon as reverse fingering. 

The mechanism of reverse fingering as shown in Fig. \ref{fig:NLS_chap5} can be explained from the spatial variation of viscosity, which is shown in Fig. \ref{fig:viscosity_concen_chap5}(b). For flow systems with such profiles, there would develop a potentially unstable region, where the viscosity increases in the flow direction, followed by a potentially stable region in the downstream direction, where the viscosity decreases in the flow direction. Thus, for non-monotonic displacements the stable zone of the viscosity profiles acts as a barrier for the forward growth of fingers, which, when viewed in a reference moving with the front, tend to propagate backwards. This feature of reverse fingering in the non-monotonic displacement is also illustrated in the analysis of the structure of optimal perturbations, shown in Fig. \ref{fig:perturb_struct_chap5}. Thus, it can be concluded that NMA successfully captures the early evolution of perturbations which is well aligned with results of NLS.

\subsection{Stability analysis for favorable end-point viscosity contrast}
\begin{figure}
	\centering
	(a)\hspace{1.5in}(b)\\
	\includegraphics[width=1.6in, keepaspectratio=true, angle=0]{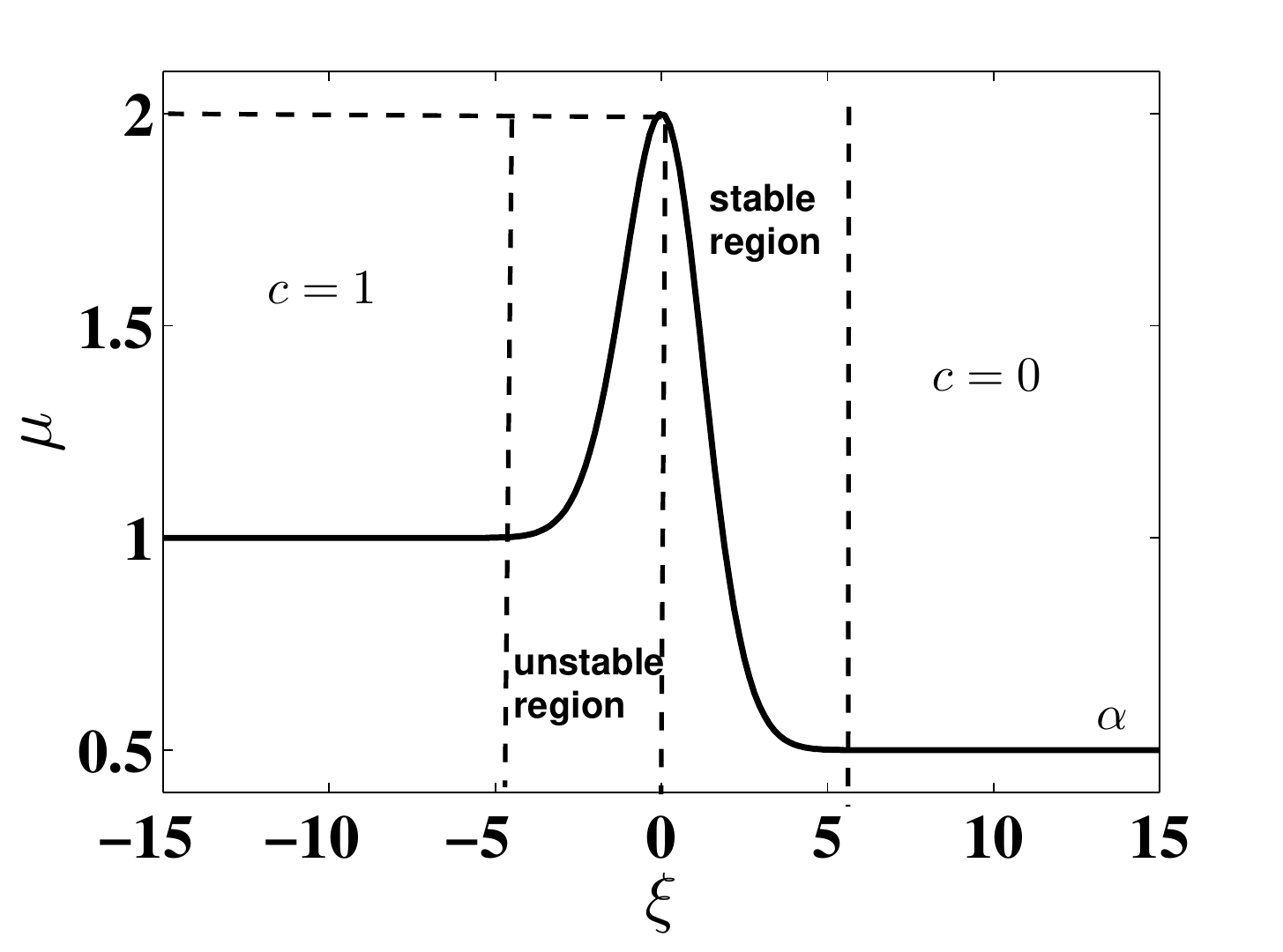}
	\includegraphics[width=1.6in, keepaspectratio=true, angle=0]{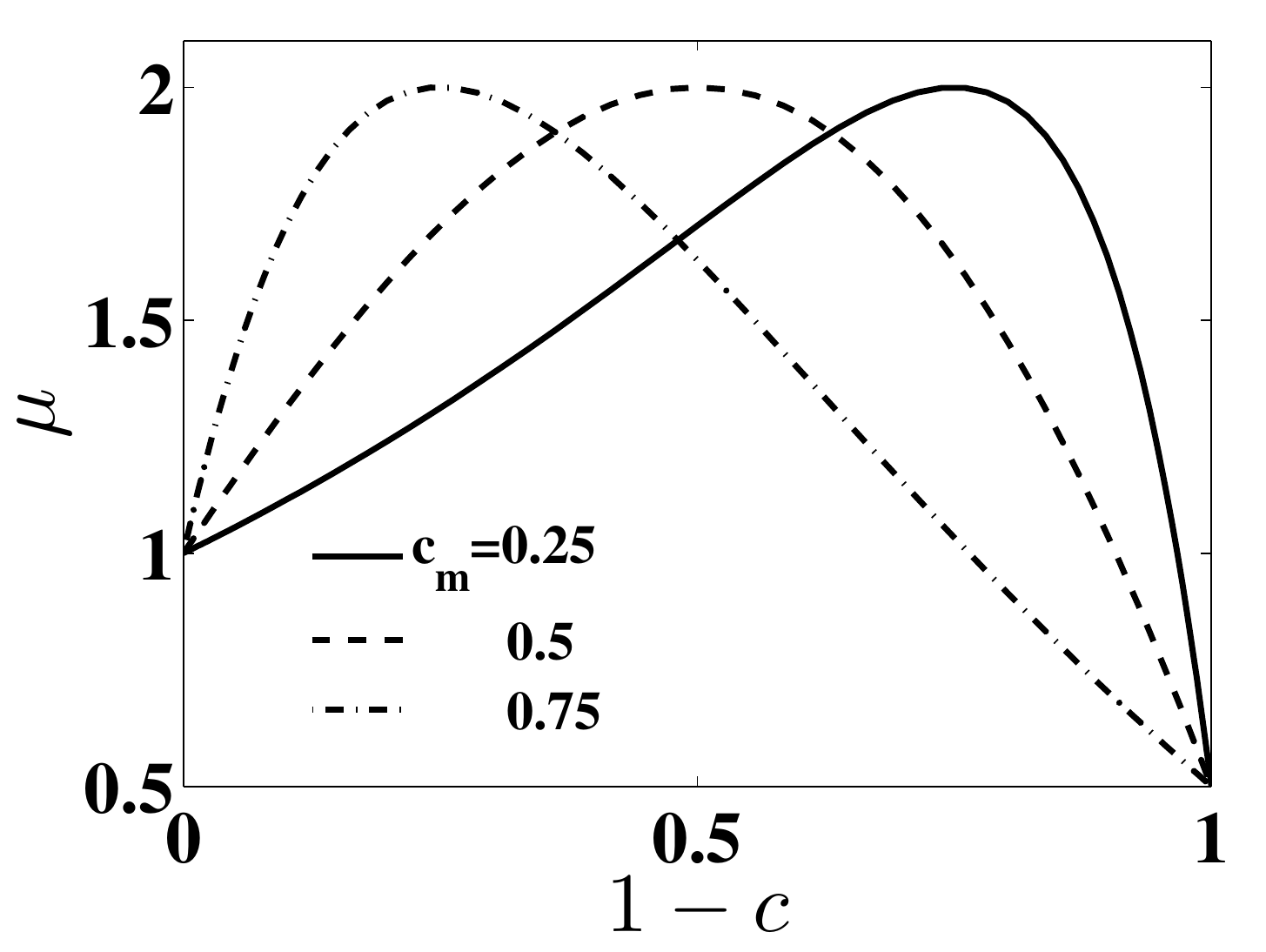}
	\caption{For $\alpha =0.5$ and $\mu_m =2$: (a) shows the spatial variation of the viscosity profile, (b) viscosity-concentration profiles for $c_m=0.25, 0.5$ and $0.75$. Across the unstable zone, viscosity increases by a factor $2$ while it decreases by a factor of $4$ through the stable zone. However, the strength of the (un)stable zone depends on $\mbox{d}\mu/\mbox{d}c$, not on the end-point viscosity ratio. Thus, for the same values of $\alpha$ and $\mu$, different $c_m$ determine the strength of the two zone. }
	\label{fig:viscosity_concen_alphaless1}
\end{figure}

In this case, we have $\alpha <1$, equivalently, $\mu_1  > \mu_2$, the displacing fluid is more viscous than displaced fluid. The spatial variation of viscosity profile and the corresponding variation with concentration are shown in Fig. \ref{fig:viscosity_concen_alphaless1}(a) and (b), respectively. For the flow systems as illustrated in Fig. \ref{fig:viscosity_concen_alphaless1} are said to have a favorable viscosity contrast as a high viscosity fluid
displaces a low viscosity fluid. As before, in order to compare our results with existing literature, we use the following parameters: $\alpha =0.5, \mu_m =2$ and $c_m = 0.25, 0.5$ and $0.75$. With this configuration, the flow has a viscosity ratio $2$ across the weaker unstable zone and a viscosity ratio of $2/0.5 = 4$ across the stable zone.

\subsubsection{Optimal amplification} \label{subsec:Optimal_amp_alphaless1}
\begin{figure}
	\centering
	(a)\\
	\includegraphics[width=2.6in, keepaspectratio=true, angle=0]{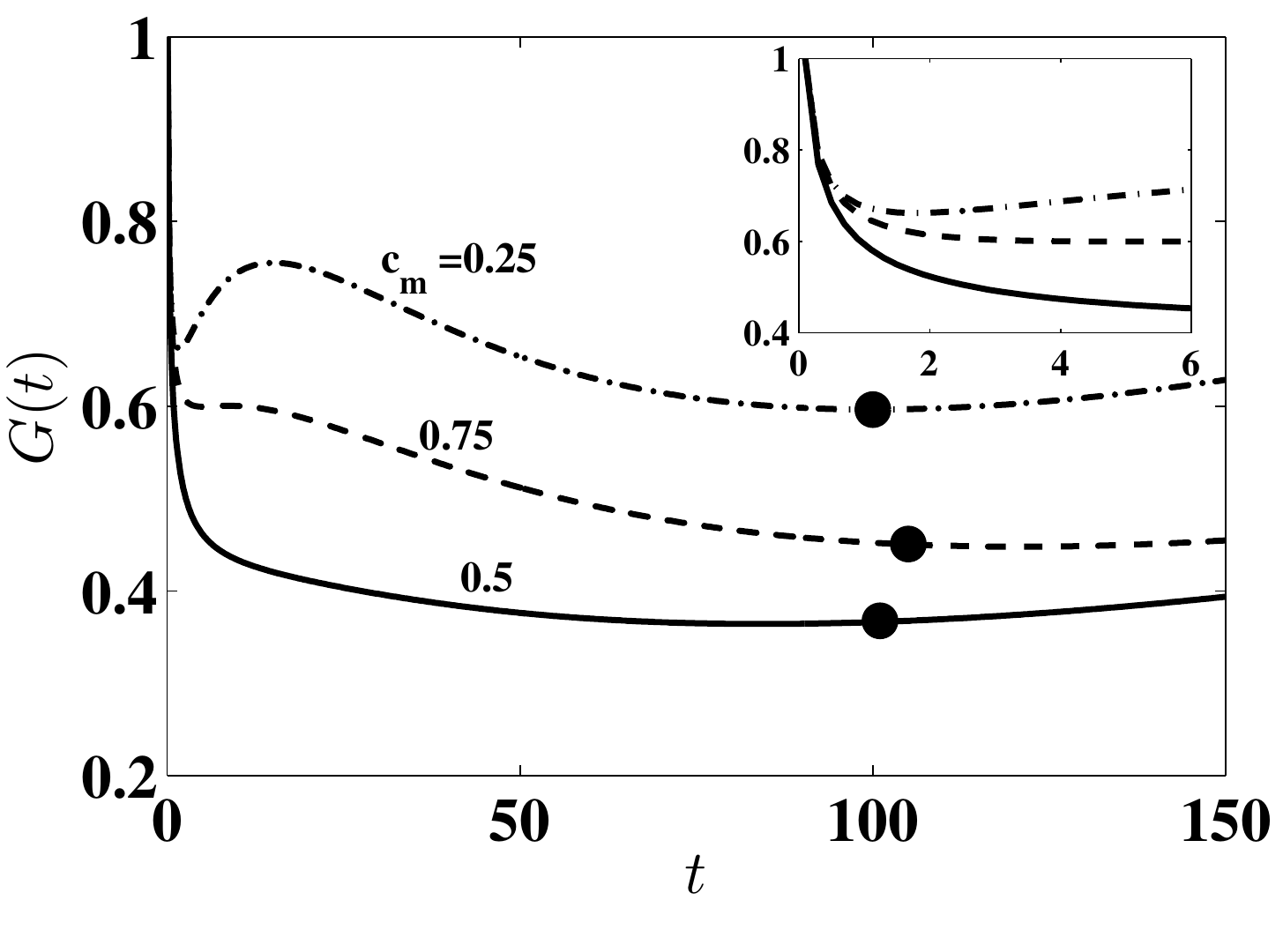}\\
	(b)\\
	\includegraphics[width=2.6in, keepaspectratio=true, angle=0]{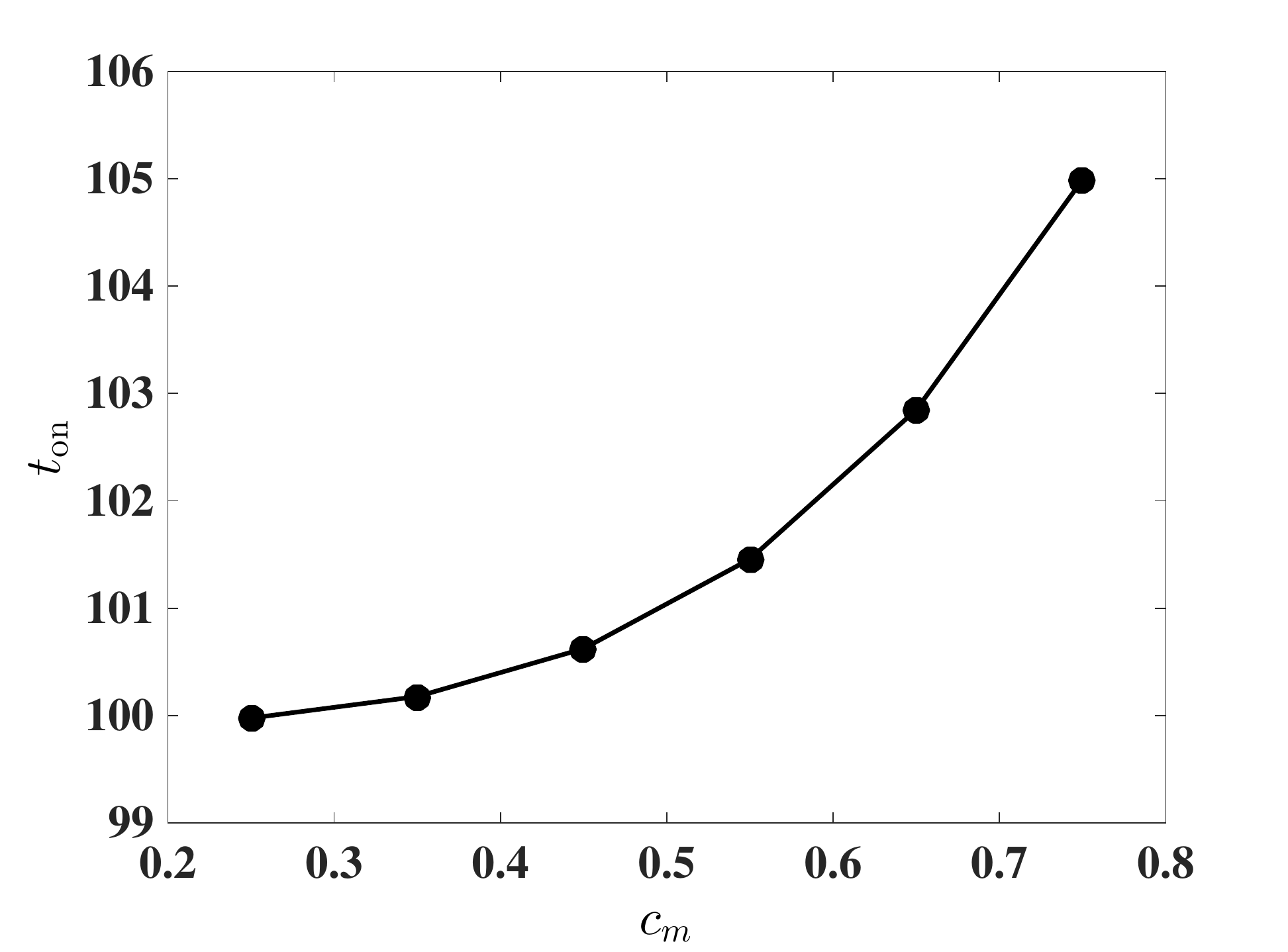} \\
	\caption{For $\alpha =0.5, \mu_m =2$: (a) Optimal amplification, $G(t)$, for $k=0.06$  with three different values of $c_m$. The black dots ($\CIRCLE$) denote the onset of instability, $t_{\rm on}$. (b) Onset time, $t_{\rm on}$, is monotonically increasing with an increase in $c_m$.}
	\label{fig:optimal_ampl_alphaless1}
\end{figure}

For $\alpha =0.5, k =0.06$ and $\mu_m =2$, Fig. \ref{fig:optimal_ampl_alphaless1} demonstrates the influence of position of maximal concentration, $c_m$ on the optimal amplification, $G(t)$. For the present case, the stability parameter $\chi$ has value $ -10.81, -1.68$ and $5.29$ for $c_m = 0.25, 0.5$ and $0.75$, respectively. For unfavorable viscosity contrast, $\chi >0$ represents the slope of the viscosity profile at the point $c = 1$ is steeper than at the point $c = 0$, whereas  $\chi <0$ denotes the reverse scenario. From Table \ref{table:stable-unstablezone}, it can be observed that the length of the stable interval is decided by the parameter $c_m$, i.e, longer stable zones with larger values of $c_m$. This affects in the delay in the onset time, $t_{\rm on}$, as $c_m$ increases.
For $\alpha=0.5$, $c_m = 0.25$ and $0.75$, an important feature resembling to the secondary instability of the optimal amplification, $G(t)$ is observed. Fig. \ref{fig:optimal_ampl_alphaless1}(a) illustrates that initially there is an influence of steeper viscosity profile [see Fig. \ref{fig:viscosity_concen_alphaless1}(b)] which helps to amplify the energy, but due to the weak unstable region, this energy only sustains for a transient period. But, when diffusion becomes weaker and the isocontours in the unstable zone overcome the stable zone then there is a growth of energy which sets the instability and we termed this as a secondary instability. This temporal evolution of the perturbation was not observed for $c_m =0.5$. Further, it is noted that the energy amplification is lowest for $c_m =0.5$ in comparison to $c_m =0.25$ and $0.75$. The reason for this secondary instability is due to steep viscosity gradient at either end at $c=0$ or $c=1$ for $c_m=0.25$ and $0.75$. This also shows that the value of $c_m$ alone is not sufficient to describe the early time temporal evolution of the disturbances.

\begin{figure}
	\centering
	(a)\hspace{1.5in}(b)\\
	\includegraphics[width=1.6in, keepaspectratio=true, angle=0]{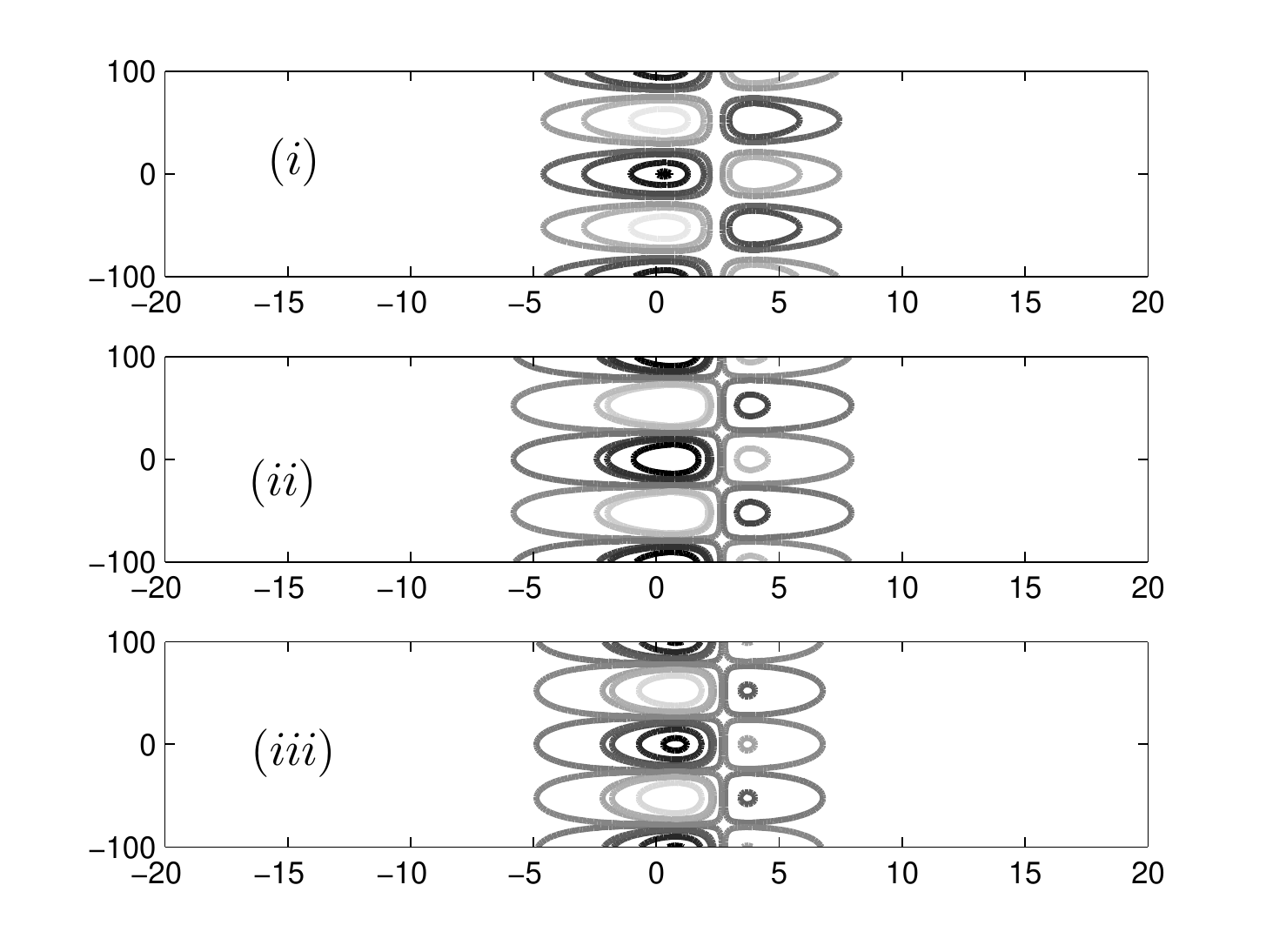}
	\includegraphics[width=1.6in, keepaspectratio=true, angle=0]{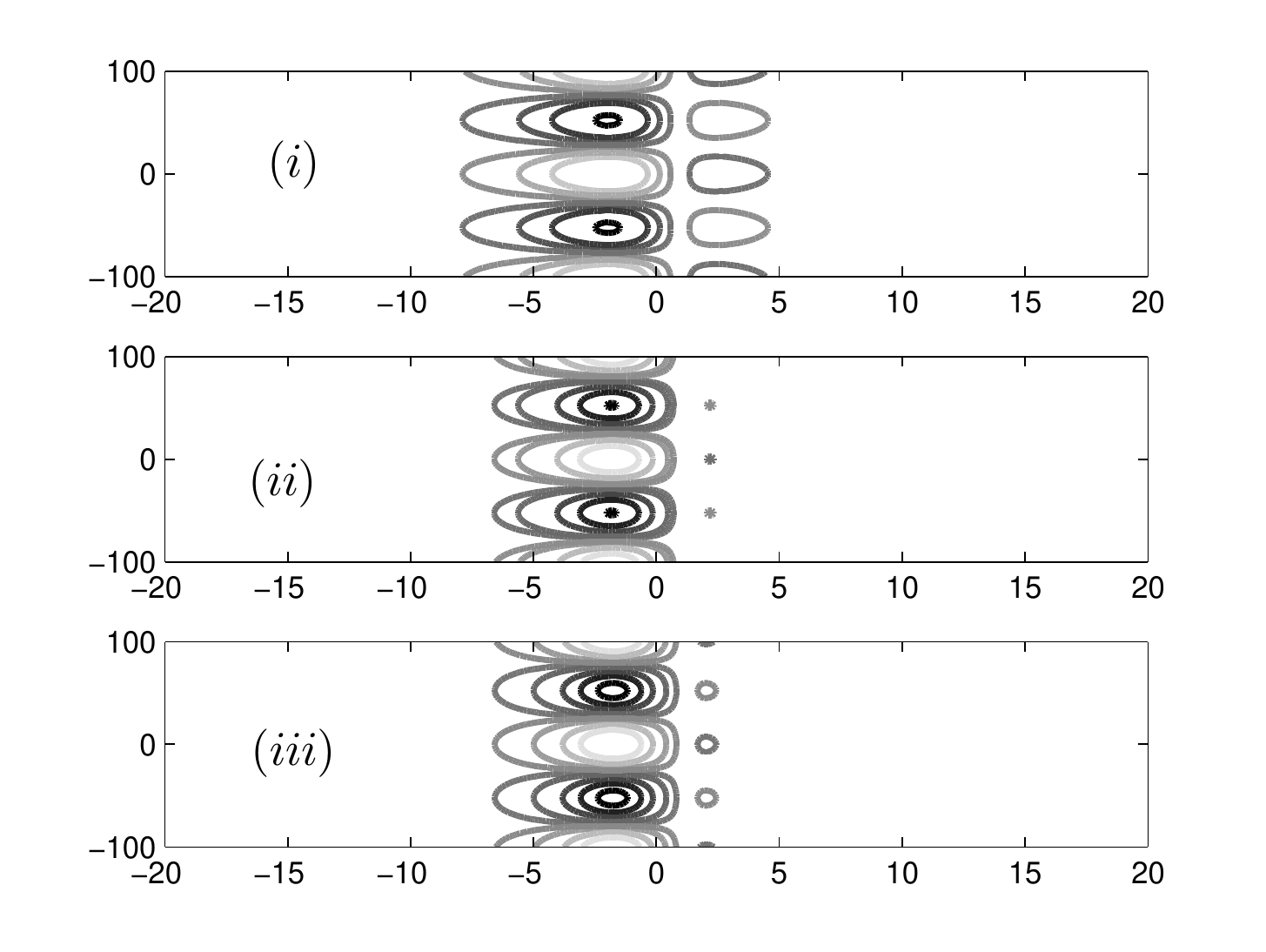}\\
	(c)\hspace{1.5in}(d)\\
	\includegraphics[width=1.6in, keepaspectratio=true, angle=0]{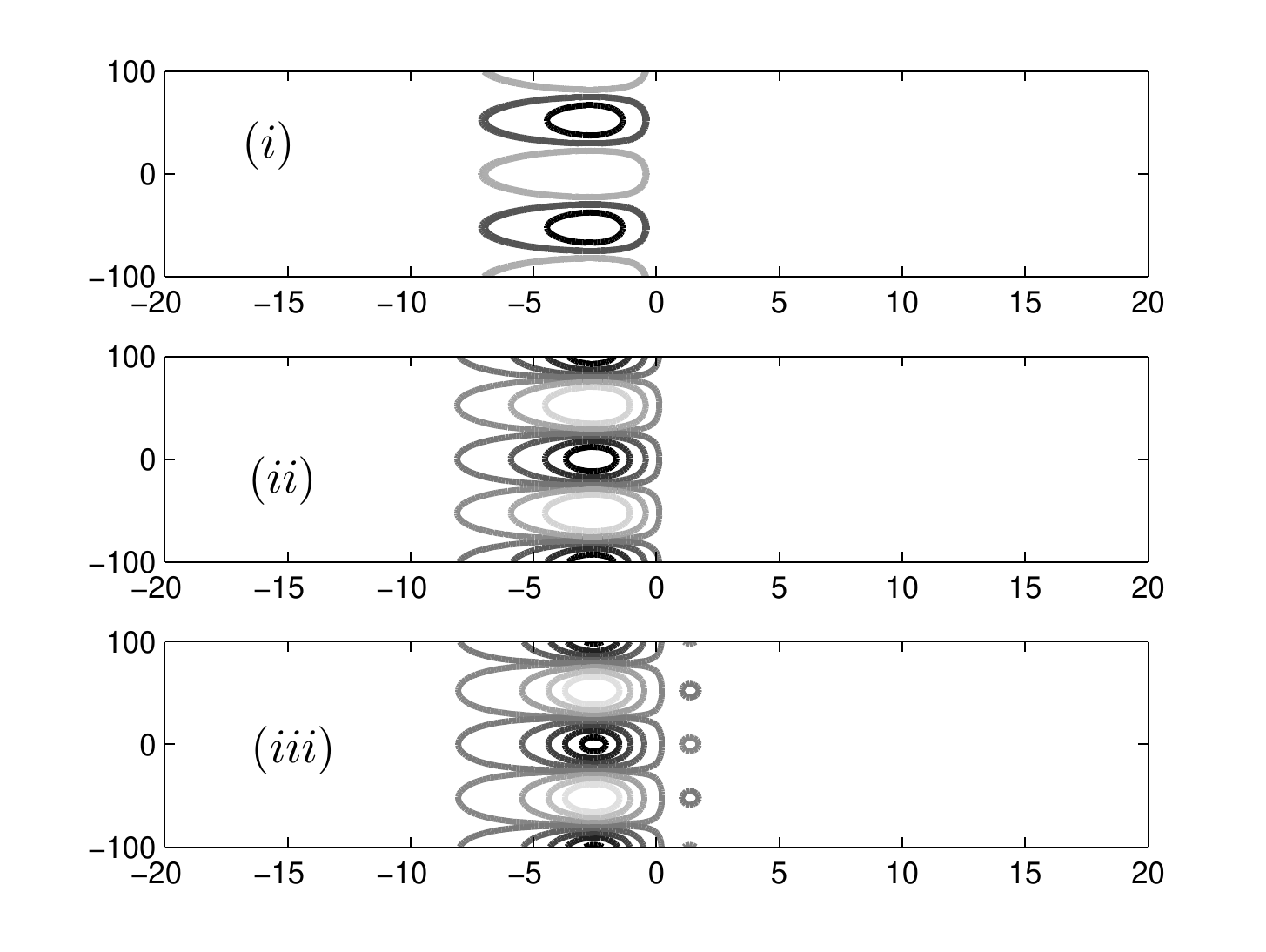}
	\includegraphics[width=1.6in, keepaspectratio=true, angle=0]{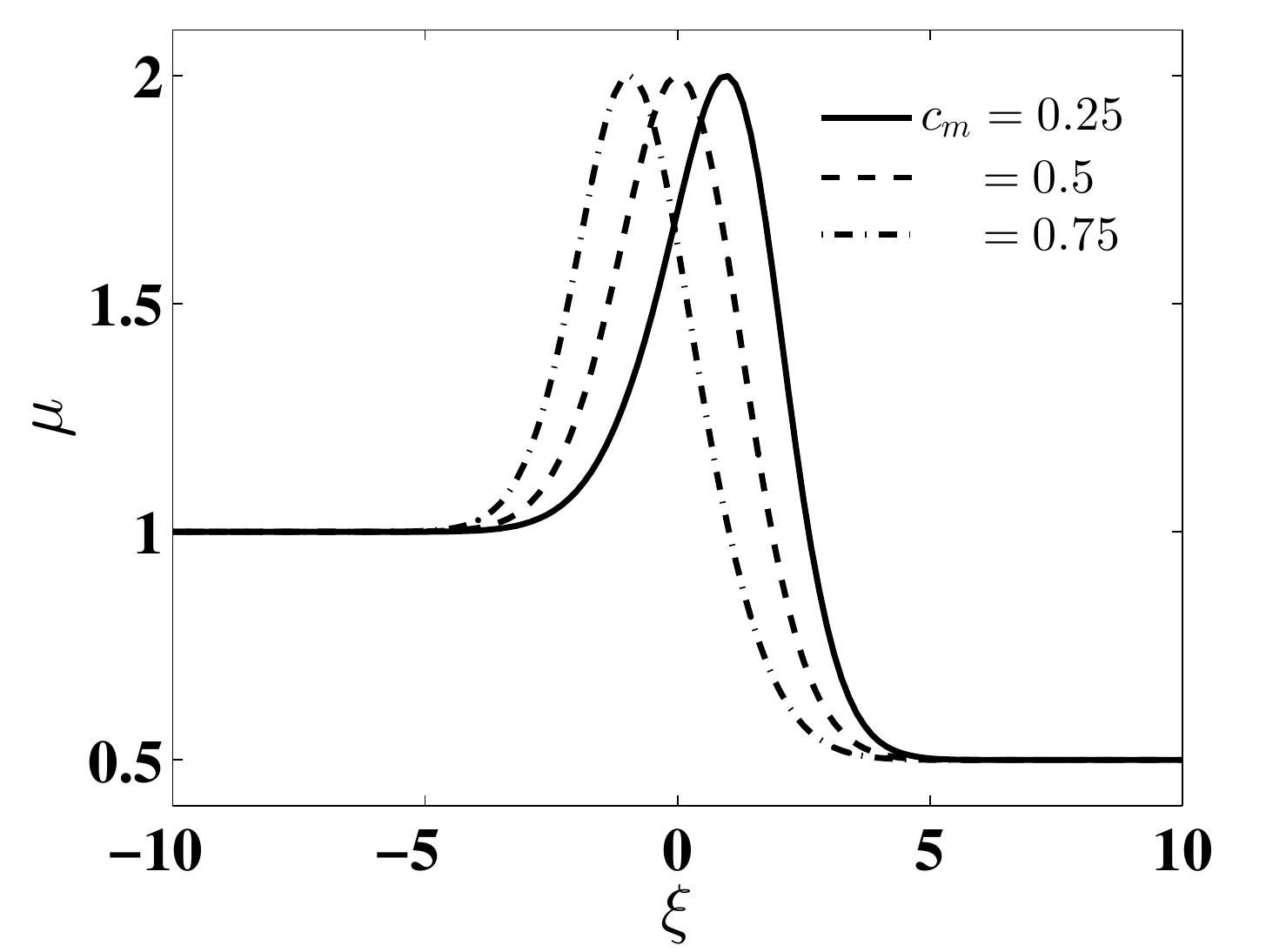}\\
	\caption{\label{fig:perturb_struct_I_alphaless1}For $\alpha =0.5, k=0.06$ and $\mu_m= 2$:(a), (b) and (c) shows the evolution of the optimal initial perturbations, $c'_p= V_{\rm opt} \cos(ky)$, for $c_m =0.25, 0.5$ and $0.75$, respectively. Here the time integration intervals are $(i)=[0.01, 50], (ii)= [0.01, 100]$ and $(iii)= [0.01, 150]$. The concentration contours shown span from its minimum to maximum values with $5$ equal increments. The initial interface is located at $\xi =0$. black lines correspond to negative contours and the grey lines correspond to positive contours. (d) shows the spatial variation of viscosity for the same values of $\alpha$ and $\mu_m$.}
\end{figure}

\subsubsection{Structure of optimal perturbations} \label{subsec:Optimal_perturb_alphaless1}
The argument based on the fact that the flow will be unstable if the destabilizing isocontours are stronger than the stabilizing one may not fully describe the physical phenomena in the non-monotonic viscosity-concentration profiles. To have a comprehensive analysis, we have studied the perturbation structures for $\alpha =0.5, k=0.06$ and $\mu_m =2$ and shown in Fig. \ref{fig:perturb_struct_I_alphaless1}. From this figure, it can be observed that the displacement is stable for all the values of $c_m$, but there is no stabilizing right isocontours for $c_m =0.75$ at time $t=50$. This is in contrast to the argument of Manickam and Homsy\cite{Manickam1993}. Whereas, the perturbation contour always have two rows of patches of opposite signs on opposite sides of the viscosity maximum for $c_m = 0.25$ and $0.5$. This can be explained from Table \ref{table:stable-unstablezone} as follows : if the recirculating fluid regions extend beyond the unstable region, the left column contours have taken the stabilizing role, which is the case when $c_m=0.75$, as shown in Fig. \ref{fig:perturb_struct_I_alphaless1}(c). Thus, we conclude that the strength of the destabilizing perturbation contours regions, their location, and their interaction with the gradients in the viscosity profile all influence the stability in case of non-monotonic viscosity-concentration profiles. Further, it is shown in Fig. \ref{fig:perturb_struct_I_alphaless1}(b) that at early times, the right side stabilizing contours are weaken at $t=100$ in compare to $t=50$, which at later time, $t=150$ again strengthened. Due to this reason the optimal amplification, $G(t)$ for $c_m =0.5$ always has the lowest energy [see the continuous line in Fig. \ref{fig:optimal_ampl_alphaless1}(a)].
\begin{figure}
	\centering
	(a)\hspace{1.6in}(b)\\
	\includegraphics[width=1.6in, keepaspectratio=true, angle=0]{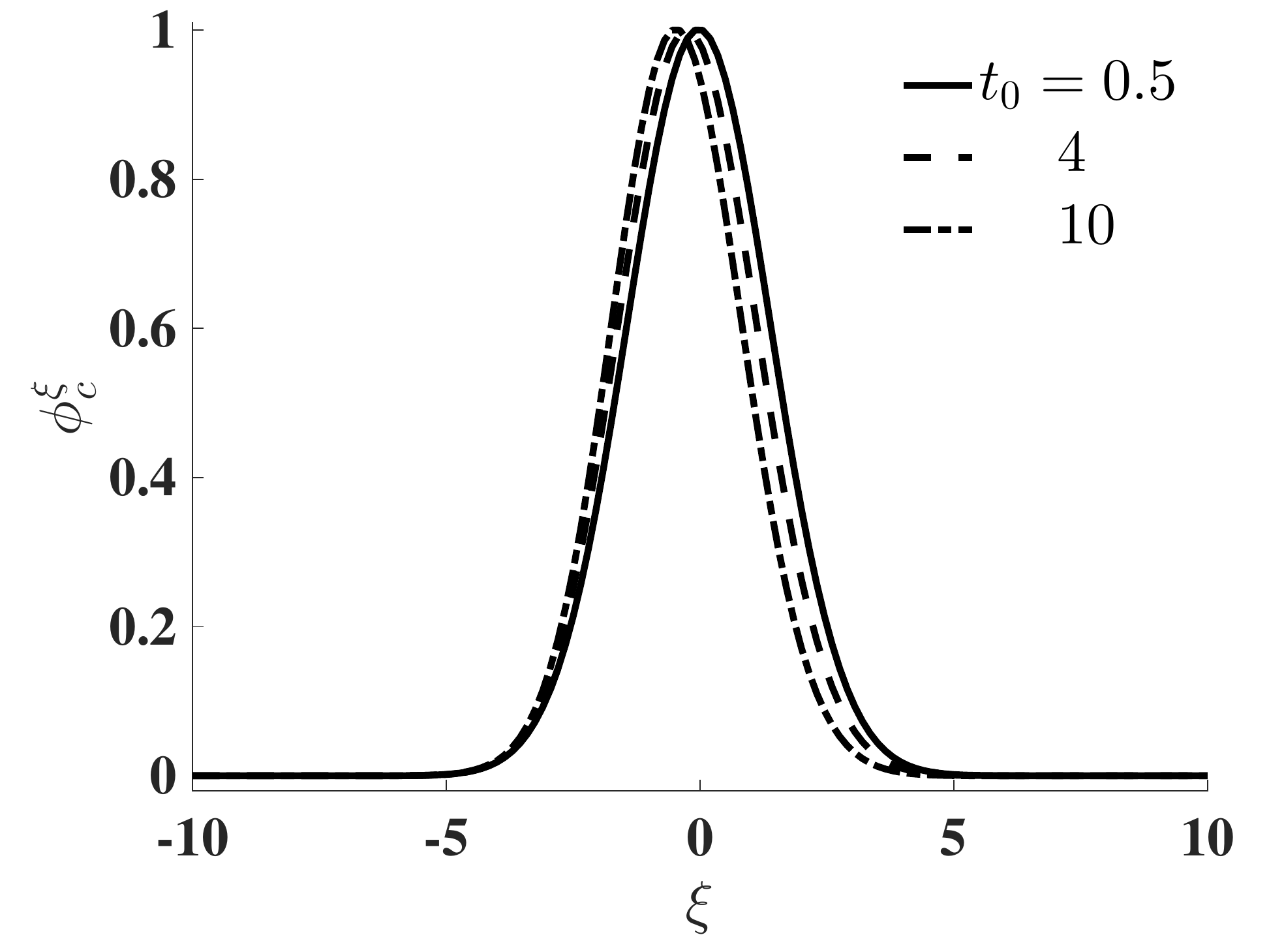}
	\includegraphics[width=1.6in, keepaspectratio=true, angle=0]{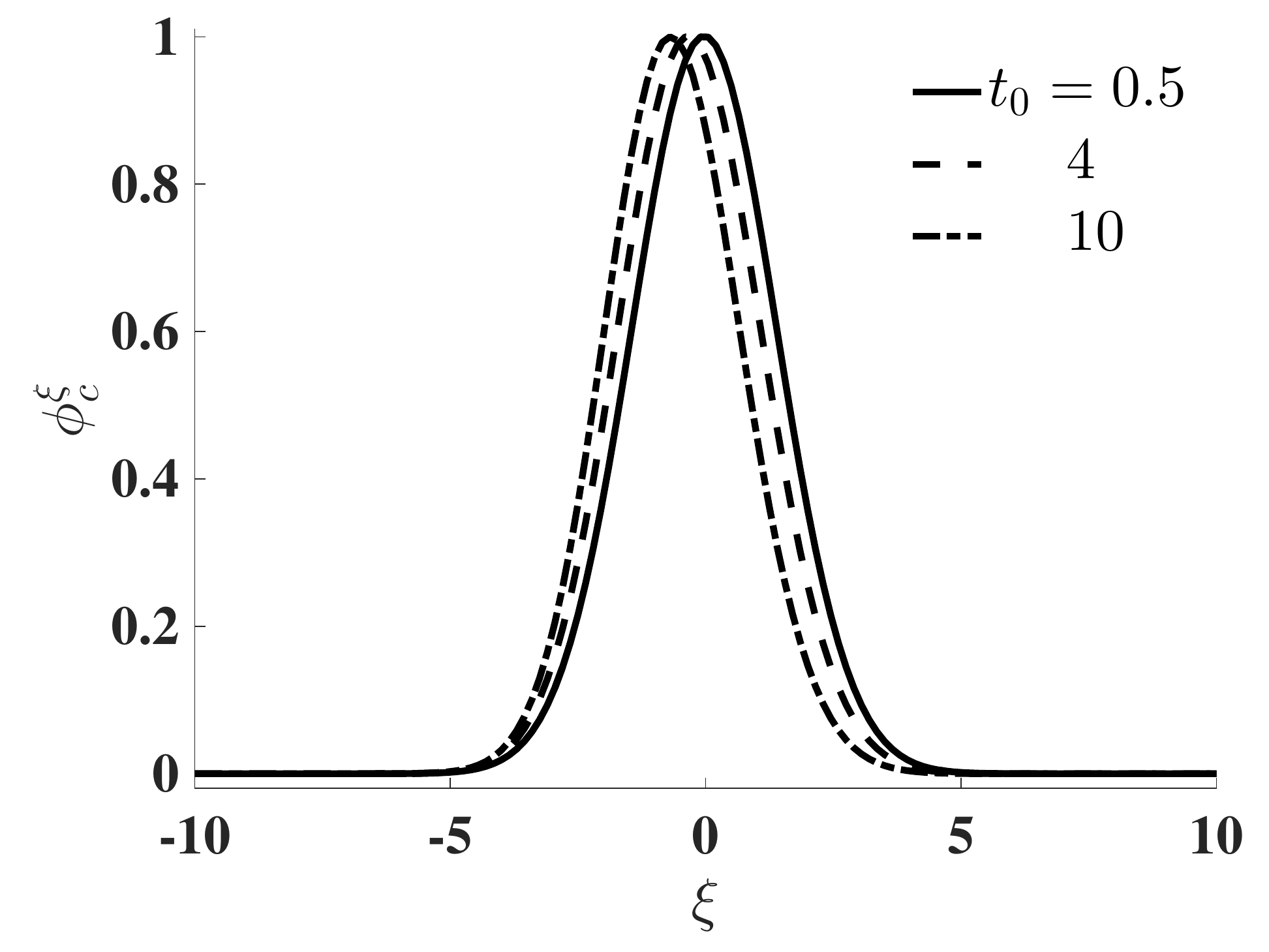}\\
	(c)\\
	\includegraphics[width=1.7in, keepaspectratio=true, angle=0]{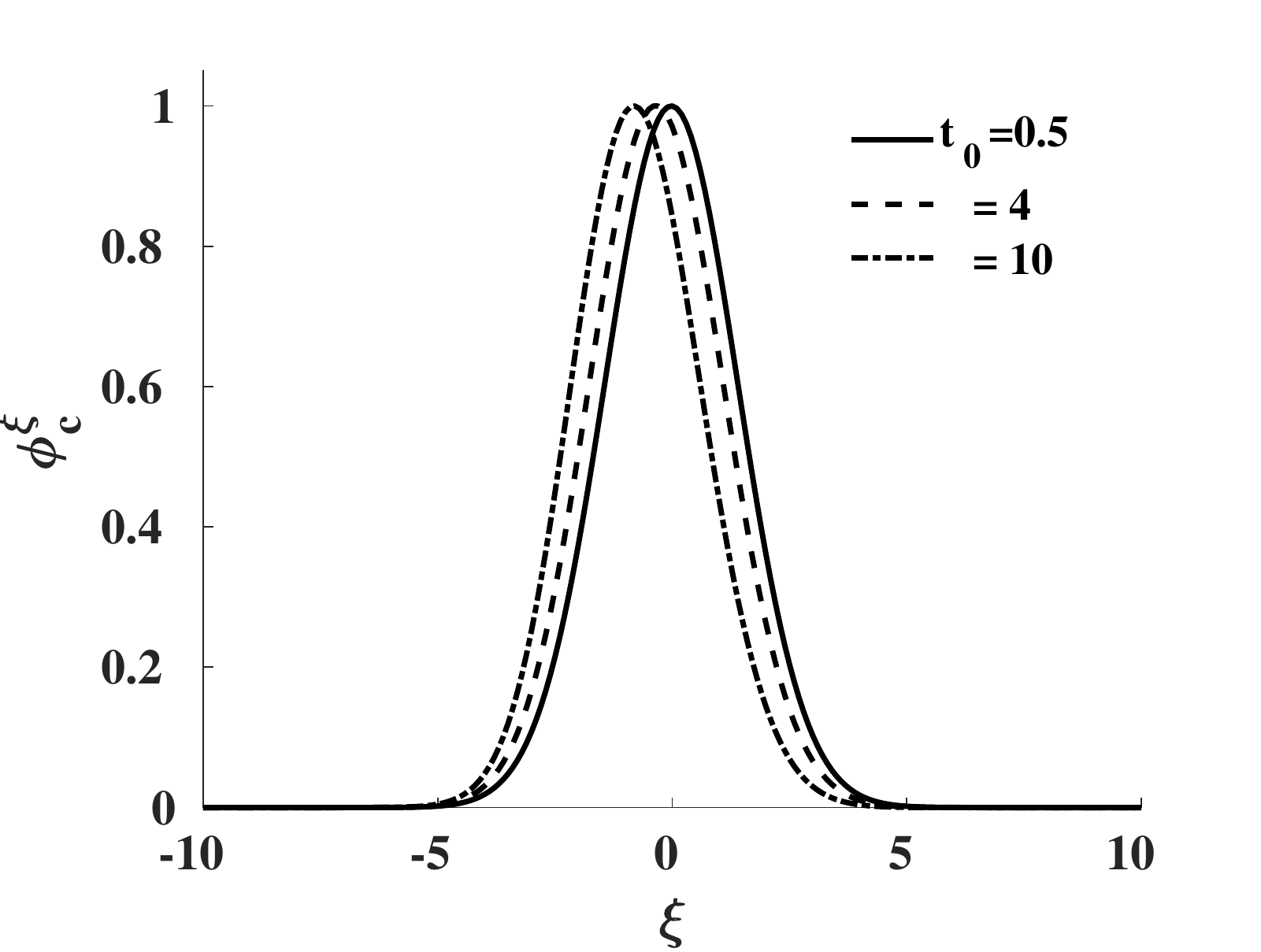}\\
	\caption{Quasi-steady eigenfunctions, $\phi_c^\xi$, of the linearized operator $\mathcal{L}$ (see equation \eqref{eq:qssa_EVP_chap5}) for $\alpha =5, \mu_m=7.5, k=0.2$: $c_m =$ (a) $ 0.25$, (b) $0.5$ and (c) $0.75$, at different frozen time.}
	\label{fig:SSQSSA_modes_chap5}
\end{figure}

\subsection{Comparison with quasi-steady-state modal analysis}\label{subsec:SSQSSA_chap5}
In this section, we present the stability analysis based on quasi-steady state approximation in the self-similar $(\xi, t)$- domain,  which we abbreviated as SS-QSSA \citep{Pramanik2013}. Kim and Choi \cite{Kim2011} used QSSA2 for quasi-steady analysis in $(\xi,t)$ domain. In SS-QSSA approach the space and time dependencies can be separated by fixing the time at $t_0$ and the disturbances quantities from equation \eqref{eq:normalmode_chap5} are assumed to be
\begin{eqnarray}\label{eq:qssa_disturbance_1_chap5}
(\phi_c, \phi_u)(\xi, t) = (\phi_c^\xi, \phi_u^\xi)(\xi)e^{\sigma(t_0)t},
\end{eqnarray}
where $\sigma(t_0)$ is the perturbation (quasi-steady) growth rate at the time $t_0$, and $\phi_c^\xi$ and $\phi_u^\xi$ are concentration and velocity perturbations, respectively. On substituting equation \eqref{eq:qssa_disturbance_1_chap5} into equations \eqref{eq:IBVP_chap5}-\eqref{eq:LSAxiteqnA_chap5} produces an eigenvalue problem for eigenvalues $\sigma$ and eigenfunctions $\phi_c^\xi$, where $\mathcal{L}$ as in equation \eqref{eq:IBVP_chap5}. It is noted that here we are presenting the temporal stability analysis, i.e. the non-dimensional wave number, $k$  is a real number, whereas, the growth rate, $\sigma(t_0)$ can be allowed to be a complex number. Numerically, the temporal stability analysis has been performed by computing the leading eigenvales, i.e, $\sigma(t_0) = \max \Re \left[ \Lambda(\mathcal{L})\right]$, the spectral abscissa of the stability matrix $\mathcal{L}(t_0)$. Here $\Lambda$ denote the set of all discrete eigenvalues of $\mathcal{L}$. It is observed that the SS-QSSA eigenfunctions are concentrated around the base state. However, these concentration eigenfunctions fail to capture the quadruple structure and consequently, only predict the temporal growth of disturbances after an initial transient period.

Fig. \ref{fig:SSQSSA_modes_chap5} demonstrates the spatial variations of SS-QSSA eigenfunctions at time $t_0 = 0.5, 4$ and $10$. The parameters used are $\alpha =5, \mu_m =7.5, k=0.2$ and $c_m = 0.25, 0.5$ and $0.75$. It is observed that SS-QSSA typically produce the dominant eigenfunctions that are qualitatively very different from the corresponding optimal initial perturbations, e.g., the typical quadruple structure of perturbations that are obtained from NMA [see Fig. \ref{fig:perturb_struct_chap5}] are not captured by SS-QSSA eigenfunctions as evident from Fig. \ref{fig:SSQSSA_modes_chap5}(a)-(c). A comparison of  quasi-steady eigenfunctions in $(x,t)$ and $(\xi,t)$ domain are discussed in Appendix \ref{App-B}.

With this substantial difference in the structure of the quasi-steady eigenfunctions in comparison to the optimal perturbations, we move to compare the growth rate determined from NMA and SS-QSSA.The growth rate in SS-QSSA can obtained by analysing the spectral abscissa, $\sigma(\mathcal{L})$ which is defined as the collection of numbers $\sigma(t_0)$ satisfying
\begin{eqnarray}\label{eq:qssa_EVP_chap5}
\mathcal{L}(t_0) \phi_c^\xi = \sigma(t_0)\phi_c^\xi,
\end{eqnarray}
where $t_0$ is the frozen time at which $\sigma(t_0)$ is determined.

\begin{figure}
	\centering
	(a)\hspace{1.6in}(b)\\
	\includegraphics[width=1.6in, keepaspectratio=true, angle=0]{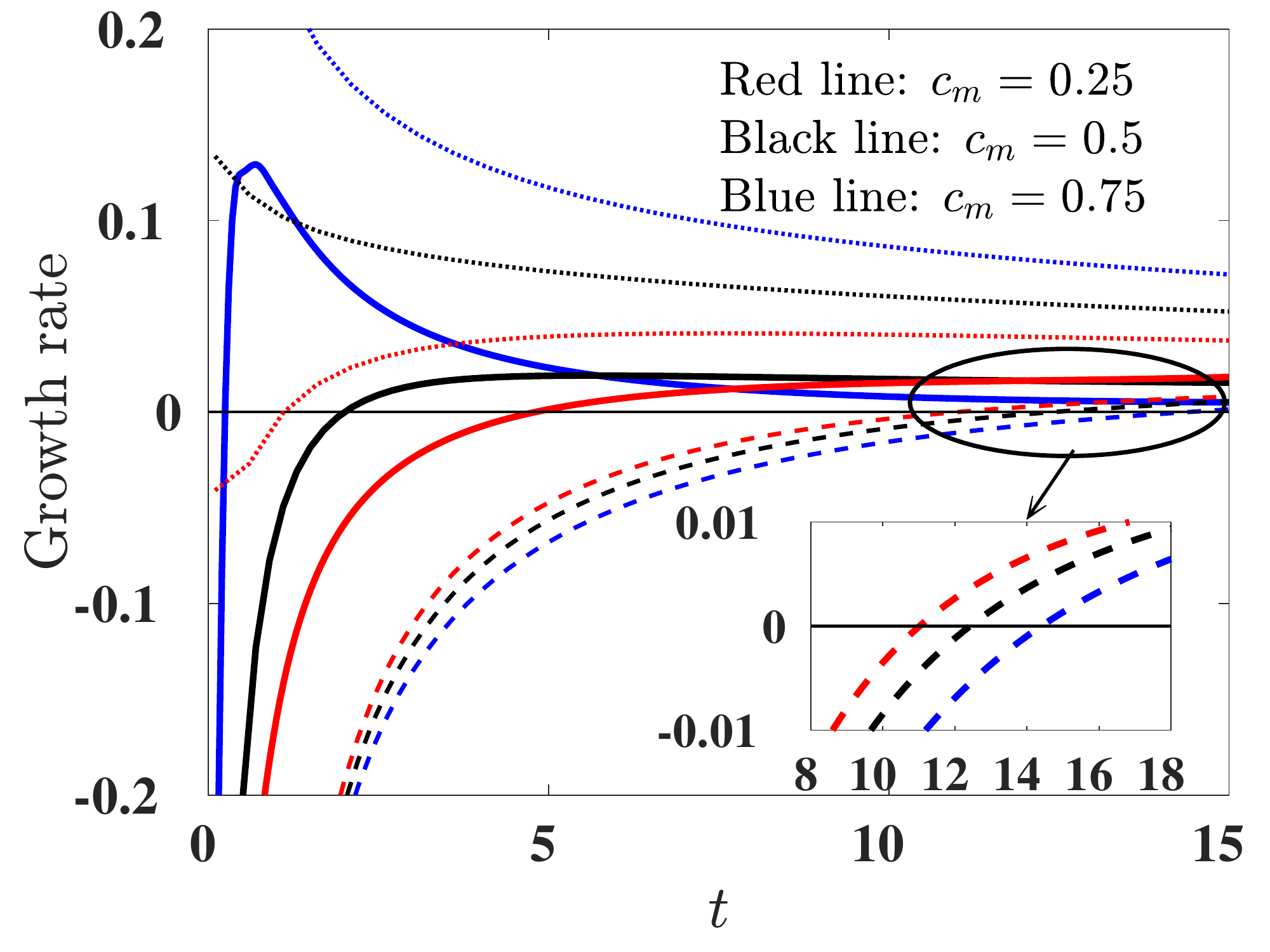}
	\includegraphics[width=1.6in, keepaspectratio=true, angle=0]{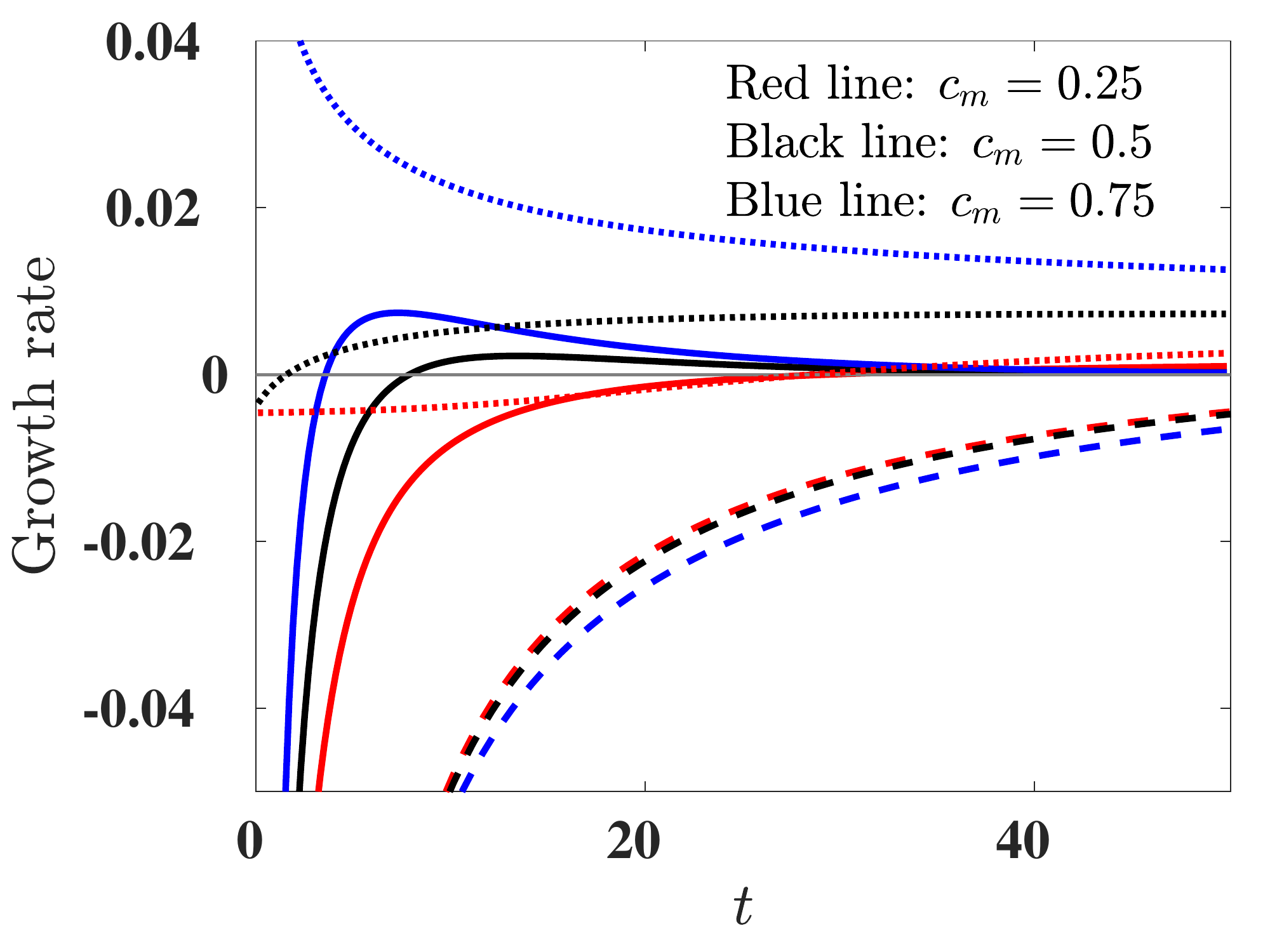}\\
	(c)\\
	\includegraphics[width=1.6in, keepaspectratio=true, angle=0]{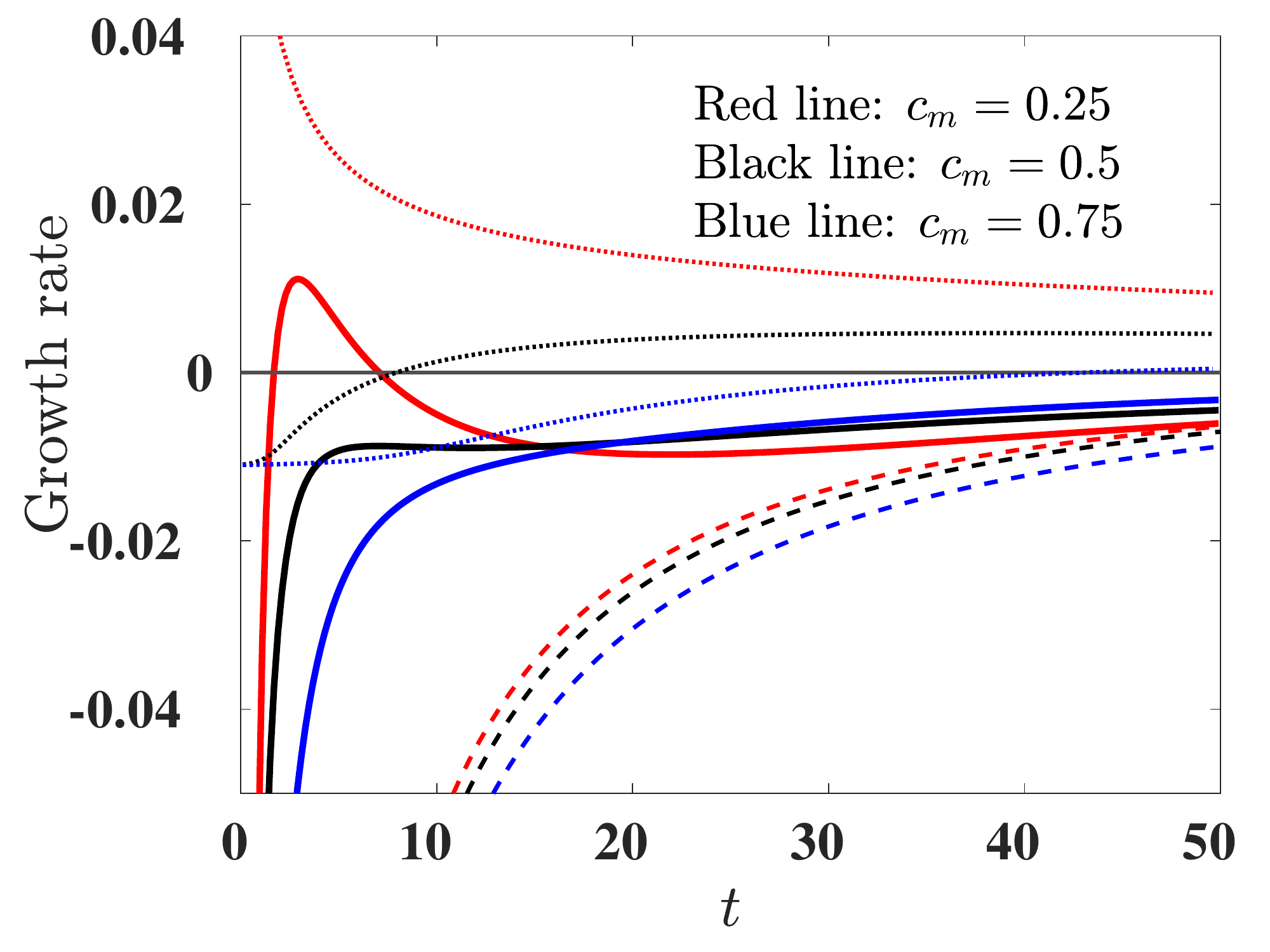}\\
	\caption{For viscosity profiles, equation \eqref{eq:visco-concen_chap5}, growth rate with (a) $\alpha =5, \mu_m =7.5$ and $k=0.2$, (b) $\alpha =1, \mu_m =2$ and $k=0.06$ and and (c)  $\alpha =0.5, \mu_m =2$ and $k=0.1$: the growth rate determined from NMA (continuous lines), SS-QSSA (dashed lines) and QSSA (dotted lines). The NMA growth rate are determined from, $\displaystyle \frac{1}{G(t)}\frac{\mbox{d} G(t)}{\mbox{d}t}$.}
	\label{fig:growth_comparison_chap5}
\end{figure}
Fig. \ref{fig:growth_comparison_chap5} illustrates the growth rate obtained from SS-QSSA, QSSA and NMA. In Fig. \ref{fig:growth_comparison_chap5}(a)-(b), for $\alpha =5$ and $1$, it is observed that the temporal evolution of growth rates determined from NMA and SS-QSSA are opposite to each other. Although both NMA and SS-QSSA predicts that the system is unconditionally stable at early times, but the later approach illustrates that the onset of instability is decreasing function of $c_m$ which is contrast to the result of NMA [see Fig. \ref{fig:optimal_ampl_chap5}(b)]. Interestingly, the growth trend of perturbations are same in the case of NMA (continuous lines) and QSSA (dotted lines). Manickam and Homsy\cite{Manickam1993} used the QSSA and suggested that when the parameter $\chi$ (see equation \eqref{eq:chi_chap5}) is positive, the flow is always unstable, and when $\chi$ is negative, the initially stable flow becomes unstable as the base flow diffuse. For the parameters given in Fig. \ref{fig:growth_comparison_chap5}(a), $\chi = -2.12, 3.58$ and $14.05$, for $c_m = 0.25, 0.5$ and $0.75$, respectively. But, the flow is stable for both $\chi > 0$ and $\chi <0$. Thus, it is observable that QSSA unpredcits the onset of instability for $c_m =0.5$ and $0.75$. Furthermore, for $\alpha =5$, the trend of onset obtained from NMA is in agreement with NLS results [see Fig. \ref{fig:NLS_chap5}], where it is shown that the onset of finger is early with the increase in values of $c_m$. This suggests that the deviation of the structure of discrete eigenfunctions and eigenvalues that of from the optimal initial perturbations and nonmodal growth, at small times, is primarily due to the the non-orthogonality of the quasi-steady eigenmodes. For $\alpha =0.5$, Fig. \ref{fig:growth_comparison_chap5}(c) shows that independent of the linear stability approach, the onset is delayed with increase in $c_m$. One important point can be noted from Fig. \ref{fig:growth_comparison_chap5}(c) is that the system can be unstable as predicted by NMA for $c_m = 0.25$ as oppose to SS-QSSA, in which system is always stable for the given parameters.  
\begin{figure}
	\centering
	(a)\hspace{1.6in}(b)\\
	\includegraphics[width=1.6in, keepaspectratio=true, angle=0]{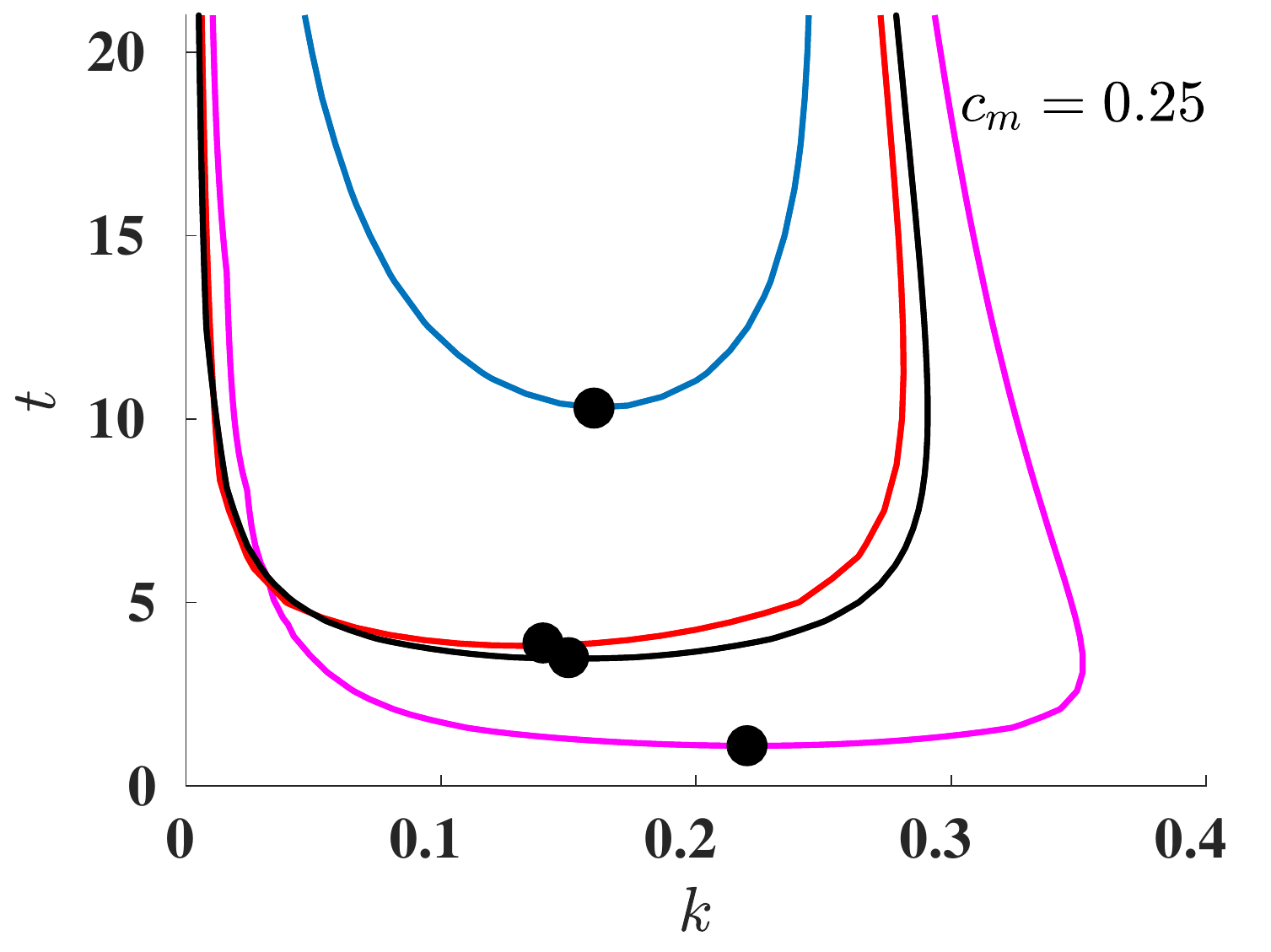}
	\includegraphics[width=1.6in, keepaspectratio=true, angle=0]{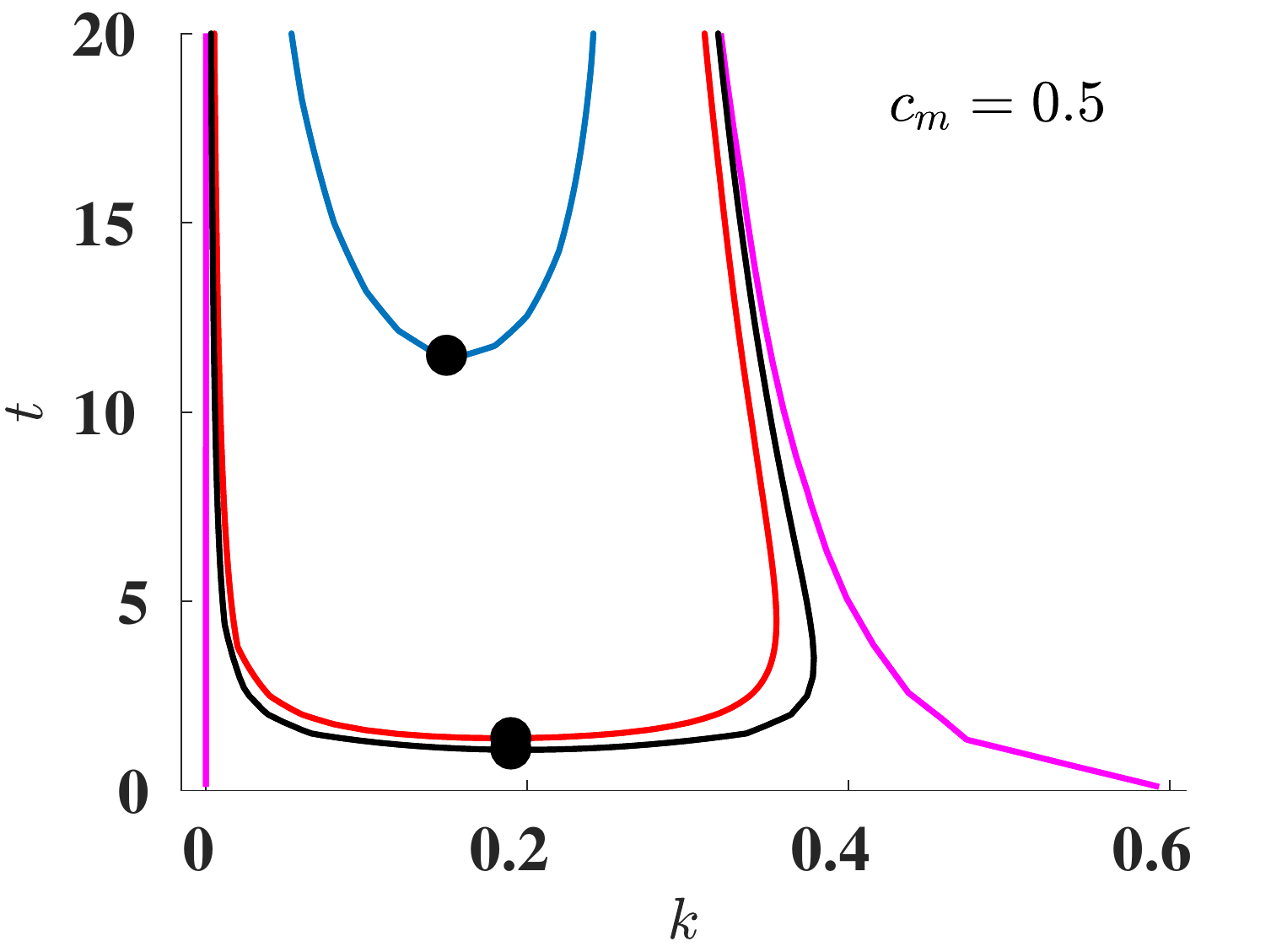}\\
	(c)\\
	\includegraphics[width=1.7in, keepaspectratio=true, angle=0]{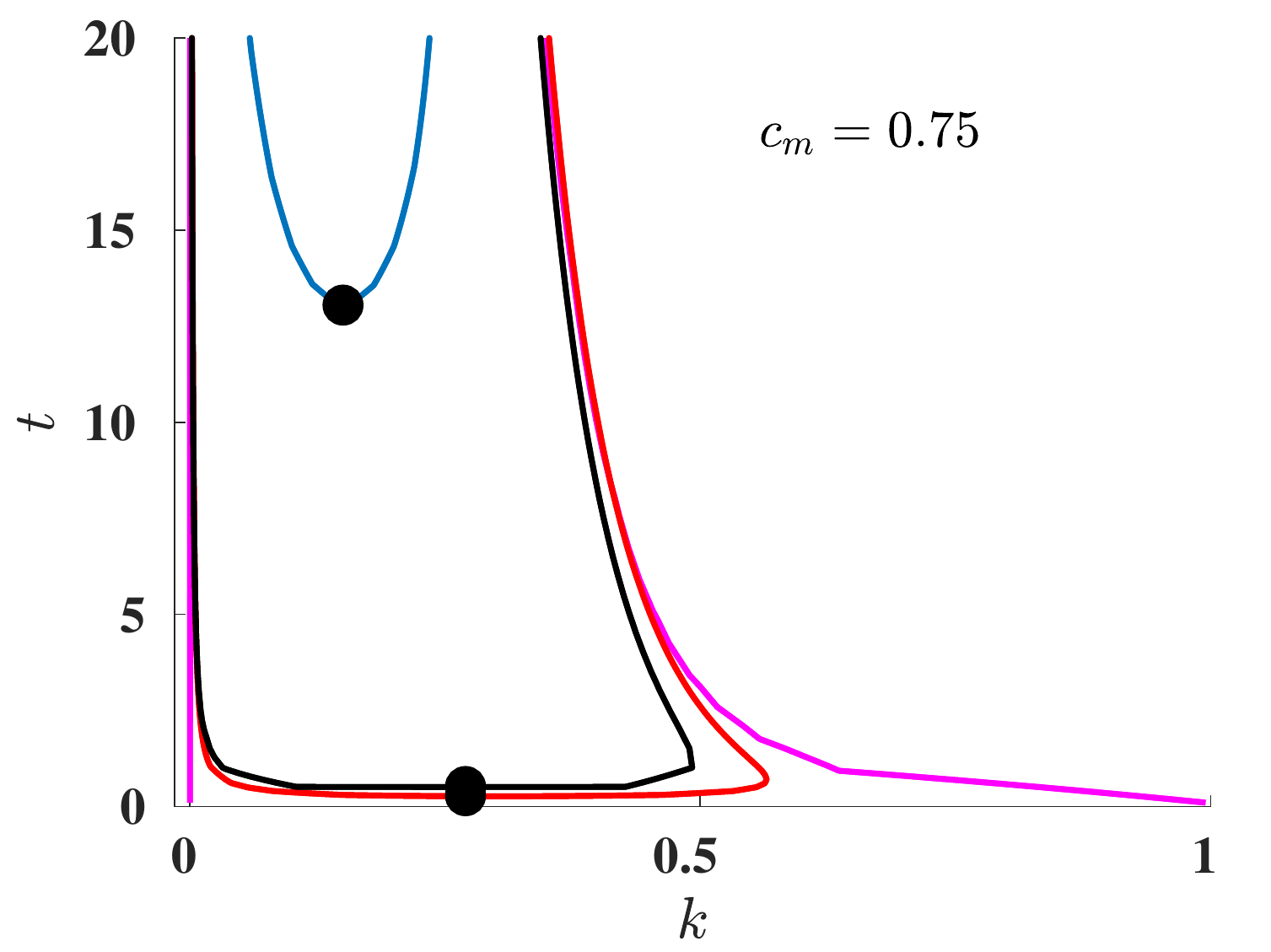}\\
	\caption{Comparison of neutral curves obtained from SS-QSSA (blue line), NMA (red line), NLS (black line) and QSSA (magenta line) for $\alpha =5, \mu_m=7.5$: $c_m =$ (a) $ 0.25$, (b) $0.5$ and (c) $0.75$. The lowest point of each of these curves, marked with the solid dots ($\CIRCLE$), corresponds to the critical  wave number, $k_c$ and critical time, $t_c$.}
	\label{fig:Neutral_curves_1}
\end{figure}

Further, in order to validate the advantage of non-modal approach, we compare the results of NMA with those of non-linear simulations (NLS). We obtained the growth rate of concentration perturbations by introducing a sinusoidal perturbations of the form 
\begin{equation}\label{eq:FSM_IC}
c'(x,y, t_0) =   \left\{
\begin{array}{ll}
\epsilon \cos(ky), & x = x_i \\
0, & \mbox{otherwise} \\
\end{array}, 
\right. 
\end{equation}
where $x_i$ is the position of unperturbed interface, $k$ is the nondimensional wave number, $t_0$ is the time when perturbations are introduced and $\epsilon$ is the amplitude of the perturbation, which is taken as $10^{-3}$. See Appendix \eqref{App-A} for more detail.

Fig. \eqref{fig:Neutral_curves_1} demonstrate the neutral curves  obtained from SS-QSSA (blue line), NMA (red line), NLS (black line) and QSSA (magenta line), in the $(k, t)$ plane. The neutral curves show the combinations of $k$ and $t$ for which $\sigma = 0$. The area above each curve determines the unstable region whereas the region below shows the stable region. The solid dots ($\CIRCLE$) mark the critical points $(k_c , t_c )$ at which perturbations initially become unstable.

In Fig. \eqref{fig:Neutral_curves_1}, we observe that the SS-QSSA analysis predict qualitatively very different behavior from rest of the neutral curves, i.e. SS-QSSA predicts with increase in the value of $c_m$, the stable region increases which in contrast to the other methods. Although QSSA analysis qualitatively agree with NLS and NMA results but some of the perturbation that are
judged unstable by QSSA turned out to be stable as shown in Fig. \eqref{fig:Neutral_curves_1} (b) \& (c). In Table \eqref{table:stable-neutral-curve} it is shown that the critical time, the unstable region, and dominant wave numbers determined from NLS and NMA shows excellent agreement.
It is also observed that NMA and NLS results shows that the critical wave number $k_c$ is an increasing function and the critical time, $t_c$ is decreasing function, of $c_m$ which is in contrast to the results obtained from SS-QSSA. It can be noted here that when discussing the physical relevance of QSSA analyses, it is important to recall that in physical systems, perturbations usually arise due to noise that excites many eigenmodes simultaneously. Consequently, modal analyses only predict which perturbations will dominate after an initial transient period.  Due to this reason we have found that QSSA, NMA and NLS results are almost identical at later time.

Hence, it can be conclude that irrespective of viscosity-concentration profile, the quasi-steady eigenvalues does not predict the accurate growth rate at early times. To analyse the early spatial and temporal evolution of perturbations, NMA is a suitable approach. Our main focus in the present article is to determine the onset of instability and describe the physical mechanism of instability. Thus, the effect of injection-driven flow and the influence of lifting in presence of non-monotonic viscosity profiles are may not directly incorporated. But, we are hopeful to explore these effects in near future.
\begin{table*}
	\centering
	\begin{tabular}{ccccc}
		\hline
		\hline
		& ~~~~~~~ &  $(k_c, t_c)$ for $\alpha =5$ and $\mu_m =7.5$& ~~ &  \\
		\cline{2-5}\\
		$c_m$ ~~ &\multicolumn{1}{c}{SS-QSSA}   ~~ &\multicolumn{1}{c}{NMA} ~~ &\multicolumn{1}{c}{NLS}   ~~ &\multicolumn{1}{c}{QSSA}\\ 
		$0.25$ ~~ & $(0.16, 10.3)$ ~~& $(0.14, 3.9)$ ~~ & $(0.15, 3.5)$ ~~ & $(0.22, 1.1)$  \\
		$0.5$ ~~ & $(0.15, 11.5)$ ~~& $(0.19, 1.4)$  ~~ & $(0.19, 1.1)$ ~~ & $(-,-)$ \\
		$0.75$ ~~ & $(0.15, 14)$ ~~& $(0.27,0.52)$ ~~ &  $(0.28, 0.27)$ ~~ & $(-,-)$ \\
		\hline
	\end{tabular}
	\caption{\label{table:stable-neutral-curve}The critical time, $t_c \equiv \min \{\tau : \sigma(\tau) \geq 0, \forall k\}$ and critical wave number, $k_c \equiv \min \{k: \sigma(t_c) =0\}$ from each of neutral curves illustrated in Fig.\eqref{fig:Neutral_curves_1}. The onset time determined from NMA and NLS are indistinguishable. Further, for $c_m = 0.5$ and $0.75$, QSSA shows that the system becomes unstable immediately.}
\end{table*}

\section{Conclusion}\label{sec:conclusion_chap5}
The influence of the non-monotonic viscosity-concentration relationship on miscible displacements in porous media is studied for the rectilinear flows. Due to the time-dependency of the stability matrix, we have used the non-modal linear stability (NMA) approach based on the singular value decomposition of the propagator matrix. This approach by construction accommodate all types of the initial conditions and hence give the optimal amplification and optimal perturbation structure. Based on the non-modal linear stability analysis, the non-monotonic viscosity-concentration relationships, proposed by Manickam and Homsy\cite{Manickam1993} are characterized by the three parameters, namely, end-point viscosity contrast, maximum viscosity, $\mu_m$ and the concentration that maximizes the viscosity, $c_m$. The stability results are interpreted in detail, based on the optimal concentration perturbations. This is in marked contrast to Manickam and Homsy\cite{Manickam1993} who used the vorticity perturbation to describe the stability mechanism. Further, the NMA results demonstrate that each of the three parameters has a significant influence on the onset of instability and the shape of eigenfunctions. We notice, for a less viscous fluid displaces a more viscous fluid, an increase in the maximum concentration, $c_m$ generally leads to a more unstable flow. This result is in contrast with earlier linear stability results based on eigen-analysis \citep{Kim2011}. Further, we have observed that the sign of the parameter $\chi$ is not a helpful to characterize the dynamics of perturbation growth. Whereas, the reverse scenario is observed when a more viscous fluid displaces a less viscous fluid. Hence, our findings suggest that the onset and the dynamics of the disturbances obtained by previous investigators using quasi-steady-approximation and eigen-analysis, can be misleading. Moreover, the present analysis describes the physical mechanism which is studied using the singular value decomposition of the propagator matrix, is in accordance with the nonlinear simulations of Manickam and Homsy\cite{Manickam1993, Manickam1994}. It can be concluded that, for non-monotonic viscosity profiles, NMA approach can describe the onset of instability and the underlying physical mechanism of instability, more accurately. Furthermore, the present linear stability analysis can be helpful to understand the effect of non-monotonic viscosity profiles in miscible reactive flows \citep{Hejazi2010} and double diffusive convection \citep{Mishra2010}.
\appendix
\section{Growth rate from Fourier Pseudo-spectral method} \label{App-A}
Stream function form for the dimensionless equations \eqref{eq:cont_eqn}-\eqref{eq:convec_diffuse} in a Lagrangian frame of reference moving with the speed $U$ in the downstream direction are
\begin{equation}
\left.\begin{aligned}\label{eq:FSM_1}
&&\nabla^2 \psi = - \frac{\mbox d \ln(\mu)}{\mbox d c} \left[ \nabla \psi \cdot \nabla c + \frac{\partial c}{\partial y}\right],\\
&&\frac{\partial c}{\partial t} + \frac{\partial \psi}{\partial y} \frac{\partial c}{\partial x} -\frac{\partial \psi}{\partial x} \frac{\partial c}{\partial y} = \nabla^2 c .
\end{aligned}
~~\right\}
\end{equation}
For the base state $\vec{u}_b=(0,0)$(equivalently $\psi_b =$ constant) and $c_b = \frac{1}{2}\bigg[\text{erfc}\left(\frac{x}{2\sqrt{t}}\right)\bigg]$, introduce an infinitesimal perturbations, $\psi = \psi_b+\psi'$ and $c = c_b + c'$ to equation \eqref{eq:FSM_1}. The boundary conditions associated are given by
\begin{equation}
\left.\begin{aligned}\label{eq:FSM_BC}
(c', \psi')(x,y,t) = (0,0),&~ \text{at}~ x=0~ \text{and}~ A\mbox{Pe},\\
(c', \psi')(x,y,t) = (0,0),&~ \text{at}~ y=0~ \text{and}~ \mbox{Pe},
\end{aligned}
~~\right\}
\end{equation}
where Pe is the P\'eclet number, $A=L/H$ is the aspect ratio, and $L$ \& $H$ are the length \& width of computational domain, respectively. We have adopted Fourier Pseudo-spectral method to solve the system equation \eqref{eq:FSM_1} subject to the boundary conditions equation \ref{eq:FSM_BC} and initial condition equation \eqref{eq:FSM_IC}. Then we obtained the spatio-temporal evolution of the perturbation quantities, $c'(x,y,t)$ and calculate the growth rates associated with concentration perturbations \citep{Kumar1999,Hota2015b},
\begin{equation}\label{eq:NLS_growth}
\sigma(t) = \frac{1}{2 E(t)}\frac{\mbox d E(t)}{\mbox d t},
\end{equation}
where the amplification measure is given by $E(t) = \int_{0}^{A\text{Pe}} \int_{0}^{\text{Pe}} (c'(x,y,t))^2\; \mbox{d}x\mbox{d}y$. Following Hota \textit{et al.}\cite{Hota2015b} we have used equation
\eqref{eq:NLS_growth} to quantify the growth rate of disturbances and the onset of instability from nonlinear simulations.


\section{Transformation of growth rate from $(\xi, t)$ co-ordinates to $(x,t)$ coordinates}\label{App-B}
We define an energy $E_1(t)$ by
\begin{eqnarray} \label{energy_xt}
E_1(t) =  \displaystyle \frac{1}{2} \parallel c_1(t)\parallel^2_2,
\end{eqnarray}
where $c_1$ is the concentration perturbation in $(x,t)$ co-ordinate and $\parallel \cdot \parallel_2$ denotes the norm on $L^2(-\infty, \infty)$, i.e., $\displaystyle \parallel f(t) \parallel^2_2 = \int_{-\infty}^{\infty} f^2(x,t)\mbox{d}x$. The growth rate corresponding to the energy $E_1(t)$ is defined as $\displaystyle \sigma_1(t) = \frac{1}{E_1(t)} \frac{\mbox{d}E_1(t)}{\mbox{d}t}$. Now, using the self-similar transformation $ \displaystyle\xi (x, t) = \frac{x}{2 \sqrt{t}}$ and the chain rule $\displaystyle \frac{\partial}{\partial t} \bigg|_{(x,t)} = \frac{\partial}{\partial t} \bigg|_{(\xi,t)} - \frac{\xi}{2t} \frac{\partial}{\partial \xi} \bigg|_{(\xi,t)}$, we have
\begin{eqnarray} \nonumber
\displaystyle \frac{\mbox{d}E_1(t)}{\mbox{d}t} &=& \frac{1}{2} \int_{-\infty}^{\infty} \frac{\partial c_1^2}{\partial t}  \mbox{d}x = \int_{-\infty}^{\infty}  c_1 \frac{\partial c_1}{\partial t} \mbox{d}x\\ \nonumber
~ & = & \int_{-\infty}^{\infty}  c_2 \frac{\partial c_2}{\partial t}  \mbox{d}\xi - \int_{-\infty}^{\infty}  \frac{\xi}{2t} c_2(\xi,t) \frac{\partial c_2}{\partial \xi}  \mbox{d}\xi,
\end{eqnarray}
where $c_2$ is the concentration perturbation in $(\xi,t)$ co-ordinate and the associated energy is given by $ E_2(t) = \displaystyle \frac{1}{2} \int_{-\infty}^{\infty} c_2^2(\xi,t)\mbox{d}\xi$. Thus, we have 
\begin{eqnarray} \nonumber
&&\displaystyle \frac{\mbox{d}E_1(t)}{\mbox{d}t} = \frac{\mbox{d}E_2(t)}{\mbox{d}t} - \int_{-\infty}^{\infty}  \frac{\xi}{2t} c_2(\xi,t) \frac{\partial c_2}{\partial \xi} \mbox{d}\xi \\ \nonumber
&&\Rightarrow \displaystyle \frac{1}{E_1(t)}\frac{\mbox{d}E_1(t)}{\mbox{d}t} = \frac{1}{E_2(t)}\frac{\mbox{d}E_2(t)}{\mbox{d}t} - \frac{1}{E_2(t)}\int_{-\infty}^{\infty}  \frac{\xi}{2t} c_2 \frac{\partial c_2}{\partial \xi} \mbox{d}\xi \\ 
&&\Rightarrow \displaystyle \sigma_1(t) = \sigma_2(t)- \frac{1}{E_2(t)}\int_{-\infty}^{\infty}  \frac{\xi}{2t} c_2 \frac{\partial c_2}{\partial \xi} \mbox{d}\xi, 
\end{eqnarray}
where $\displaystyle \sigma_2(t) = \frac{1}{E_2(t)} \frac{\mbox{d}E_2(t)}{\mbox{d}t}$ is the growth rate in $(\xi, t)$ co-ordinate.

\begin{figure}
	\centering
	\includegraphics[width=2.in, keepaspectratio=true, angle=0]{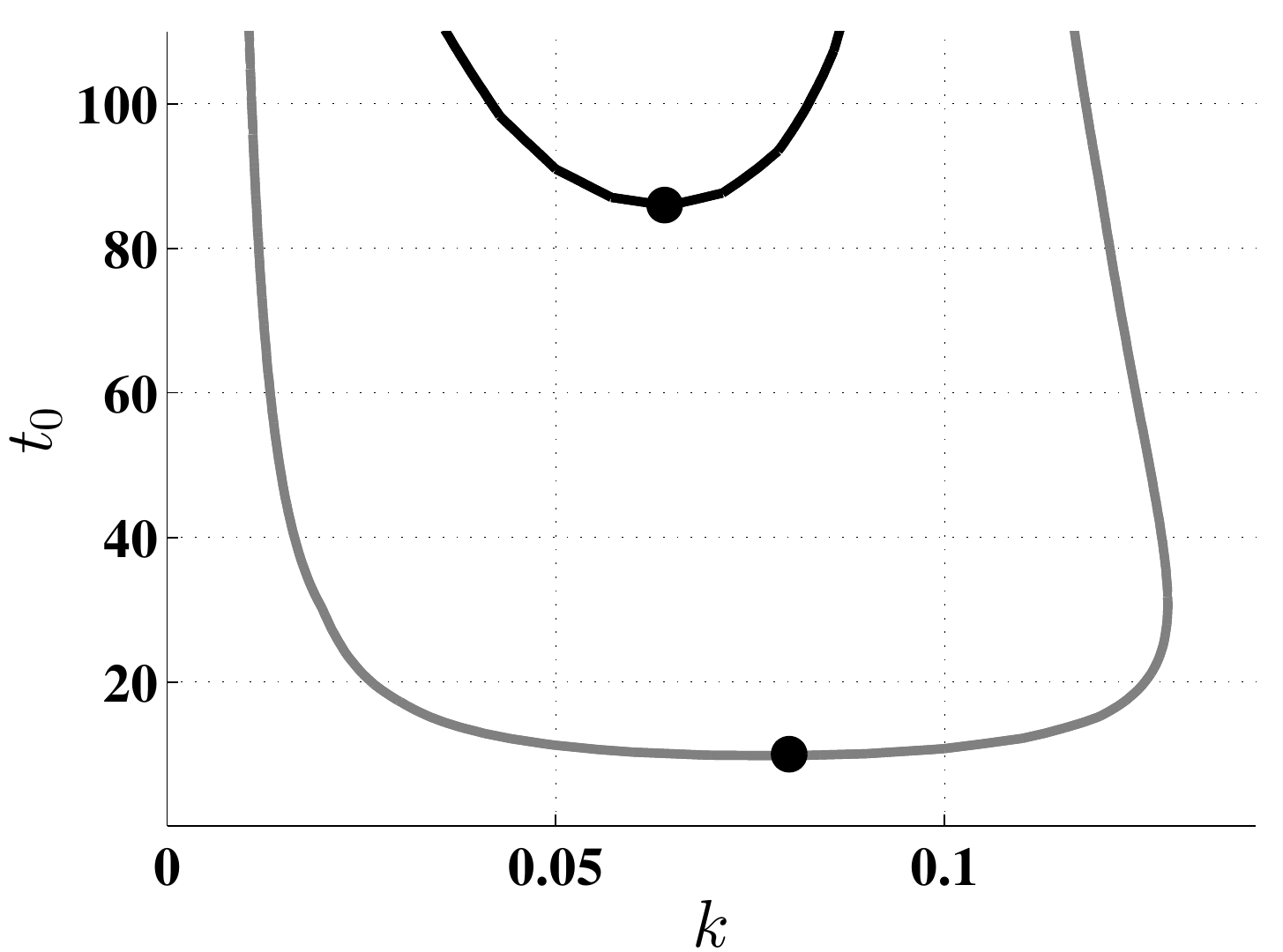}
	\caption{Comparison of the SS-QSSA (black line) and QSSA (grey line) neutral curves for $\alpha =1, c_m =0.4$ and $\mu_m =2$. The critical points $(0.08, 10)$ and $(0.064, 86)$, shown as solid dots ($\CIRCLE$) are obtained respectively from QSSA and SS-QSSA. It is illustrated that both the onset of instability and the corresponding critical wave number are significantly different for QSSA and SS-QSSA.}
	\label{fig:Neutral_curves_2}
\end{figure}

Fig. \ref{fig:Neutral_curves_2} demonstrates the neutral curves, i.e. $\sigma_i =0, i =1,2$, for $\alpha =1, c_m =0.4$ and $\mu_m =2$. The lowest point of each of these curves, marked with the solid dots ($\CIRCLE$), corresponds to the critical time, $t_{c,i}$ and critical wave number, $k_{c,i}$. The dissimilarities between the critical wave numbers (critical time) $k_{c,1}$ and $k_{c,2}$(similarly $t_{c,1}$ and $t_{c,2}$) are apparent. 

\section{Quasi-steady eigenmodes of linear stability matrix $\mathcal{L}(t)$ }\label{App-C}
\begin{figure}
	\centering
	(a)\hspace{1.5in}(b)\\
	\includegraphics[width=1.6in, keepaspectratio=true, angle=0]{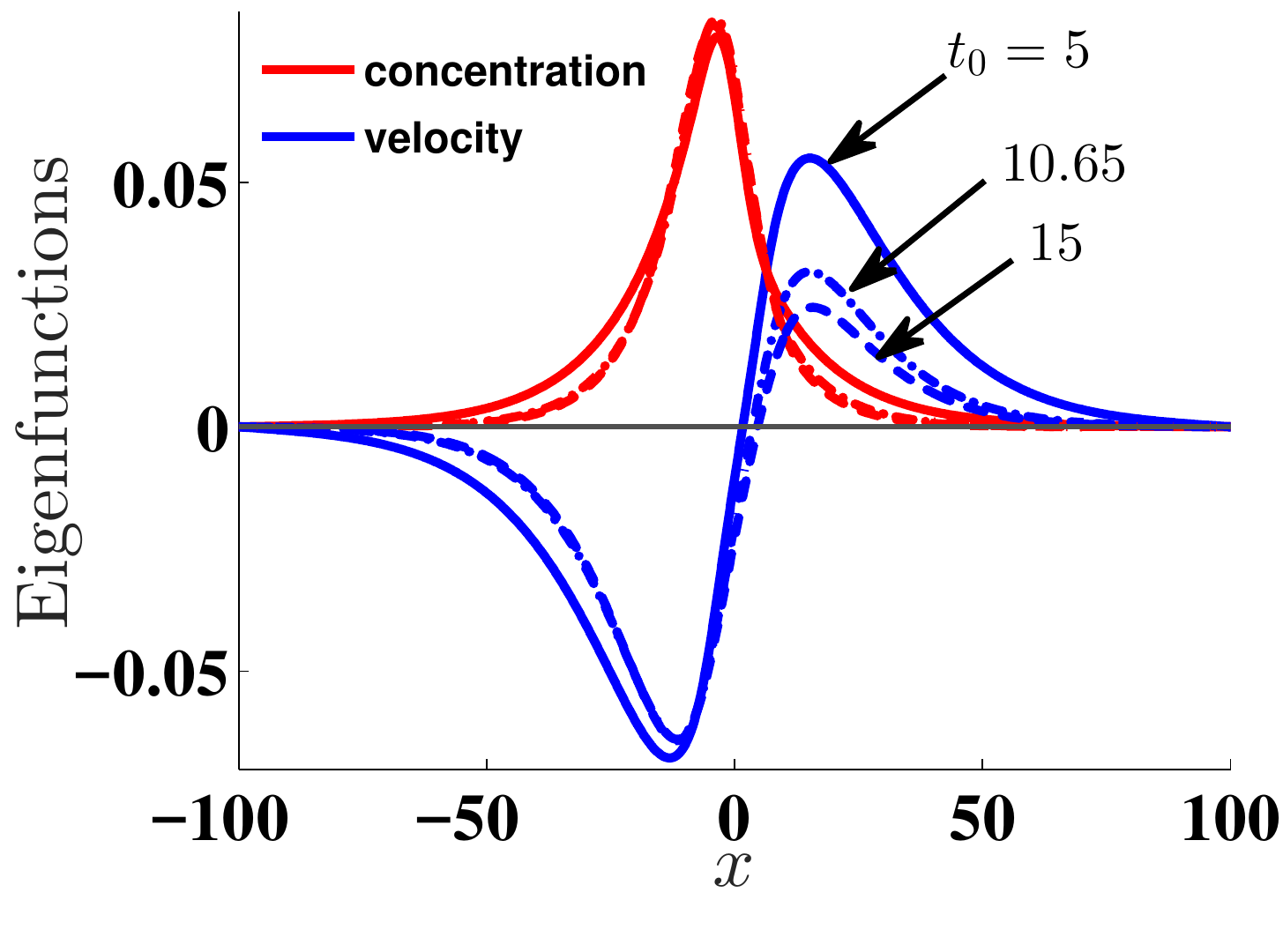}
	\includegraphics[width=1.6in, keepaspectratio=true, angle=0]{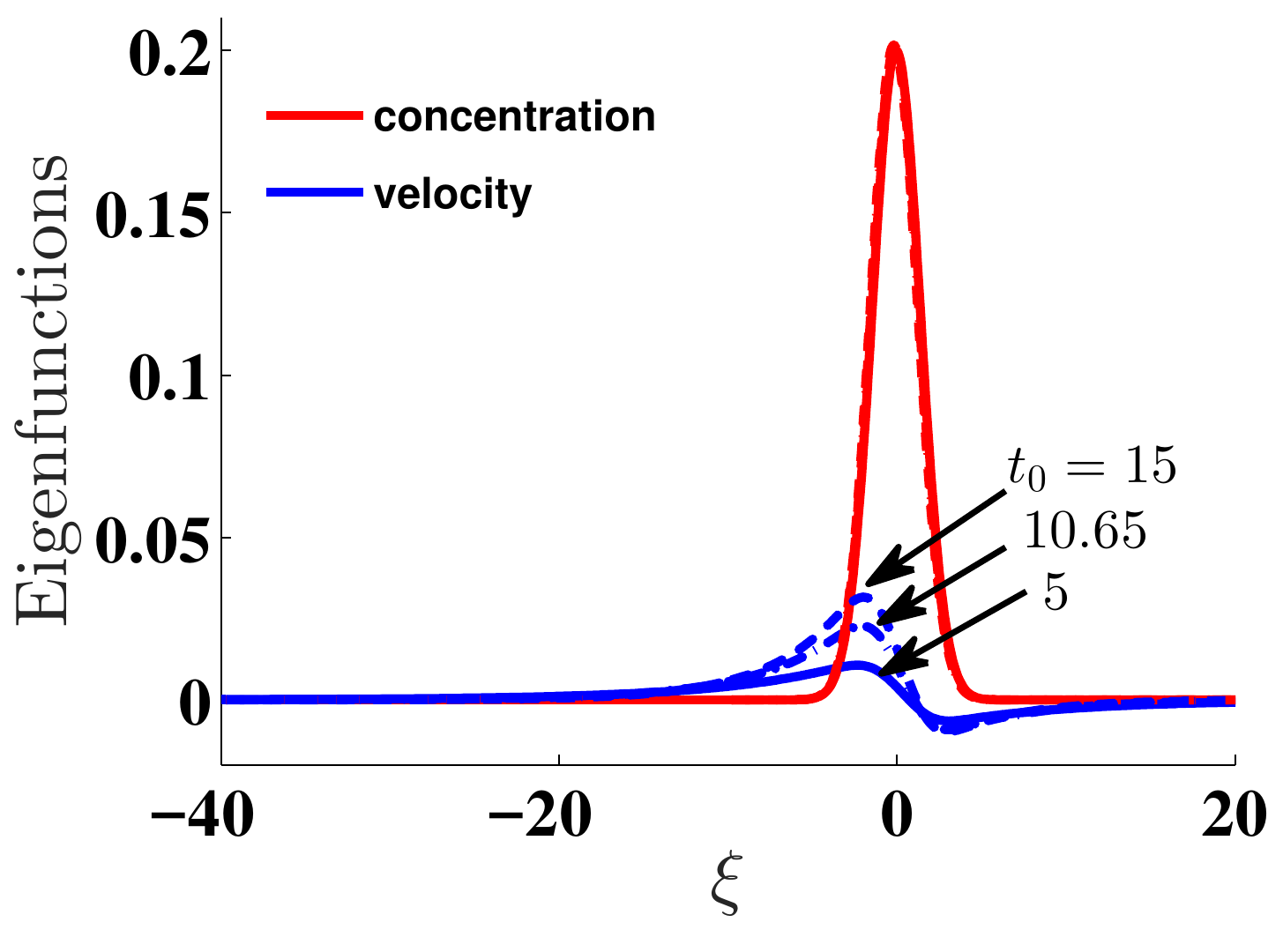}\\
	\caption{For the viscosity profile $\alpha =1, \mu_m =2, k=0.1$ and $c_m=0.4$, quasi-steady eigenfunctions obtained from (a) QSSA  and (b) SS-QSSA, for the least stable eigenvalue at different time, $t_0$. }
	\label{fig:eigenfunctions_modal_comparison}
\end{figure}

\begin{figure}
	\centering
	(a)\hspace{1.6in}(b)\\
	\includegraphics[width=1.6in, keepaspectratio=true, angle=0]{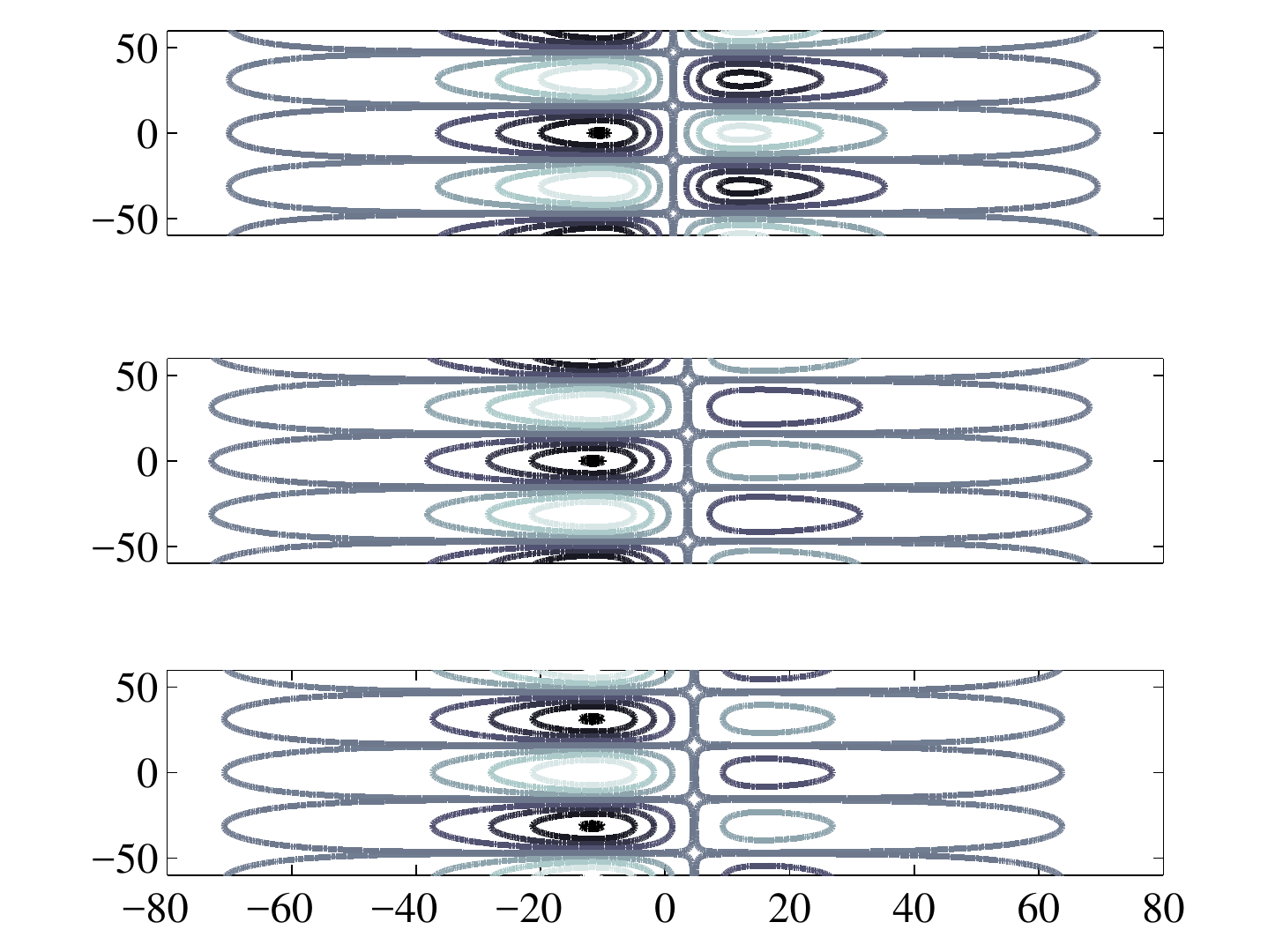}
	\includegraphics[width=1.6in, keepaspectratio=true, angle=0]{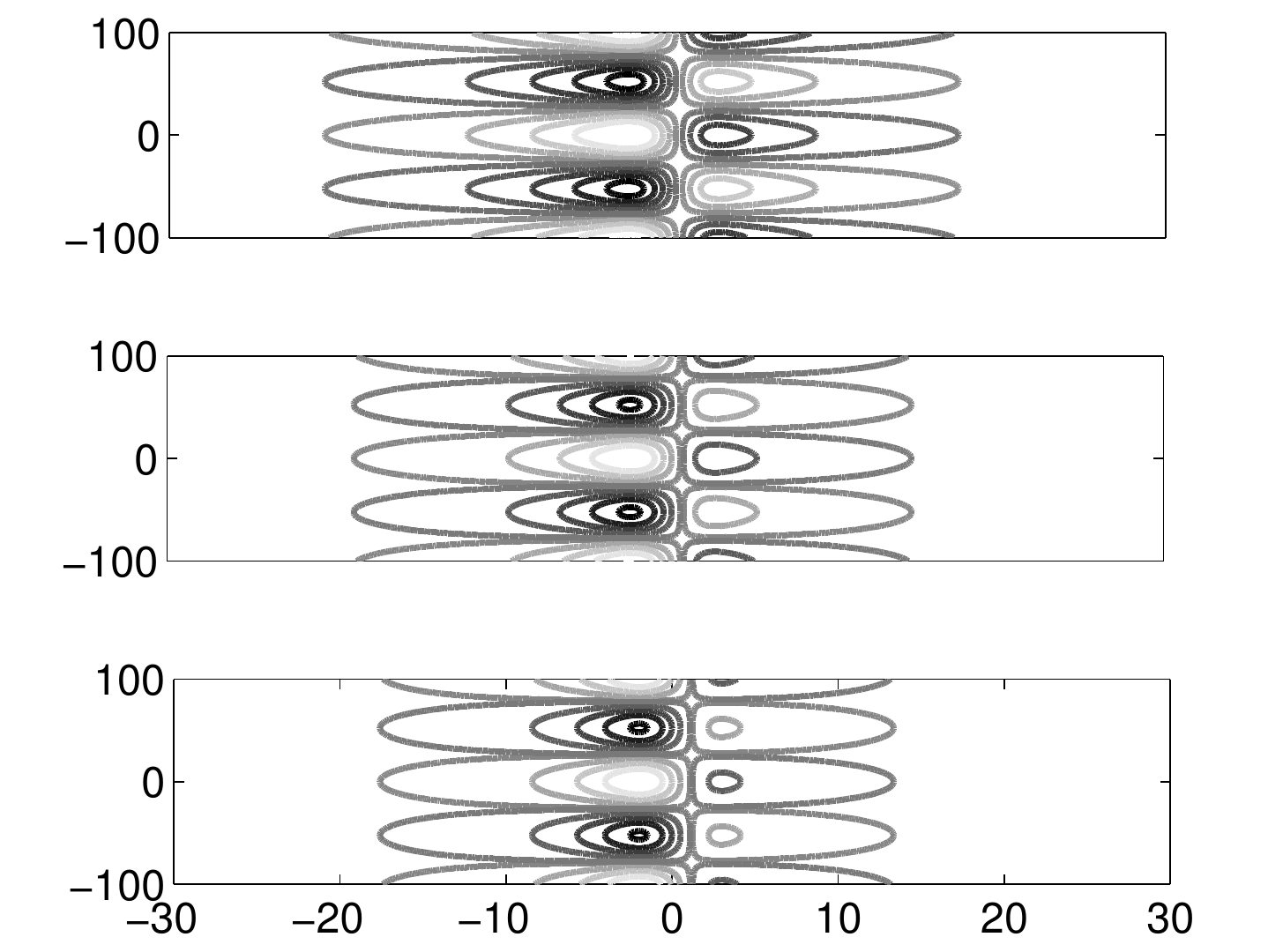}\\
	\caption{For the viscosity profile $\alpha =1, \mu_m =2, k =0.1$ and $c_m=0.4$, contours of quasi-steady velocity eigenfunctions obtained from (a) QSSA  and (b) SS-QSSA, for the least stable eigenvalue at different time, $t_0$. From top to bottom: $t_0=5, 10.65, 15$. The positive perturbations are plotted with black color lines and the negative perturbations with grey color lines. The velocity contours shown span from the minimum values of to the maximum values of velocity with four equal increments.}
	\label{fig:contours_modal_comparison}
\end{figure}

In order to study the physical destabilizing mechanism involved in non-monotonic viscosity profiles, Manickam and Homsy\cite{Manickam1993} examine the evolution of the eigensolutions of  $\mathcal{L}(t)$ at freezing times $t_0$ known as quasi-steady-state-approximation (QSSA). In contrast to Manickam and Homsy\cite{Manickam1993}, we have analyzed the evolution of eigenmodes in self-similar coordinate, $(\xi,t)$. To compare SS-QSSA eigenmodes to that obtained by Manickam and Homsy\cite{Manickam1993} we choose the following parameters: $k=0.1, \alpha =1, \mu_m =2,$ and $c_m=0.4$. The eigenfunctions associated with concentration and velocity perturbations obtained from QSSA and SS-QSSA are shown in Fig. \ref{fig:eigenfunctions_modal_comparison} and the velocity contour plots are illustrated in Fig. \ref{fig:contours_modal_comparison}. From these two figures it can be concluded that the eigenfucntions in $(\xi, t)$ coordinates are localized around the interface whereas eigenfunctions in $(x,t)$ are spanned all over the whole spatial domain, i.e. these eigenfunctions are global modes. This is the reason why some of the profiles that are judged unstable by QSSA analysis could turn out to be stable in SS-QSSA or in NMA. Further, the physical mechanism of fingering instability at early time can be studied by analyzing the velocity eigenfucntion instead of concentration eigenfunctions. 


\section{Quantifying the non-orthogonality of quasi-steady eigenmodes }\label{App-D}
\begin{figure}
	\centering
	\includegraphics[width=3.2in, keepaspectratio=true, angle=0]{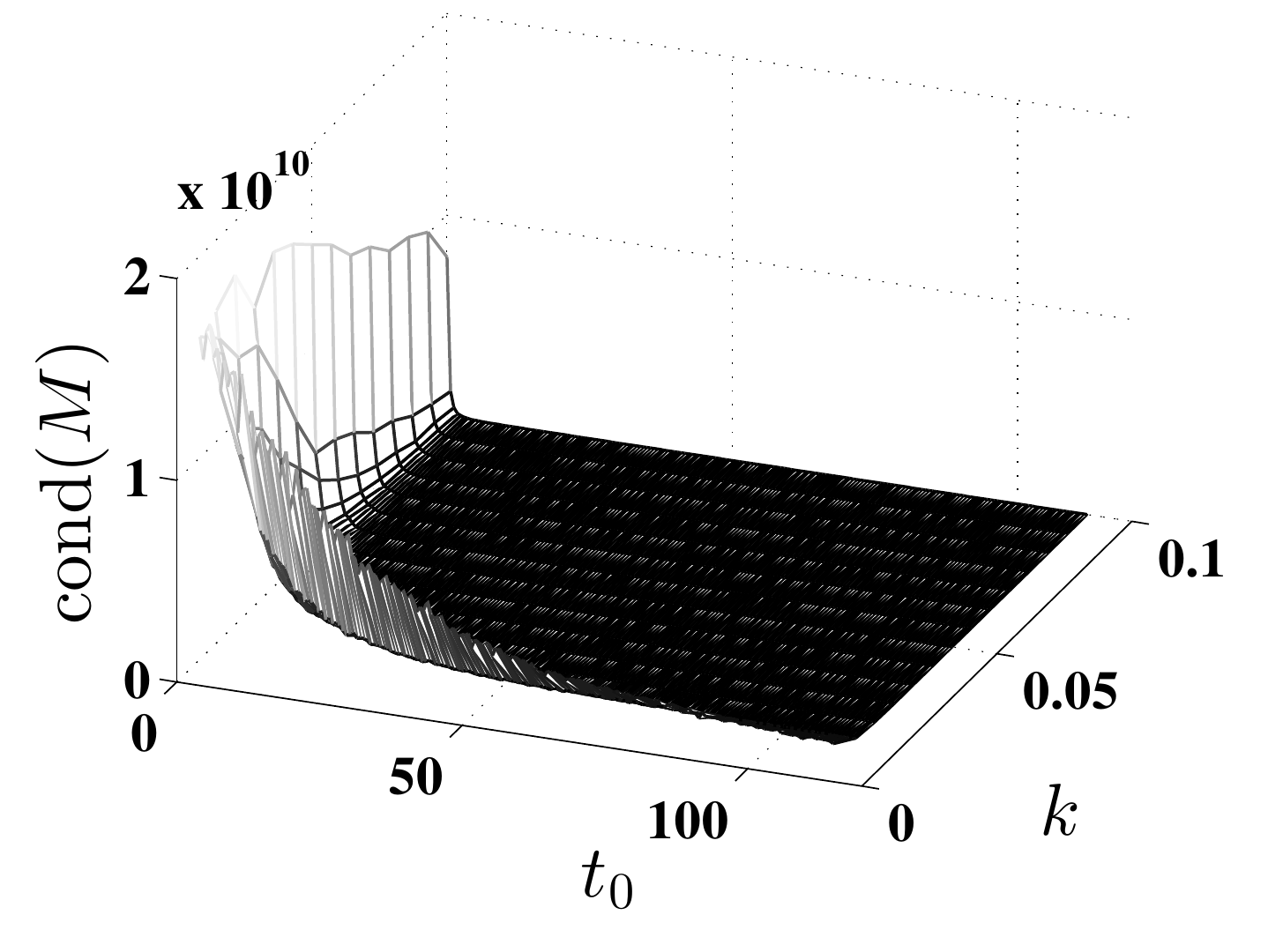}
	\caption{\label{fig:Gramian_velocity}For the viscosity profile $\alpha =1, \mu_m =2$ and $c_m=0.4$, condition number of Gramian matrix $M$ considering first $6$ quasi-steady velocity eigenfunctions obtained from SS-QSSA.}
	
\end{figure}
The extended duration of transient period can be illustrated by analyzing the interaction of non-orthogonality of quasi-steady eigenfunctions. To measure the non-orthogonality, let us consider the Gramian matrix, $M$ which is defined as \citep{Bhatia1997}
\begin{equation}
M=\begin{pmatrix}
\langle f_1, f_1 \rangle  & \langle f_1, f_2 \rangle &\ldots &\langle f_1, f_n \rangle\\
\langle f_2, f_1 \rangle  & \langle f_1, f_2 \rangle &\ldots & \langle f_1, f_n \rangle\\
\vdots & \vdots & \vdots &\vdots\\
\langle f_1, f_1 \rangle  & \langle f_1, f_2 \rangle &\ldots & \langle f_1, f_n \rangle
\end{pmatrix},
\end{equation}
where $\langle f_i, f_j \rangle = \int_{-\infty}^{\infty} f_i(x,t) \overline{f_j}(x, t) \mbox{d}x$, $\overline{f_j}$ denote the complex conjugate of the vector and $\{f_j: j = 1,2, \ldots, n \}$ represents either concentration or velocity-perturbations. It is clear that if the set of vectors $\{f_j: j = 1,2, \ldots, n \}$ forms an orthogonal set, then $M$ is a unitary matrix and the condition number of $M$, denoted by $\mbox{cond}(M)$ must be $1$ or nearly $1$. But if any two eigenfunctions are non-orthogonal (they may be linearly independent), then $\mbox{cond}(M)$ can be a very large number. In such cases, the stability analysis investigated from the eigemodes is either incorrect or suboptimal \citep{Schmid2007}.

Fig. \ref{fig:Gramian_velocity} illustrates the change in $\mbox{cond}(M)$ with respect to $t_0$ and $k$ for $\alpha =1, \mu_m =2$ and $c_m=0.4$. It is observed that at early times the condition number is as large as of order, $\mathcal{O}(10^{10})$. This shows that at early times, the velocity eigenfucntions are not orthogonal, which leads to the disagreement between the onset of instability determine from NMA and SS-QSSA as depicted in Fig. \ref{fig:growth_comparison_chap5}(b). Moreover, the non-orthogonality tends to persists for longer period of time for small wave numbers.\\


\end{document}